# Optical Effects and Applications Associated with Photonic Crystal Materials

Alex Lonergan[1] and Colm O'Dwyer[1,2,3,4]*

[1]*School of Chemistry, University College Cork, Cork, T12 YN60, Ireland*

[2]*Micro-Nano Systems Centre, Tyndall National Institute, Lee Maltings, Cork, T12 R5CP, Ireland*

[3]*AMBER@CRANN, Trinity College Dublin, Dublin 2, Ireland*

[4]*Environmental Research Institute, University College Cork, Lee Road, Cork T23 XE10, Ireland*

## Abstract

The ability to selectively redirect specific wavelengths of light has attracted a lot attention for photonic crystal materials. Presently, there is a wealth of research relating to the fabrication and application of photonic crystal materials. There a number of structures which fall into the category of a photonic crystal; 1D, 2D and 3D ordered structures can qualify as a photonic crystal, provided there exists ordered repeating lattices of dielectric material with a sufficient refractive index contrast. The optical responses of these structures, namely the associated photonic bandgap or stopband, are of particular interest for any application involving light. The sensitivity of the photonic bandgap to changes in lattice size or refractive index composition creates the possibility for accurate optical sensors. Optical phenomena involving reduced group velocity at wavelengths on the edge of the photonic bandgap are commonly exploited for photocatalytic applications. The inherent reflectivity of the photonic bandgap has created applications in optical waveguides or as solar cell reflector layers. There are countless examples of research attempting to exploit these facets of photonic crystal behavior for improved material design. Here, the role of photonic crystals is reviewed across a wide a variety of disciplines; cataloguing the ways in which these structures have enhanced specific applications. Particular emphasis is placed on providing an understanding of the specific function of the tunable optical response in photonic crystals in relation to their application.

# 1 Introduction: Nanostructure, Biomimetics and Structural Coloration

Nanostructures are widely used in materials research to optimise the performance or functional characteristics associated with a particular material in a system. Nanostructures used in microelectronics led to the development of nanoelectronics with more components per chip, improved performances, lower costs and power consumption[1]. Many other fields of research have also exploited the structural properties of nanomaterials for enhanced material performances in areas such as photocatalysis[2], energy storage[3] and biomedicine[4]. A common advantage of nanomaterial structures is often the increased specific surface of the material compared to bulk counterparts, creating favourable geometries or increased surface reactions in many systems[2,5,6]. Precise control over nanostructure feature sizes and geometries allows certain material responses to surface plasmon, electrical and optical properties to be governed to some extent by the existence of the nanoscale order[7,8,9].

The optical properties of a structure are particularly susceptible to variation based on the particular ordering involved in the composition of the nanostructure. The symmetry, spacing and type of repeating structure are all important considerations for the optical appearance of the material. Nature provides many different and vivid examples of nanostructured materials and the interesting optical phenomena associated with structural order. Structural color is present throughout the natural world, where the colors associated with bird feathers, butterfly wings and beetle shells arise from nanostructured materials as opposed to varying material composition or dyes[10,11,12]. Studying the lattices and finer structure of these materials provides an insight into the appearance and striking coloration associated with naturally occurring nanostructured materials which can exist as one-, two- or three- dimensional repeating structures[13]. The variation in structural color, even between similar species, can be understood from the different types of nanostructured order present in the material[14]. Iridescence observations, where certain viewpoints or angles appear to influence a material's color appearance, can also be explained for materials featuring structural coloration through consideration of the light interaction and diffraction effects in these ordered materials[15,16].

In addition to structural color, certain naturally occurring phenomena can be understood by direct observation of the nanostructure associated with the material. One prominent example is the lotus effect[17].

The specific spacing of the material nanostructure heavily influences the wettability of many types of plant leaves[18] and other biological structures[19][20]. In the case of lotus leaves, a strong structural hydrophobicity is achieved through the nanostructure of the leaf in combination with the hydrophobic properties of the waxes on the surface of the leaf[21]. Figure 1 displays several examples of prominent nanostructured materials present in the natural world. The hydrophobicity of the lotus leaf is shown from the high water contact angle of a droplet sitting atop a lotus leaf surface in Fig. 1 (b). Structural coloration is demonstrated via the vibrant colors and iridescence effects shown for butterfly wings, peacock feathers, beetle shells, sea mouse spines, opal gemstones and begonia plant leaves. Microscopy examples of the fine structure of materials are displayed alongside images showing natural structural order. Figure 1 showcases just a sampling of the many instances of structural color occurring in the natural world.

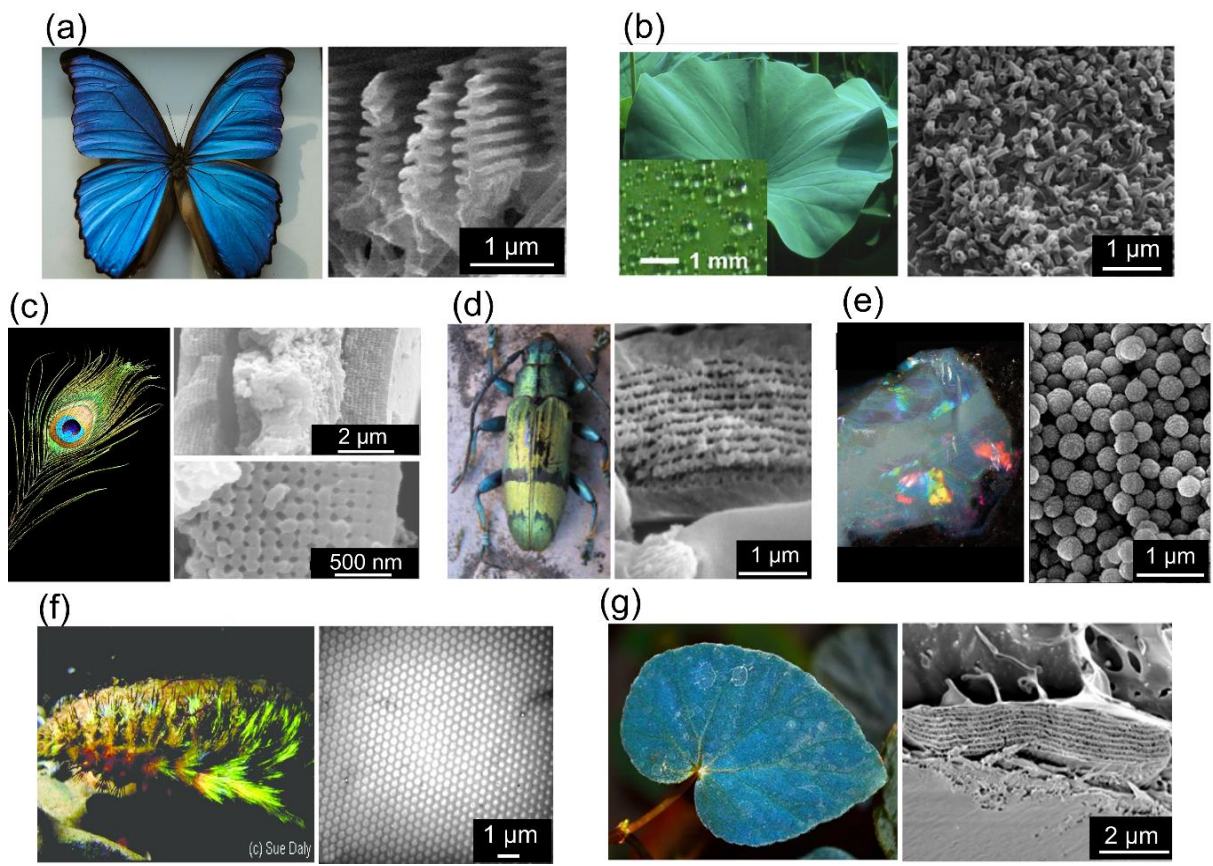

**Fig. 1** Stuctural colors and phenemona arising in nature are attributed to micro and nanoscale features of the surface. (a) Blue wings of the morpho didius butterfly wing[22] with a ridged structural order[23]. Reproduced from ref.[22]. Copyright 2018, Nature. Adapted from ref.[23]. Copyright IOP Publishing. (b) Hydrophobic properties of the lotus leaf[24] attributed to the waxy tubule structures on the upper leaf surface[25]. Adapted from ref.[24]. Copyright 2014, MDPI. Adapted from ref.[25]. Copyright 2011, Beilstein Institute for the Advancement of Chemical Sciences. (c) Bright colors of the eye of peacock feathers are associated with mealnin rods connected by keratin[26]. Adpated from ref.[26]. Copyright 2003, United States National Academy of Sciences. (d) Green shell of the Tmesisternus isabellae beetle composed of scales with alternating dielectric layers[27]. Adapted from ref.[27]. Copyright 2009, OPTICA. (e) Bright iridescent colors of the opal gemstone[28] with ordered spheres of silica creating the effect[29]. Reproduced from ref.[28]. Copyright 2018, Mineralogical Society of America. Adapted from ref.[29]. Copyright 2008, Mineralogical Society of America. (f) Colorful hairs of the seamouse[30] arise from structural holes filled with sea water[31]. Reproduced from ref.[30]. Copyright 2001, CSIRO Publishing. Adapted from ref.[31]. Copyright 2003, Elsevier. (g) Blue iridescence of begonia plant leaves is created from stacked layered structures in iridoplasts[32]. Adapted from ref.[32]. Copyright 2016, Nature.

Material designs or applications with inspirations from naturally occurring phenomena or structures can be classified as biomimetic systems. The field of biomimetics is concerned with understanding and emulating systems found in nature for the purposes of establishing biologically inspired designs, adaptions or derivations[33]. There are many examples of practical applications of natural biological structures. Biomimetic inspirations can be found in biological engineering[34], photonics[35], surface science[36], medical tissue engineering[37] and functional material designs[38]. Through a combination of surface morphology, nano/micro structural design and the chemical composition of materials, many different effects can be imitated from biological constructs. Controlled hydrophilicity, surface adhesion, mechanical strength, anti-reflection, thermal insulation, aerodynamic lift and fluid drag reduction are among some of the biomimetic properties of commercial interest[33].

For a specific mechanical structural application of biomimicry, nacre layers composed of calcium carbonate and nanostructurally integrated biomolecules (proteins and polysaccharides) create ceramic materials with high mechanical strength in bivalves such as molluscs[34,39] and are inspiring designs for biomedical implants due to their strong integration with bone tissue and excellent mechanical properties[40]. Many emerging super hydrophobic or self-cleaning material surfaces feature structural design morphologies with prominent biomimicry of the lotus leaf effect[24,41,42]. The fine nanostructure patterns found in moth eyes, a series of patterns and sub-wavelength bumps on the outer corneal lens, inspired the biomimetic development of anti-reflective coatings used to reduce reflections in surfaces such as those for solar cells[43,44,45]. Figure 2 demonstrates the close relation between the natural structure of materials and the biomimetic nanostructured materials they have inspired for some of the effects discussed here. Many more examples of biomimetic designs exist in the literature with nanostructure designs in the natural world providing a blueprint for imitation and application of specific effects in emerging technologies.

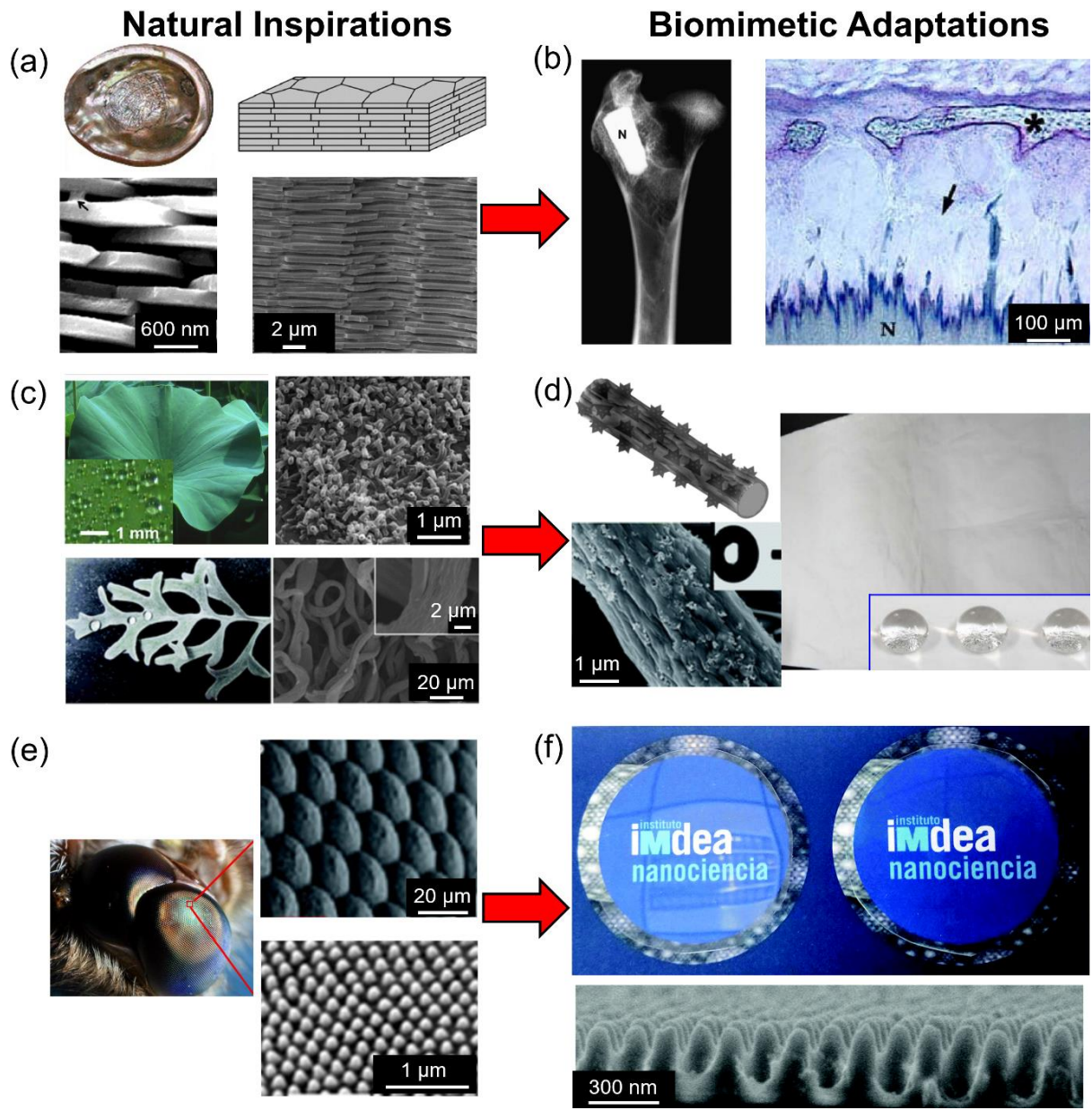

**Fig. 2** Examples of natural structural biomimetic applications. (a) The stacked structure of nacre, shown in a schematic and from SEM images[46 47], found in the red abalone. Adapted from ref[46]. Copyright 2007, Elsevier. Adapted from ref[47]. Copyright 2004, the American Chemical Society. (b) Biomimetic inspired and durable bone implant in sheep based on nacre structures, nacre is marked as N in the radiograph[48] and optical microscope image[49]. Reproduced from ref[48]. Copyright 1999, Elsevier. Adapted from ref[49]. Copyright 2005, Elsevier. (c) Nano-sized bumps on the lotus leaf[25] and the curved fibrous microstructure of the silver ragwort leaf[41]. Adapted from ref[25]. Copyright 2011, Beilstein Institute for the Advancement of Chemical Sciences. Adapted from ref[41]. Copyright 2011, Royal Society of Chemistry. (d) Leaf-inspired polystyrene fibers coated with silica nanoparticles to create a polystyrene fiber mat with excellent hydrophobic properties[41]. Adapted from ref[41]. Copyright 2011, Royal Society of Chemistry. (e) Cone-shaped micro and nanostructures found on moth eyes[50 51] create an anti-reflective effect. Adapted from ref[50]. Copyright 2008, SPIE. Adapted from ref[51]. Copyright 2018, Nature. (f) Moth eye inspired $TiO_2$-PMMA nanocomposite structural film[52] showing improved anti-reflective properties (right disc) versus a non-structured PMMA film (left disc). Adapted from ref[52]. Copyright 2018, Royal Society of Chemistry.

The concept of ordered structures and associated optical properties, such as structural color, are found in materials classified as photonic crystals (PhCs). From the first suggestions of alternating dielectric slabs and their applications to light manipulation[53,54], PhCs have been explored for their inherent optical abilities. The periodic structural order present in PhCs acts to attenuate specific photon frequencies from propagating through the structure. This control over photon frequencies is known as the photonic bandgap or photonic stopband of the structure. Depending on the degree of frequency depletion, complete photon inhibition is most commonly described by a complete or omnidirectional photonic bandgap[55] whereas partial or incomplete photon depletions are described by pseudo-photonic bandgaps or photonic stopbands[56,57]. The photonic bandgap is often presented as an optical analogue to the electronic bandgap associated with allowed electron energy levels in semiconductor materials[58]. The placement of atoms or molecules in semiconductor materials form a crystal lattice with a periodic potential which dictates the energy of electrons allowed to propagate through the semiconductor. Gaps in the energy band structure of the crystal lattice create forbidden electron energies in the material. For PhCs, the system of a periodic potential of atoms or molecules is replaced by a periodic dielectric function where different dielectric contrasts of larger ordered structures create similar effects for photons as seen with electrons. Much like in the case of semiconductor materials with band structures used to predict the electronic bandgaps of materials, the photonic bandgap can be predicted for materials through solutions of Maxwell's equations with specific knowledge of the lattice structure[58,59].

In addition to those occurring naturally, artificially structured PhCs are often manufactured and explored in relation to their ability to modulate certain frequency ranges of light. Artificial PhCs are architected in a variety of different structural forms depending on the application. In terms of the repeating nanostructure lattice, it is possible to form 1D (e.g. stacked dielectric slabs[60]), 2D (e.g. arrays of dielectric rods with an ordered arrangement or patterned dielectric slabs with air perforations[61,62]) and 3D PhC structures. Fully 3D PhCs can be designed as dielectric spheres in an air background (often called opals based on the arrangement of silica particles in an opal gemstone[63]), inverse opal[64] (IO) air spheres in a high dielectric background, a dielectric slab with holes drilled through at specific angles called a Yablonovite structure[65] or layered stacks of dielectric rods forming a woodpile structure[66]. Some of these common PhC constructs and designs are seen depicted in Fig. 3. All structures display a periodic dielectric contrast in their composition, an essential component for establishing the photonic bandgap property of the material. Similarities between some of the natural nanostructure designs seen in Fig. 1 can be seen for many of the artificially designed PhC structures.

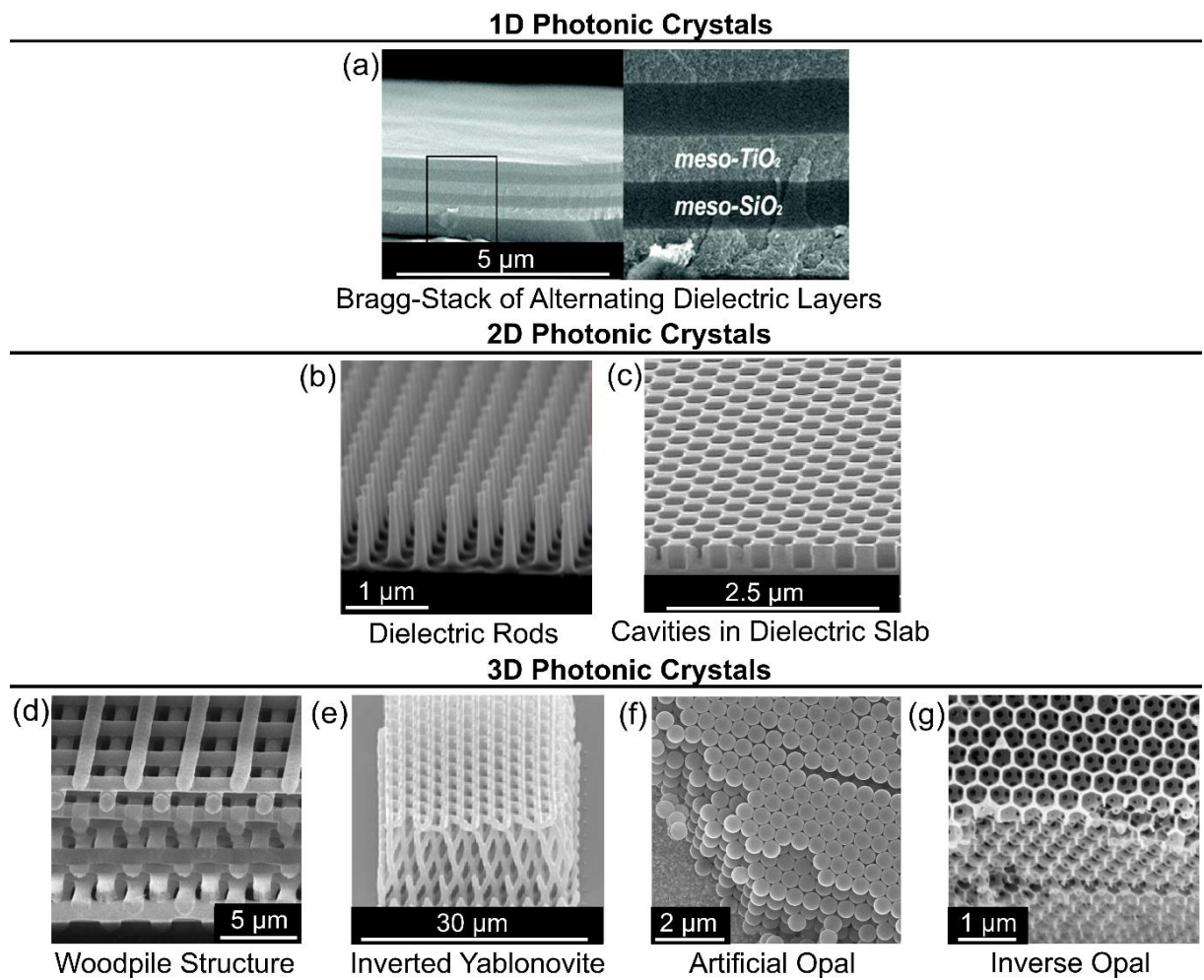

**Fig. 3** Examples of artificial PhC structures. (a) 1D PhC with alternating $TiO_2$ and $SiO_2$ layers prepared via spin coating[67]. Adapted from ref[67]. Copyright 2006, American Chemical Society. (b) 2D PhC of dielectric rods of silicon in air background[68]. Adapted from ref.[68] Copyright 2010, OPTICA. (c) 2D PhC consisting of ordered air cavities in a silicon background[69]. Adapted from ref[69]. Copyright 2006, Nature. (d) 3D PhC of a stacked woodpile configuration of a polymer mixture of poly(acrylic acid) and poly(ethylenimine)[70]. Adapted from ref[70]. Copyright 2006, Wiley-VCH. (e) 3D PhC of an inverted Yablonovite structure of silica[71]. Adapted from ref[71]. Copyright 2012, Springer. 3D PhCs showing ordered stacks in an FCC configuration for (f) artificial polystyrene opals and (g) a $TiO_2$ IO structure.

Light interactions with specific and ordered PhC planes can be predicted based on light diffraction, using the Bragg-Snell model to project diffracted wavelengths from the photonic stopbands of structures. Using this model as a basis, maximum wavelengths of reflection from the structure are predicted for the maximum constructive interference condition in the reflected wave. The Bragg-Snell treatment combines Bragg's law of diffraction for x-ray diffraction in crystal lattices with Snell's law of refraction for light incident on media with different refractive indices. The Bragg-Snell law used for PhC structures with alternating dielectrics of refractive indices $n_1$ and $n_2$, a light incidence angle on the entire structure of $\theta_1$, an interplanar spacing of $d$, a resonance order of $m$ and wavelength of maximum diffraction $\lambda$ can be stated as follows:

$$m \lambda = 2 d \sqrt{n_{eff}^2 - n_1^2 \sin^2\theta_1} \qquad (1)$$

A more detailed explanation of the origins of certain parameters and the derivation of this equation can be found in Appendix I. Looking at the Bragg-Snell equation, the maximum wavelength of diffraction is dependent on many factors associated with the structural composition of the PhC. The interplanar spacing ($d$), factors in the periodicity of the material and the overall geometry of the crystalline structure. The refractive indices ($n_1$ & $n_2$) incorporate the optical properties of the alternating dielectric materials and control the magnitude of $n_{eff}$, the effective refractive index. By this relation, changes to any of these parameters should result in changes to the wavelength of maximum diffraction from the PhC structure. This is important for both interpreting and understanding the optical properties of the nanostructured materials and also design and development of an optical application using PhCs. Control over the size or constituent materials of the structure enables the optical response of the material to be tuned over specific wavelength ranges[72][73], if desired. The tuning of the optical response of the material is an important consideration in the development of many artificial PhC structures for use in high sensitivity systems such as PhC display screens[74][75] or sensors[76][77]. Knowledge of the relation between nanostructure and optical behavior has furthered the understanding of many natural observations such as material iridescence or color changes associated with liquid infiltration[20][78]. Figure 4 displays some of the changes observed to structural coloration for both natural and artificial PhCs which can be understood through changes to the wavelength of maximum diffraction of the Bragg-Snell relation.

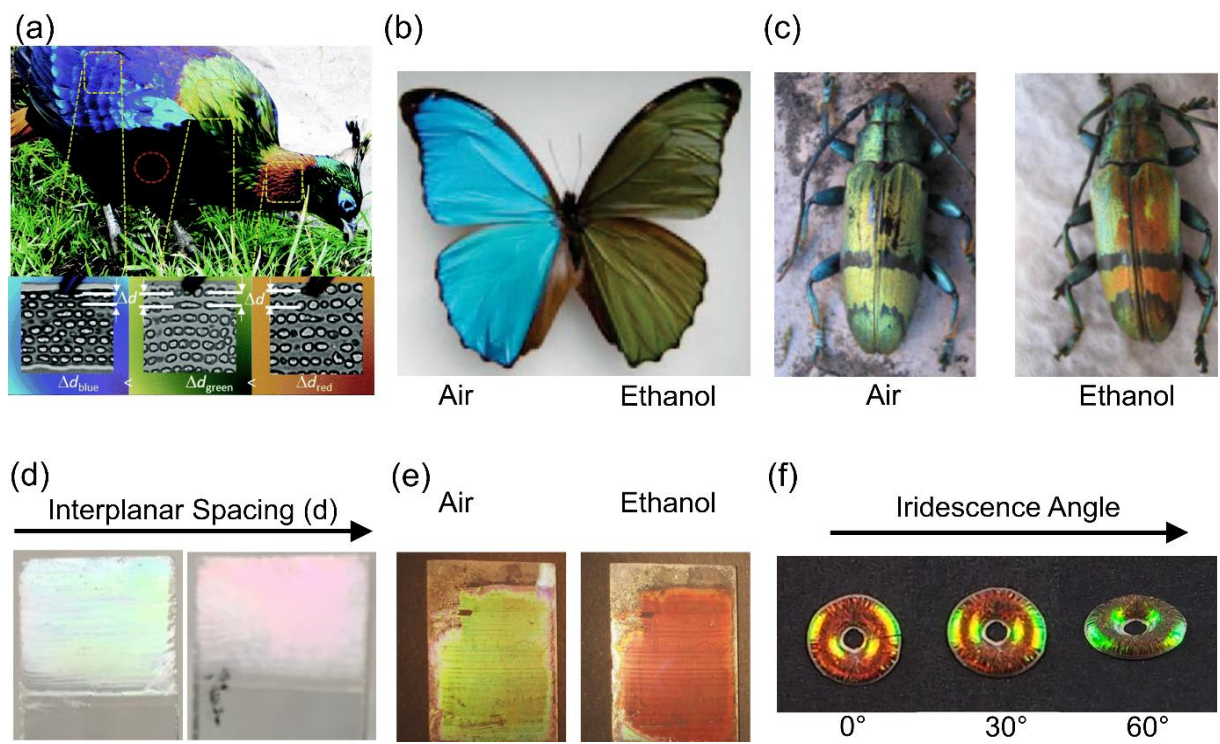

**Fig. 4** Changes to structural coloration associated with periodicity, material composition and viewing angle for several PhC structures. Shorter structural color wavelengths are reflected for smaller interplanar spacing periods shown for (a) the red, green and blue feathers of the Himalayan monal[79] and (d) $TiO_2$ IOs of diameters 384 nm (pink/red) and 278 nm (green). Adapted from ref.[79]. Copyright 2020, Royal Society of Chemistry. Shifts to longer wavelength structural colors are displayed for the replacement of air with ethanol in (b) Morpho menelaus butterfly wings[80] changing from blue to green. Adapted from ref.[80]. Copyright 2009, Nature. (c) Tmesisternus isabellae shells[27] changing from yellow/green to orange. Adapted from ref.[27]. Copyright 2009, OPTICA. (e) $ZrO_2$ IOs (287 nm diameter coated with 5 wt% Au) shifting from a green to orange color[81]. Adapted from ref.[81]. Copyright 2018 American Chemical Society. (f) Irisdescence is observed for 285 nm diameter polystyrene spheres coated with thin (2.5 nm) polydopamine shells[78], showing a shift to shorter wavelengths with increasing incidence angle. Apapted from ref.[78]. Copyright 2016, Nature.

In terms of the optical response of these materials, understanding and predicting the light interaction process is critical when developing applications centred around the structural color of the material. Advancements in the construction of colloidal particle templates has allowed for greater order and flexibility in the design of PhCs based on colloidal particle templating, such as opals and IOs[82] [83]. Feature sizes, defects and material choice all act to influence the signature optical response of these materials. Tuning the optical response of prepared PhCs post-production has also been studied extensively, showing that factors such as stretching deformation[84] or solvent infiltration[85] among many others have important considerations for the structural color of the material.

Knowledge of the flexibility of the photonic stopband and the associated structural color of the material creates specialised applications involving the wavelength of light reflected from the structure. Electrical tuning of an opal template with infiltration of a metallo-polymer substance has been demonstrated to tune the wavelength reflected from the structure and create a tunable photonic response, with potential applications in PhC color displays[86] for substances dubbed photonic inks[87]. This wavelength shift is attributed to expansion and contraction of layer spaces in the material due to the swelling of the infiltrated material with an applied voltage. The magnetic response of the photonic bandgap of superparamagnetic colloidal nanoparticles has been exploited to develop anti-counterfeit PhC labels where the structural color of the anti-counterfeit strip is dependent on the applied magnetic field[88] [89]. Figure 5 illustrates the electrical and magnetic tuning effects of the structure utilised in these applications. In these examples, the application of the PhC is directly linked to the structural color and the visual appearance of the material emphasising the importance of understanding the effect of the photonic stopband and the factors which affect wavelengths reflected from the film surface.

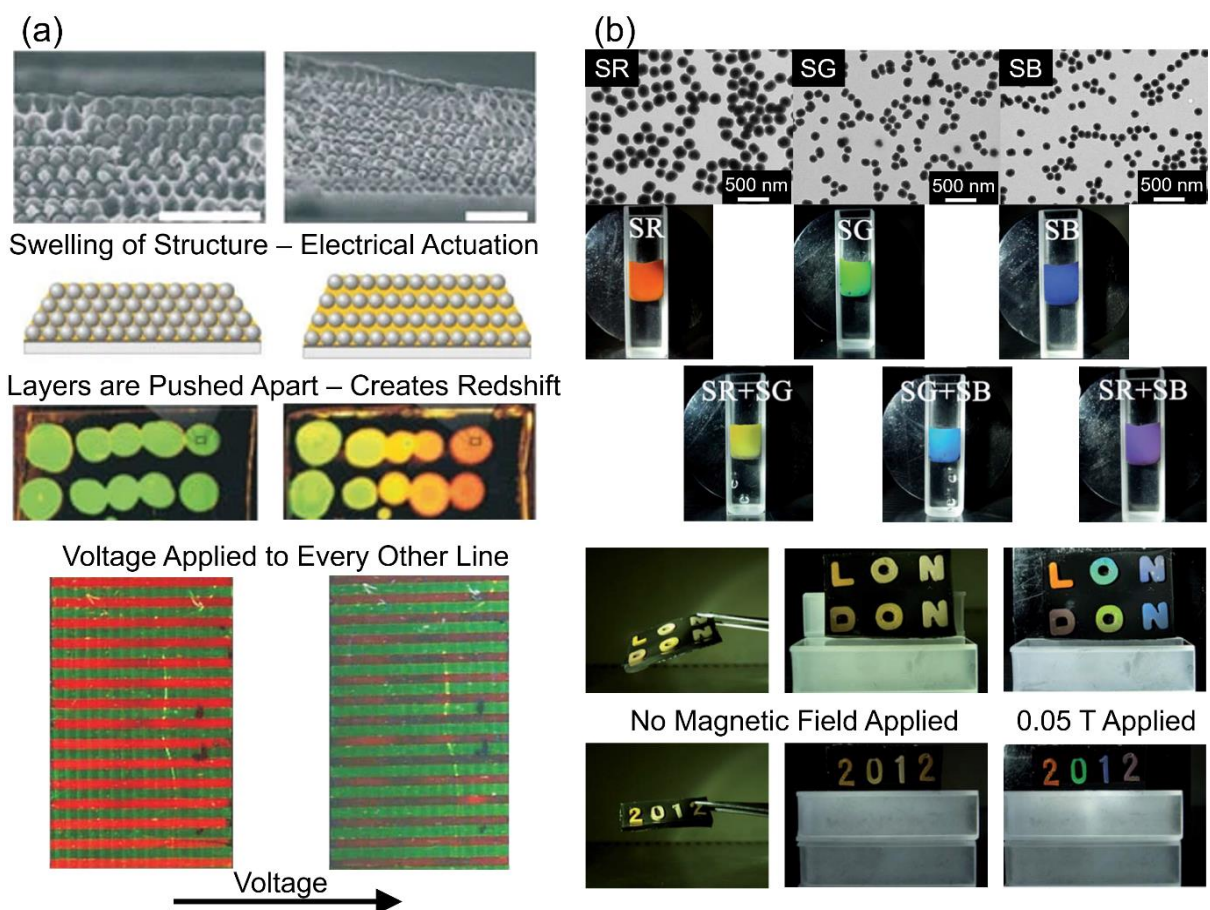

**Fig. 5** Applications of PhCs centred around the visual appearance of structural color. (a) Electrical actuation of silica spheres in a cross-linked network of polyferrocenylsilane is shown to induce a redshift for structural swelling and the application of this technique to color display screens is demonstrated via a voltage induced redshift to the structural color of the PhC composite on indium tin oxide layers[86]. Adapted from ref.[86].

Copyright 2007, Nature. (b) Magnetic structural color tuning of superparamagnetic colloidal nanoparticles, achieved through magnetic field particle ordering, is shown for an applied magnetic field of 0.1 T for particles sizes SR = 200 nm, SG = 140 nm and SB = 125 nm with structural color tuning demonstrated for anti-counterfeiting purposes where an applied magnetic field is a requisite for a visible color display[88]. Adapted from ref.[88]. Copyright 2012, Royal Society of Chemistry.

Research into applications involving the structural color of PhC films is continually innovating with new advancements steadily emerging from the literature. The concept of tuning the light reflected from the film surface creates a fascinating application where the structural color detected visually can selectively be altered for an observer. Figure 6 details some of the recent studies which utilise the structural color effect of PhC films in their primary application. The work shown in Fig. 6 (a) is a prime example of biomimicry where the dermal layers of a gecko inspired the design for a temperature dependent PhC film with tunable color and transmittance intensity by incorporating thermochromic and electrochromic pigments into the film. The environmental temperature was found to affect the color of the film and an applied voltage affected the transmittance of the film to light, with a suggested application to smart windows with a variable degree transmittance linked to an applied voltage[90]. Controllable PhC film iridescence, where structural color appears to vary with viewing angle, has been explored for images printed with colloidal photonic inks, as depicted in Fig. 6 (b). For $SiO_2$ particles dispersed in acrylate-based resins, the relative degree of viscosity of the dispersion resin was found to impact the vibrancy and iridescence of structural color for the printed images with broader reflection peaks and less iridescent images found to arise from higher viscosity resins[91].

Automotive paint coatings based on $SiO_2$ photonic crystals in a solvent medium spray coated onto surfaces and re-coated with a commercial clear coat are shown in Fig. 6 (c). The resultant PhC films displayed vibrant colors which could be tuned by the $SiO_2$ particle size or by using combinations of different sizes to achieve films of mixed colors. Films were resistant to mechanical and chemical stress tests and exhibited high structural color iridescence making them attractive candidates for automobile paint[92]. Figure 6 (d) illustrates the design for a finger motion-sensing device based on a 1-D PhC design of an interpenetrated hydrogel network block copolymer with alternating water absorbable and non-absorbable lamellae which was found to be sensitive to relative humidity[93]. The expansion of the hydrogel network in response to relative humidity caused a red-shift in the structural color of the film and the natural humidity of the human finger was capable of being detected by the PhC film depending on the relative degree of proximity and motion. Arrays of these PhC films could even be used to detect the motion of a human hand using the same principle, creating a motion sensor based on the visual structural color appearance of the PhC film.

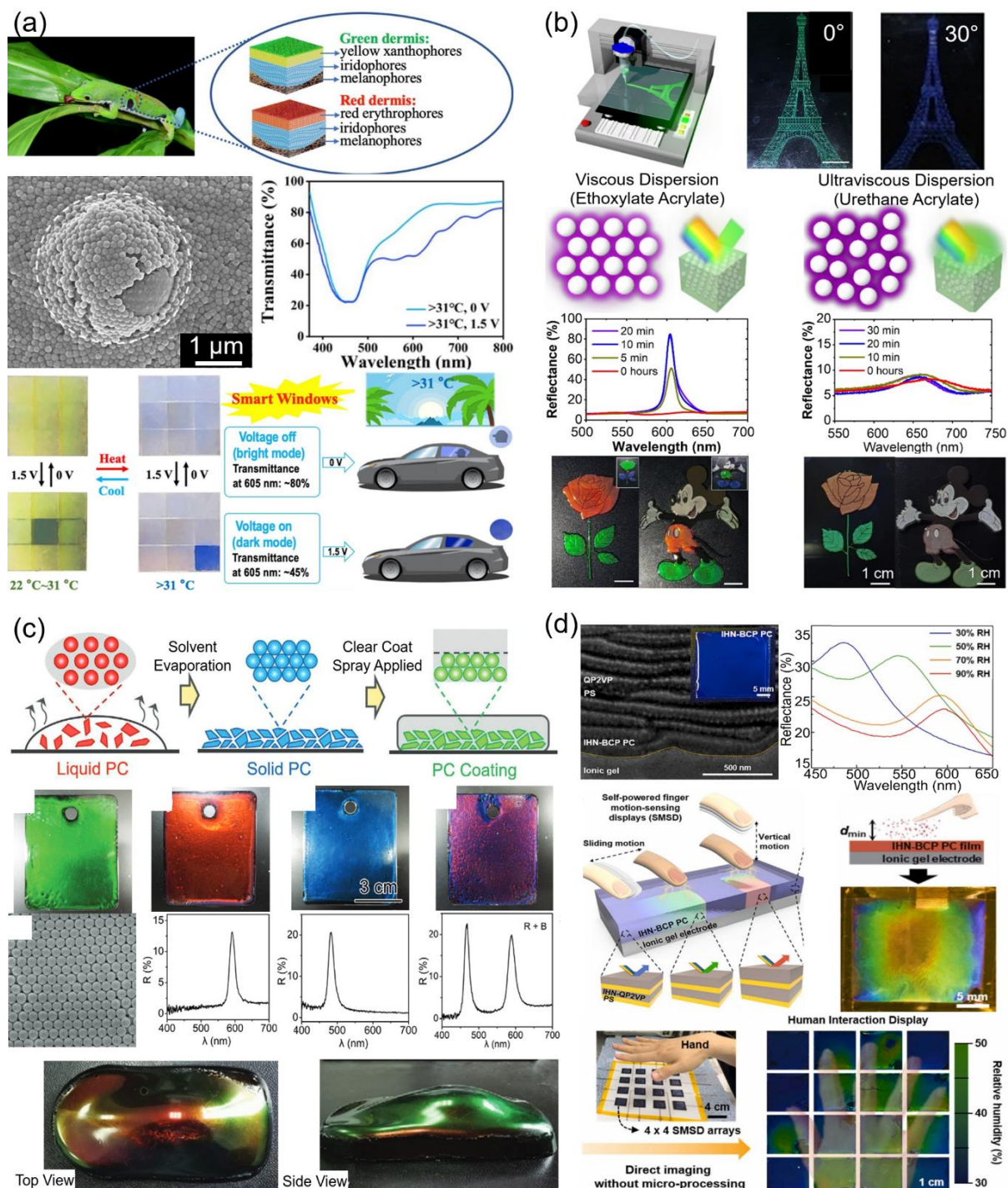

**Fig. 6** Emerging applications of PhCs with a dependence on their structural color. (a) Temperature-dependent thermochromic microcapsules incorporated into a PhC film inspired by layers in gecko dermis[90]. Structural color was dependent on temperature and film transmittance was affected by applied voltage, with a proposed application for smart windows. Adapted from ref[90]. Copyright 2022, Elsevier. (b) Colloidal photonic inks used to print images with controllable degrees of iridescence depending on the relative viscosity of the particle dispersion medium[91]. Adapted from ref[91]. Copyright 2021, American Association for the Advancement of Science. (c) Easily spreadable and durable PhC films deposited by spray coating $SiO_2$ particles dispersed in solvent with a proposed application for automotive coatings[92]. Adapted from ref[92]. Copyright 2021, Wiley-

VCH. (d) A 1-D PhC film of a block copolymer with an interpenetrated hydrogel network showing structural color sensitivity to relative humidity for use a finger motion-sensing display by utilising the natural humidity of a finger[93]. Adapted from ref[93]. Copyright 2022, Elsevier.

There are, of course, other emerging applications involving PhC films which hinge on the observation of a structural color. The vivid colors and iridescence association with PhC films makes them particularly appealing for decorative designs for textiles[94,95,96,97], photonic glasses[98,99,100] and even wood surfaces[101,102]. The freedom to tune the photonic response and structural color in PhCs also makes them suitable for encoding information, often through controlled wettability[103,104,105,106] using a solvent, for applications involving anti-counterfeiting measures[107,108,109]. Aside from the applications involving direct observation of a change to the structural color of a PhC, many research fields have specifically adopted PhCs for their ability to direct and alter the optical properties of structured materials. Nanostructures designed and applied to materials have been exploited for the photonic effects induced on the structure.

The proceeding sections will detail and review the mechanism of operation and application of PhC structures as applied to common research interests which specifically exploit and utilise the optical properties of the photonic bandgap or stopband in their application. Of course, other physical properties of these structures are often cited for their uptake and application to certain research areas and these will be discussed also. These would include an increased exposed surface area for improved electrochemical performance[110,111] or photocatalysis reactions[112,113] and the advantages of the scaffold architecture of these ordered structures to biomedical research areas such as tissue engineering or regenerative medicine[114,115]. For this work, emphasis is placed on the importance of the optical response of these ordered photonic materials by selectively examining applications with fundamental relations to the optical performance of the structured material. More specifically, the optical properties of PhCs will be examined in relation to their implementation in material sensor devices, solar cell technology, photocatalytic materials and waveguides in optical fibers. Principally, the design, understanding and valid interpretation of the optical response of these materials will be shown to be crucial for the incorporation of PhCs in these areas of research.

## 2 Photonic Crystal Materials as Optical Diagnostic Sensors

One of the most prevalent applications of the optical behavior of PhC materials involves their ability to act as optical sensors, assessing a range of different material properties using the signature optical response of the ordered structure. The Bragg-Snell relation, as seen in eqn 1, allows wavelength shifts in the maximum wavelength position of diffracted light from the material to be predicted and interpreted. Optical sensors exploit the wavelength shift of the photonic bandgap or stopband to identify and assess changes to the material or material environment using the optical response of the structure. Typically, wavelength shifts in these structures are brought about by changes to the effective refractive index ($n_{\text{eff}}$) or the interplanar spacing ($d$) of the structure. In alternating dielectric materials of refractive indices $n_1$ and $n_2$ occupying volume fractions $\varphi_1$ and $\varphi_2$, the effective refractive index $n_{\text{eff}}$ of the structure is commonly estimated by[73][81][116][117]:

$$n_{\text{eff}} = n_1\, \varphi_1 + n_2\, \varphi_2 \qquad (2)$$

Combining this relation with eqn 1, an expression for all the factors influencing the maximum wavelength position of the photonic stopband is given by:

$$m\, \lambda = 2\, d\, \sqrt{(n_1\, \varphi_1 + n_2\, \varphi_2)^2 - n_1^2 \sin^2\theta_1} \qquad (3)$$

From eqn 3, the position of the photonic stopband is dependent on the interplanar spacing between layers, the angle of incidence and both of the refractive indices of the alternating dielectric materials. Porous materials, where the alternating dielectric consists of a high index material in a low index air background, are particularly suited to exploit this wavelength dependence on both the high index material ($n_2$) and the low index background voids ($n_1$) for optical sensor applications. Wavelength shifts associated with the photonic stopband can be used to identify materials which can freely occupy void space in the structure, effectively changing the refractive index contrast of the PhC. Likewise, engineered changes to the interplanar spacing or refractive index of the high index material can also be used to diagnostically assess changes to the material system.

In terms of acting as an optical sensor, perhaps the simplest application of this analysis arises from material characterisation for artificial opal and IO structures. For a known material, the wavelength position of the photonic stopband can be used to determine the size dimensions of the ordered structure from the interplanar spacing. This optical dimensional analysis has been frequently shown to accurately determine the size of the repeating structure, as compared with microscopy measurements[73][118][119]. The appearance of the photonic stopband can also be used as an indicator of the degree of monodispersity present in the sample, with stronger scattering and lower optical transmittance associated with higher disorder and lower monodispersity[120][121].

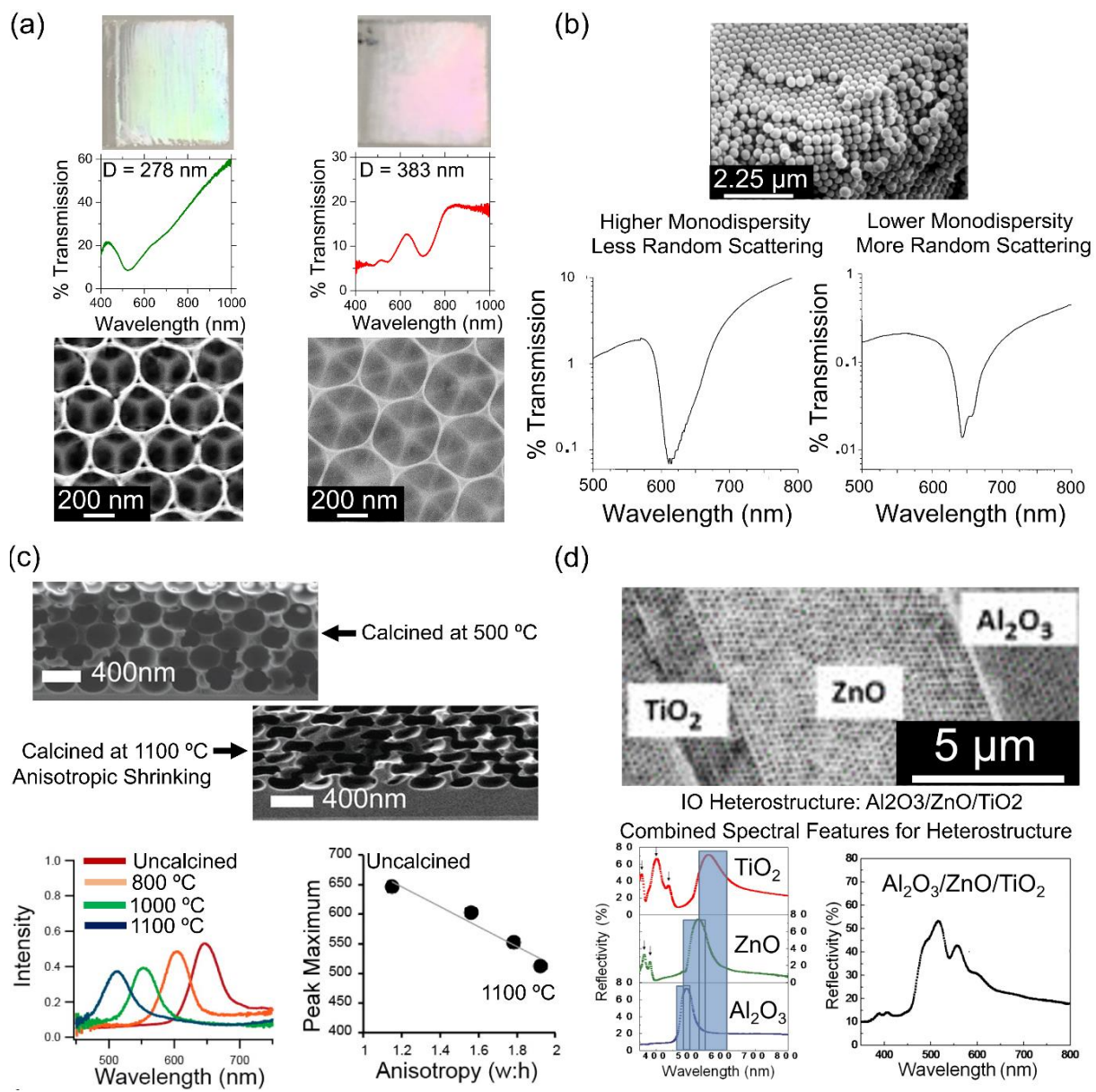

**Fig. 7** Examples of material characterisation and assessment of PhCs using the optical stopband. (a) Two TiO$_2$ IOs with measured centre-to-centre pore distances of 278 nm (green) and 383 nm (red) can be distinguished by the different positions of the optical stopband[122]. Adapted from ref[122]. Copyright 2018, American Institute

of Physics. (b) The degree of scattering and transmittance intensity of the optical spectrum of two identically thick $SiO_2$ opals are shown to indicate relative monodispersity in the sample[121]. Adapted from ref[121]. Copyright 2001, Wiley-VCH. (c) The optical stopband position indicates anisotropic shrinking of an $SiO_2$ IO related to the calcination temperature upon inversion[123]. Adapted from ref[123]. Copyright 2014, American Chemical Society. (d) The optical spectrum of an $Al_2O_3$/ZnO/$TiO_2$ IO heterostructure is shown with identifying features of the optical stopband arising from each individual IO structure[124]. Adapted from ref[124]. Copyright 2009, American Institute of Physics.

IOs prepared using sol-gel methods from artificial opal templates, are known to feature reduced lattice constants upon heating and inversion relative to the initial opal template[117][125][126]. An optical assessment of the IO dimensions can provide a quick means to quantify any shrinking upon inversion. For 3D PhC structures, the information from the photonic stopband is relevant for the entire 3D surface interacting with incident light, whereas scanning electron microscopy assessment is largely limited to surface information. The photonic stopband is therefore ideal for non-destructively assessing any 3D structural anisotropy which might be present in the material[123]. For known size dimensions, the optical information of the photonic stopband can, in principle, be used to assist in identifying the material comprising the PhC network. The refractive indices of materials, as determined from the optical stopband analysis, have been frequently used, in conjunction with other analytical methods, to determine the crystalline material of IO structures, even for composite or heterostructure materials[124][127][128]. Figure 7 provides an overview of some of the optical sensing and material characterisation related to photonic stopband analysis.

In a broader sense, the optical sensing ability of PhC materials extends far beyond simple material characterisation of the structures themselves. With proper calibration and knowledge of the optical spectrum, PhCs have been demonstrated to operate as excellent sensors for a range of different chemical substances. The photonic stopbands of $SiO_2$, $ZrO_2$, $TiO_2$, $SnO_2$, $CeO_2$ and $Al_2O_3$ IOs have exhibited a predictable and reversible shift of the principle diffraction peak upon addition of various solvents to occupy voids in the structures[73][116][117][129][130]. The refractive index of the solvent and the Bragg-Snell relation proved to function as an excellent predictor of the wavelength shift observed for the photonic stopband. The transmission dips associated with the stopband red-shifted to longer wavelengths with increasing solvent refractive index. This spectral shift with solvent refractive index creates a sensing application, where the wavelength position of the photonic stopband in a properly calibrated structure can be used to determine the refractive index of a liquid infiltrating the PhC.

The sensitivity of this type of approach is particularly appealing for identifying mixtures or solution concentrations. $TiO_2$ IOs have been developed that are sensitive to various concentrations of ethanol solutions[131], with reported variations in refractive indices smaller than 0.002 yielding shifts in the stopband position greater than 1 nm[132]. Similar work for ethanol solutions in $WO_3$ IOs confirm the ability to optically sense the concentration of ethanol in binary ethanol/water solutions[133]. Solvent/water mixtures and their concentrations are particularly well studied for photonic crystal detection systems[131,134,135]. More recently, some works have focused on developing photonic crystal sensors which offer a visual structural color indicator for the presence and concentration of particular solvents[136,137,138], with even the optical image intensity and peak absorbance linked to ethanol concentrations for $SiO_2$ IOs[139]. Similarly, carbon IOs have demonstrated promising oil sensing capabilities with excellent response times and reversibility of the photonic stopband[140]. The visual structural color of oil/gasoline and ethanol/gasoline mixtures has been exploited to indicate mixture concentrations in $SiO_2$ IO coated with hydrophobic fluoroalkylchlorosilanes[141]. In this case, the appearance of the structural color is dependent on the wettability of the coated IO with respect to different concentrations of oil/gasoline or ethanol/gasoline.

Photonic stopband shifts are not limited to just liquid detection. Gas and vapor sensors have been developed using the principle of the spectral shift in PhC materials. Visual detection of volatile aromatic hydrocarbon vapors was achieved using a 1-D photonic crystal in which layers would swell, inducing a shift in structural color, when detecting compounds like benzene, toluene or xylene[142]. $SiO_2$ IOs coated with polyvinyl alcohol display a spectral red shift with respect to the relative humidity in water vapor[143]. The polyvinyl alcohol acts to swell in the presence of water vapor, altering the space between $SiO_2$ layers and leading to the red-shift. A visual spectral red shift of the photonic stopband of over 50 nm was recorded for $SiO_2$ IOs infiltrated with tetraphenylethene polymer in the presence of organic vapors of tetrahydrofuran and acetone[144]. This red shift is attributed to an increase in the effective refractive index of the structure caused by adsorption of the organic vapor by the polymer substance. Similar applications of polymer infiltrated $SiO_2$

IOs have been developed for detecting aromatic volatile organic compounds such as xylenes[145]. Likewise, spectral red shifts are proposed to arise from increases to the effective refractive index of the structure caused by adsorption of organic vapor by the polymer. Figure 8 illustrates the stopband sensing capabilities for various IO structures to different analytes.

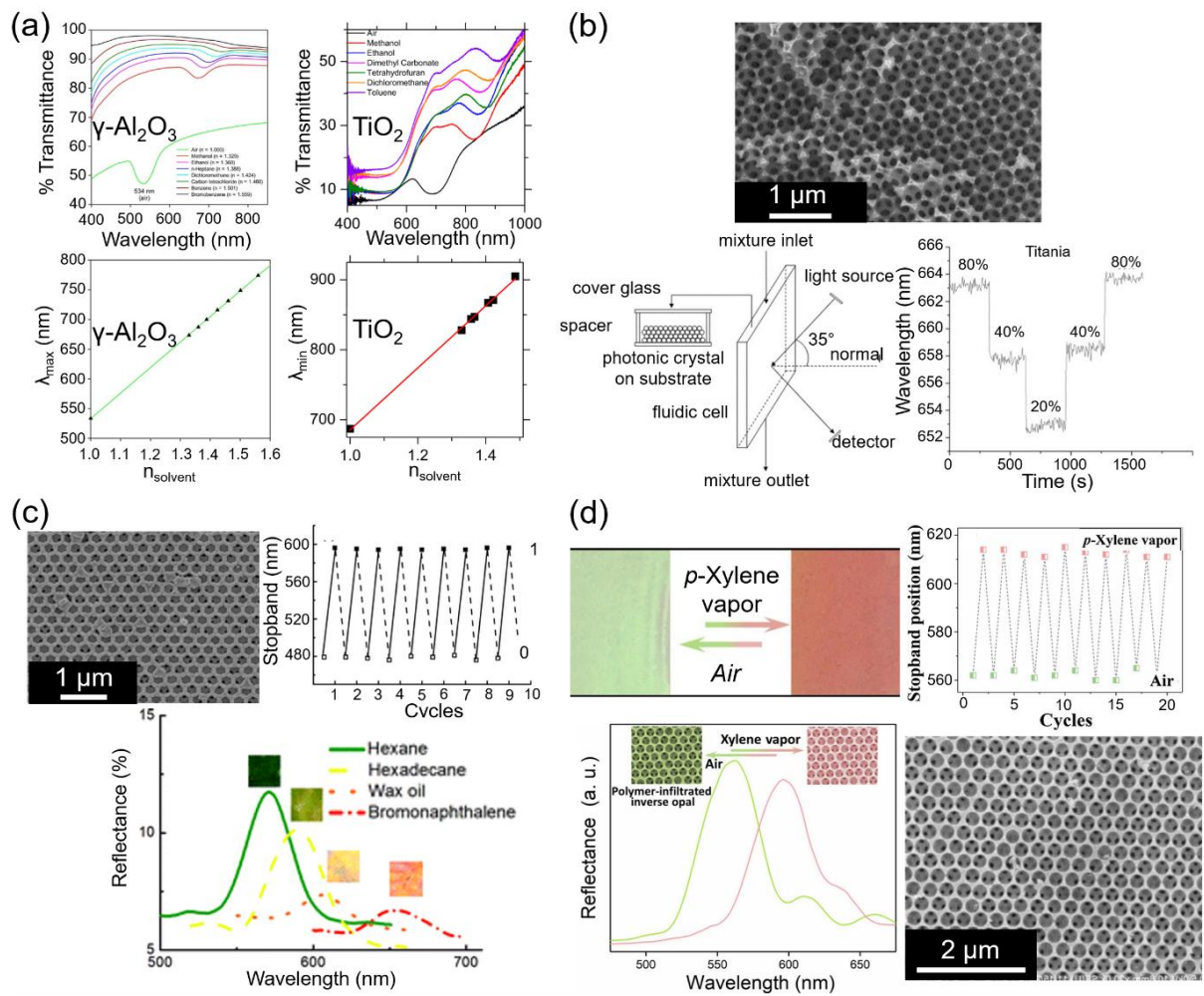

**Fig. 8** Applications of the photonic bandgap in solvent and vapor sensing. (a) The optical stopbands for a γ-$Al_2O_3$[117] and $TiO_2$[129] IO red-shift continually with the addition of higher refractive index solvents. Adapted from ref[117]. Copyright 2015, American Chemical Society. Adapted from ref[129]. Copyright 2020, American Physical Society. (b) Design of a $TiO_2$ IO ethanol sensor where the shift in stopband can be related to ethanol concentration in ethanol/water mixtures[132]. Adapted from ref[132]. Copyright 2007, Elsevier. (c) A carbon IO using the photonic stopband position to distinguish between different oils with strong recyclability between test cycles[140]. Adapted from ref[140]. Copyright 2008, Royal Society of Chemistry. (d) A polymer infiltrated $SiO_2$ IO calibrated to display a visual colorimetric response of the photonic stopband in the presence of xylene vapor with promising recyclability of the sensor across detections[145]. Adapted from ref[145]. Copyright 2019, Elsevier.

Specialised applications of PhC structures are present in PhC optical fibers, commonly used as sensitive refractive index sensors. PhC optical fibers and waveguides will be covered in more detail in Section

5. For now, emphasis will be placed on their ability to distinguish between small variations in refractive index. 2D PhC slabs with air holes etched in a square or triangular lattice into a slab of dielectric material, such as seen depicted in Fig. 3 (c), are commonly used as the basis for a photonic waveguide. In simple terms, a waveguide is created when certain rows of the periodic structure are removed, e.g. air holes are selectively not-etched into specific row(s) of the dielectric material. Light propagating in this region is confined by the photonic bandgap of the surrounding periodic structure in the plane of the PhC and total internal reflection effects of the optical fiber out of the plane of the plane of the slab. Defects introduced to the PhC regions act to create allowed modes for photon propagation in the photonic bandgap region of the structure. Defects can be introduced by removing holes or changing their radii in the repeating structure. The newly allowed photon propagation frequencies create an optical cavity in the defect region, introducing an optical confinement effect. Spectrally, these optical cavities appear as resonances with sharp dips/peaks on the optical spectrum of the PhC fiber. The wavelength position of these spectral features is strongly linked to the refractive index of the surrounding medium, through evanescent waves of the optical cavity interacting with the neighbouring material of the fiber.

The sensitivity of the resonant mode frequencies to the surrounding material has been exploited in PhC fibers for refractive index sensing, as seen in Fig. 9. One or more of the air cavities in the PhC structure can be filled or continually pumped with an analyte which can be assessed via changes to the resonant mode wavelength. Calibration of a particular sensor design for multiple refractive indices should allow for accurate determination of analyte materials based on their refractive indices. The design of the optical cavity and waveguide will determine the spectral properties of the sensor. A seven-cavity (L7) wide defect in a silicon-on-insulator wafer has been demonstrated to feature a 12.65 nm shift in the resonant mode wavelength position when switching between water and ethanol, constituting a significant wavelength shift for a small refractive index difference of just 0.027[146]. Salinity PhC fiber sensors, designed for detecting salt concentration in sea water, have been shown to detect salt concentrations in water via shifts in the resonant wavelength of confinement losses corresponding to minor shifts in the refractive index of salt water solutions ranging from 1.3326 to 1.3505 (0 – 100%)[147].

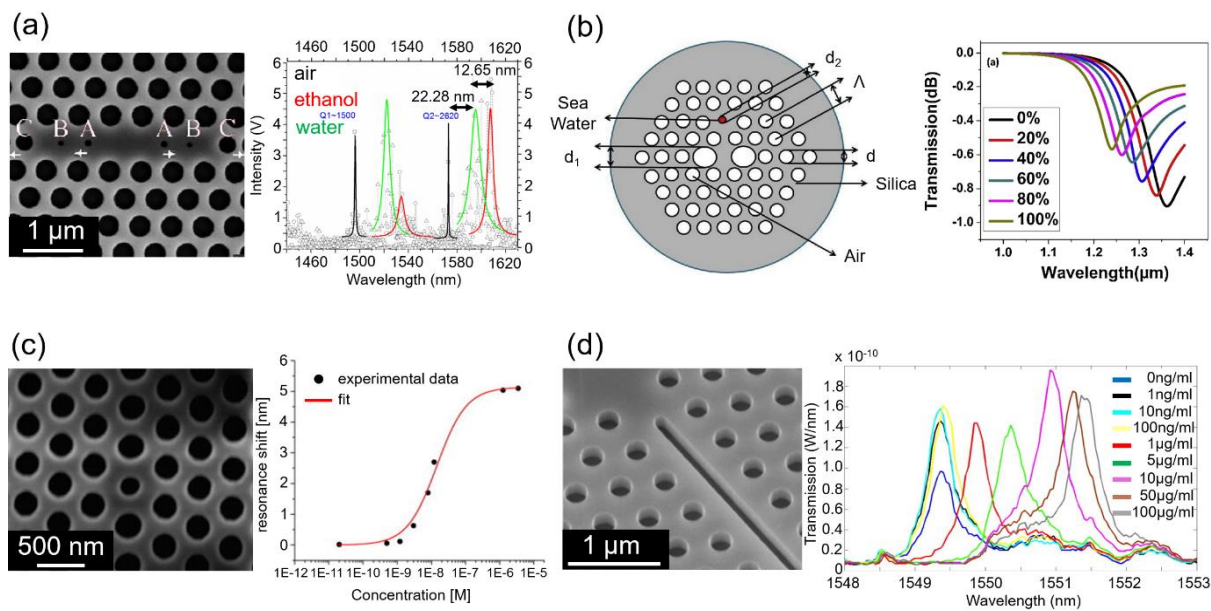

**Fig. 9** PhC fiber sensor designs for detecting a range different substances. (a) A seven cavity wide microcavity in a silicon-on-insulator wafer displaying a wavelength shift of 12.65 nm for a refractive index difference of 0.027 between water and ethanol[146]. Adapted from ref[146]. Copyright 2012, OPTICA. (b) Air holes patterned into a silicon fiber sensor used to detect relative salinity in seawater[147]. Blue-shifts in the confinement loss spectra correspond to higher salinity levels. Adapted from ref[147]. Copyright 2018, Elsevier. (c) A silicon-on-insulator PhC fiber with a smaller central cavity for light confinement is used to monitor anti-biotin levels in bovine serum by measuring the shift in resonance position with concentration[148]. Adapted from ref[148]. Copyright 2009, Elsevier. (d) A slotted PhC cavity on a silicon-on-insulator substrate monitors dissolved avidin concentrations using the shift in cavity resonant wavelengths[149]. Adapted from ref[149]. Copyright 2011, Elsevier.

Biological processes such as protein binding can be monitored over time using this change in refractive index of the resonant modes. The binding of anti-biotin to biotinylated-bovine serum albumin was monitored over time using the shift in resonance wavelength of the PhC fiber as an indicator of the anti-biotin concentration with detection sensitivity estimated at just 20 pM of anti-biotin concentration[148]. Similarly, PhC cavities functionalised with biotin were used to detect dissolved avidin concentrations by monitoring resonant wavelength shifts with distinct peaks shifts observed at concentrations of 1 μg/mL[149]. PhC sensors can also be specifically designed to detect a plethora of other material parameters, including temperature, curvature, pressure, strain or electric field[150][151].

PhC hydrogels are increasingly common sensors designed to monitor and track changes to specific polymer infused PhC structures with a wide array of applications, particularly in biological and medical fields. Conceptually, the operation of these hydrogel systems is similar to the process of electrical actuation depicted in Fig. 5 (a), where applied voltages increased the separation between PhC layers. Hydrogel systems used for optical sensing applications are typically designed such that the signature optical wavelength of the photonic stopband shifts in response to a swelling or contraction of the polymer medium upon exposure to certain chemicals or environmental conditions. Principally, it is the shift in the photonic stopband that allows a monitoring of the hydrogel material. The hydrogel polymer is selected to react with specific chemicals or biomolecules, creating a distinct optical response indicating the specific presence of reactant species.

Chemically, PhC hydrogels have been used as sensors to detect a number of different system properties. As previously discussed, humidity sensors were developed using polyvinyl alcohol embedded in $SiO_2$ layers[143]. Polyvinyl alcohol hydrogel sensors have also been developed for pH detection, displaying a 350 nm wavelength shift of the photonic stopband position for pH values varying between 3.3 and 8.5[152]. The hydrogel responds to increasing pH with an increase in the osmotic pressure of the polymer created when carboxyl groups are ionised and counter ions become immobilised in the hydrogel. The increase in osmotic pressure acts to swell the hydrogel, increasing the separation between polystyrene sphere layers and red-shifting the photonic stopband. This method of pH sensing was also found to apply to polyacrylamide hydrogels, where the swelling and shrinking of the gel caused by osmotic pressure was indicative of the pH and ionic strength of the solution[153].

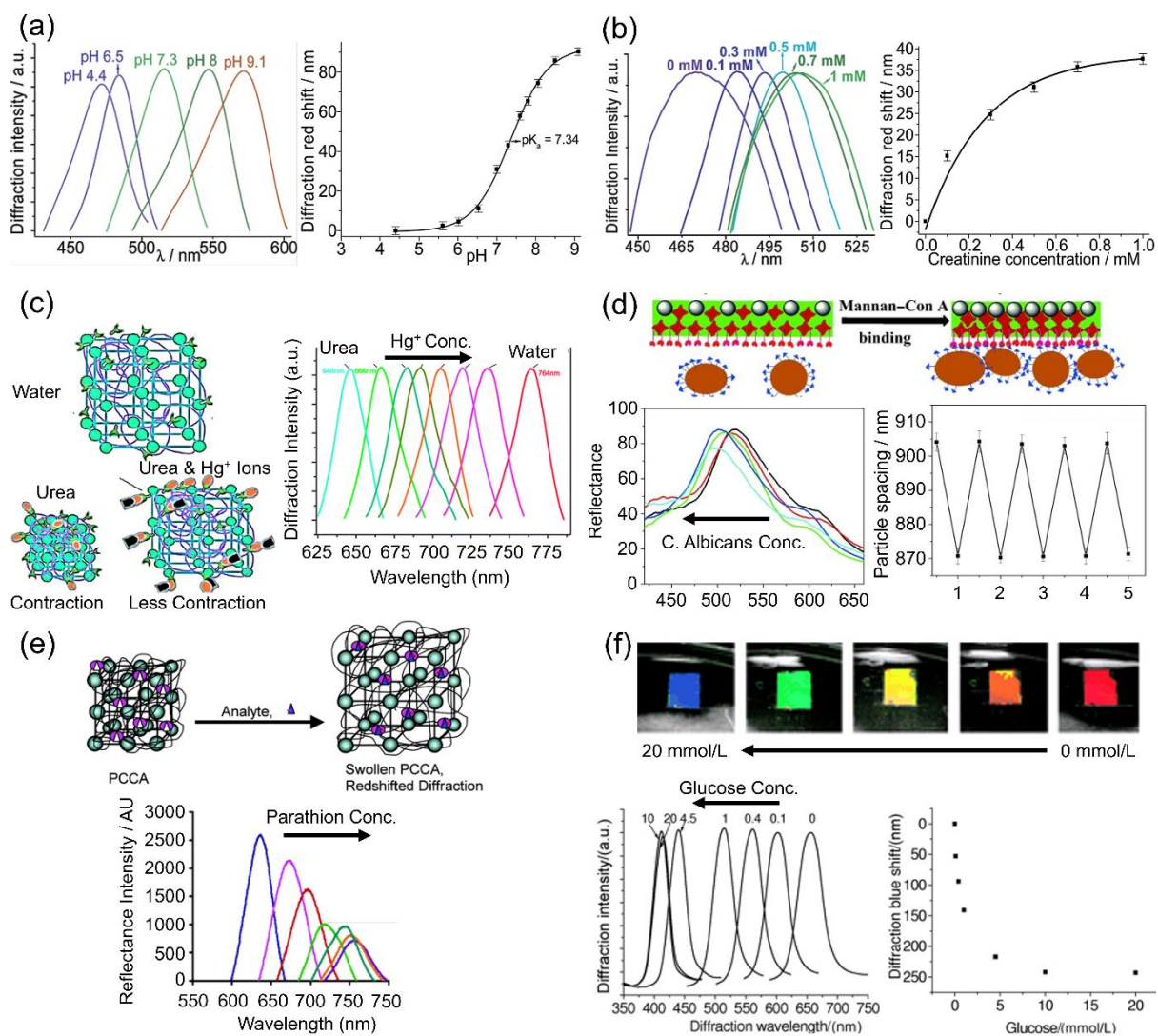

**Fig. 10** Hydrogel PhC sensor applications. (a) The photonic stopband position of a polyacrylamide hydrogel functionalised with 2-nitrophenol surrounding a polystyrene PhC is demonstrated to be sensitive to pH[154]. Adapted from ref[154]. Copyright 2004, American Chemical Society. (b) The pH sensitivity of this hydrogel is further shown to act as an optical indicator for creatinine concentrations[154]. Adapted from ref[154]. Copyright 2004, American Chemical Society. (c) Mercury ions concentrations in water are shown to inhibit the contraction of urease hydrogel networks when exposed to urea, reducing the observed stopband blue-shift[155]. Adapted from ref[155]. Copyright 2011, Royal Society of Chemistry. (d) The presence of candida albicans in a Concanavalin A functionalised hydrogel is illustrated to contract the hydrogel network, creating a stopband blue-shift for optical monitoring[156]. Adapted from ref[156]. Copyright 2015, Wiley-VCH. (e) A red-shift in the photonic stopband position indicates the presence of parathion species, an organophosphate nerve agent, via an induced swelling of the hydrogel[157]. Adapted from ref[157]. Copyright 2005, American Chemical Society. (f) Glucose concentrations in polyacrylamide-polyethylene glycol hydrogel systems are detected via a blue-shift in the stopband position with increasing concentration[158]. Adapted from ref[158]. Copyright 2004, American Association for Clinical Chemistry.

Metal ions, such as $Pb^{2+}$, $Ba^{2+}$ and $K^+$, could also be detected in similar systems by attaching crown ether molecules to the hydrogel networks which complexed with the metal cations and increasingly red-shifted the photonic stopband through increased osmotic pressure created with higher concentrations of metal ions binding to the crown ether[159]. This hydrogel sensing method has been adapted for detecting various metal ions in systems, e.g. toxic mercury ions in water[155] or toxic divalent beryllium ($Be^{2+}$) ions in seawater[160]. Immobilised urease in a hydrogel system was found to react with urea to produce $NH_4^+$ and $HCO_3^-$ ions which shrink the hydrogel, creating a blue-shift of the photonic stopband relative to the system in pure water. $Hg^{2+}$ ions inhibit the urea reaction from proceeding inducing a relatively smaller blue-shift depending on the concentration of ions.

Outside of chemical applications, biomedical studies have adapted PhC hydrogels to respond to biomolecules, proteins and viruses for detection of monitoring of concentration levels. Polyacrylamide hydrogels functionalised with a creatinine deiminase enzyme and a 2-nitrophenol embedded in a polystyrene PhC structure have been demonstrated to function as a concentration sensor for creatinine levels in blood samples, a useful indicator in monitoring renal dysfunction[154]. A red-shift in the photonic stopband was linked to higher concentrations of creatinine, explained as an increase in pH resulting from hydroxide released from creatinine metabolism and an osmotic swelling of the hydrogel. Fungal microbes have also been detected by hydrogel systems. Candida albicans, a fungal microbe in which overgrowth in humans can lead to illnesses like pneumonia, was detected via a blue-shift in the photonic stopband related to a shrinking of the Concanavalin A protein hydrogels in a 2D PhC arrangement[156].

Hydrogel designs for detecting harmful organophosphate nerve agents, such as parathion, have been shown to detect ultra-trace concentrations of organophosphorus compounds using the enzyme acetylcholinesterase[157]. The magnitude of the photonic stopband red-shift was directly linked to parathion concentration in aqueous solutions arising from a swelling of the hydrogel network when the parathion molecules bind to the enzymes in the hydrogel. Glucose monitoring in polyacrylamide-polyethylene glycol hydrogel designs was achieved through a blue-shift in the photonic stopband arising from increased hydrogel crosslinking related to the formation of a supramolecular complex created by glucose molecules self-assembling boronic acid and polyethylene glycol function groups on the hydrogel surface[161]. Monitoring glucose levels is an important aspect of managing diabetes, with further works proposing non-invasive methods of monitoring of glucose in human tear fluid[158,162] and urine[163,164] or a minimally invasive methods using a microneedle wearable patch[165][166] utilising this colorimetric hydrogel detection method. Figure 10 contains examples of the optical sensing capabilities associated with hydrogel PhC sensors.

There are many other emerging hydrogel PhC detection systems being researched with consistent innovation in the literature for a wide array of different detectable substances. Hydrogel PhCs have been developed as mechanochromic sensors with a colorimetric response to stretching or mechanical for use in wearable devices[167] or as underwater motion detectors[168]. The range of specific chemical compounds which can be detected by PhC hydrogel networks is always expanding with new reports of hydrogen peroxide[169], aldehydes[170], alcohol concentration in beverages[171] or for drug delivery[172], thiol biomolecules[173] and adenosine in DNA[174] detection among recent works. Health-monitoring PhC hydrogel systems continue to be developed for detecting harmful residual antibiotics present in livestock milk[175], gram negative bacteria such as E.coli[176] or Pseudomonas aeruginosa[176][177] present in milk or drinking water, alkaline phosphatase enzyme levels as a marker for liver or endocrine diseases[178] and metabolites such as L-kynurenine in blood as markers for immune suppressant disorders[179]. Notably, all of these applications emphasise the importance of sensitivity of the photonic stopband response and the presence of the structural color of the PhC as a visual indicator for changes to the ordered system.

## 3 Photonic Crystals in Slow Light Generation and Photocatalysis

The photonic bandgap or stopband associated with structured PhC materials can inhibit the transmission of certain frequencies of light, but we can also exploit the interesting optical phenomenon of slow light propagation occurring at frequencies close to the photonic band edges. Compared to bulk or non-structured films, it is the specific presence of the photonic bandgap/stopband that introduces slow photon effect. For a simple definition, the slow photon effect is a reduction in the group velocity of light as the light increasingly approaches the properties of a standing wave at energies close to the photonic stopband[180]. This reduction in the group velocity ($v_g$) of light can be visualised by a decrease in slope when approaching the bandgap/stopband in the dispersion relation diagram. Photonic band structure diagrams can be modelled for different types of PhC designs with various refractive indices or material filling fractions to predict the existence and energies of photonic bandgaps/stopbands[181 182 183].

Figure 11 (a) displays band structure diagram for a $TiO_2$ IO, showing the existence of photonic stopband in the Γ – L direction, spanning the approximate wavelength range 380 – 450 nm[112]. These band structure diagrams illustrate the allowed photon energies (ω(k)) at different wavevectors (k). For light propagating through an isotropic, non-structured material, the dispersion relation can be written as[184];

$$\omega(k) = \frac{c_0}{n} |\boldsymbol{k}| \tag{4}$$

The group velocity at any point on these band diagrams can be found as:

$$v_g = \frac{\partial \omega}{\partial k} \tag{5}$$

The bending of the photonic band structure to accommodate for the photonic bandgap/stopband frequency range creates frequencies where the magnitude of this derivative is reduced, introducing the slow photon effect. The changes to the dispersion relation bands arising from a simple 1D PhC slab are shown in Fig. 11 (b)[184].

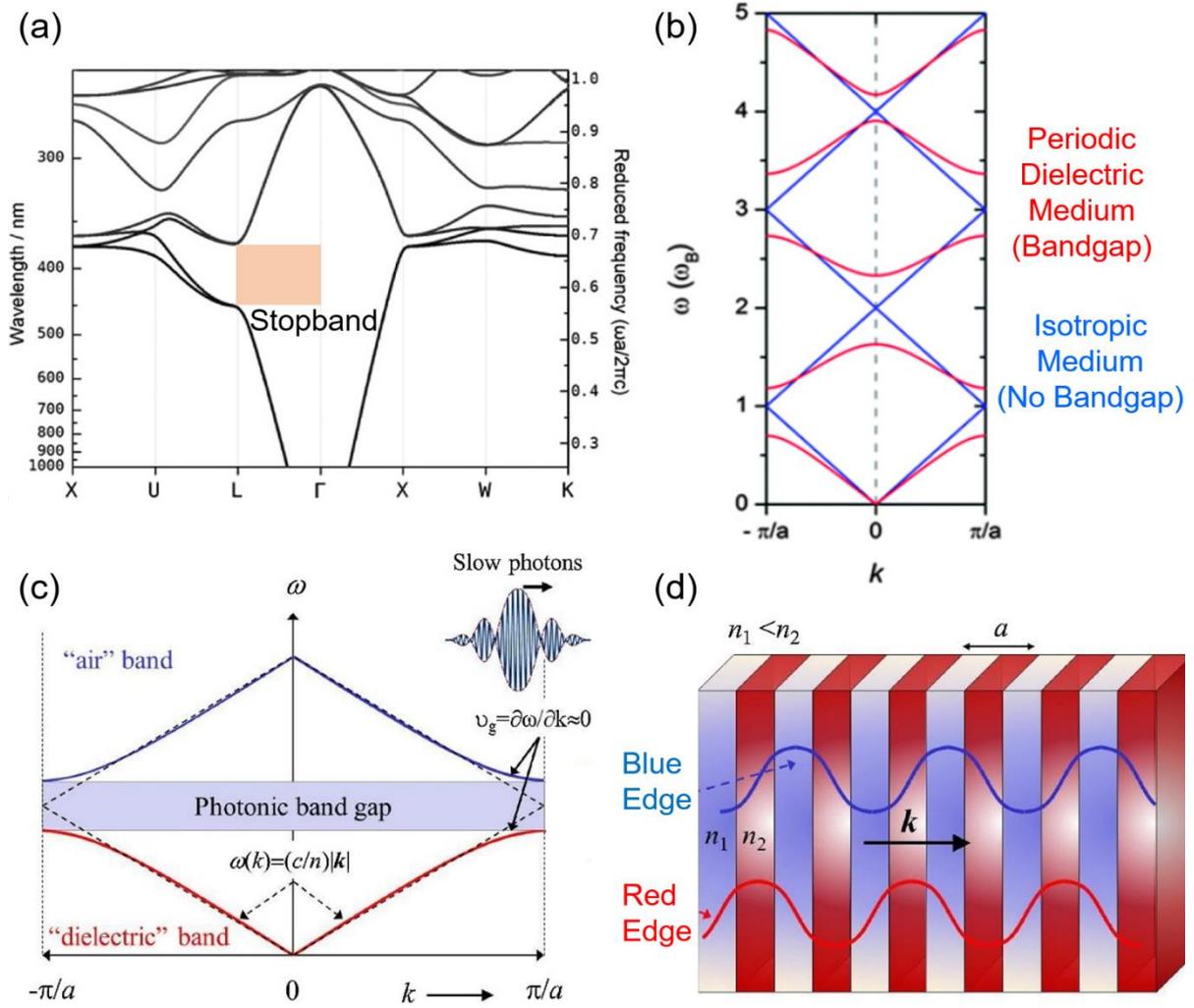

**Fig. 11** (a) Photonic band structure for a $TiO_2$ IO in air[112] with spherical cavities of diameter 180 nm and n = 2.5 for $TiO_2$. The stopband in Γ – L direction ([111] direction) is highlighted. Adapted from ref[112]. Copyright 2015, American Chemical Society. (b) Simulated dispersion relation bands showing photonic bandgap formation in a 1D PhC with material slabs of thickness 200 nm and alternating refractive indices of 1 and 3.5 for a periodic dielectric (in red) compared to an isotropic slab with artificial periodicity (in blue)[184]. Adapted from ref[184]. Copyright 2013, Royal Society of Chemistry. (c) A photonic bandgap in the dispersion relation diagram for a 1D PhC illustrating the reduction to the group velocity at the band edges[185]. Adapted from ref[185]. Copyright 2018, Elsevier. (d) Schematic representation of light localisation in a 1D PhC slab ($n_2 > n_1$) showing a localisation in the higher index region at the red-edge and vice-versa[185]. Adapted from ref[185]. Copyright 2018, Elsevier.

The reduction in the group velocity, as shown in the dispersion diagram, creates a scenario where the light, which is quickly approaching a standing wave, becomes more strongly localised in different dielectric regions of the material[180]. Slow light effects appear at either edge of the photonic bandgap/stopband region and the dispersion bands are often referred to as a dielectric (red edge) and an air (blue edge) band[185], as shown labelled in Fig. 11 (c). With the slow photon effect, it is thought that the maximal amplitudes of the standing wave become more localised in different dielectric media when approaching from the red edge compared to the blue edge. Stronger localisation of the light in the higher dielectric material is anticipated for the red edge of the photonic bandgap/stopband[180 185 186]. Conversely, light localisation for the lower dielectric material (air) is expected at the blue edge. A schematic representation of this effect can be seen in Fig. 11 (d).

For PhC materials, the group velocity attenuation is dependent on the magnitude of the contrast between the dielectric materials, the structural quality and thickness of the film[186]. The slow photon effect in photon crystal materials has attracted much attention for applications where reduced group velocities and a targeted localisation of light in specific materials is beneficial, particularly in photocatalysis applications. The direct relationship between photonic bandgap/stopband position and the slow photon effect allows the frequencies of the slow photon region to be tuned alongside the photonic bandgap/stopband, as depicted in Fig. 12.

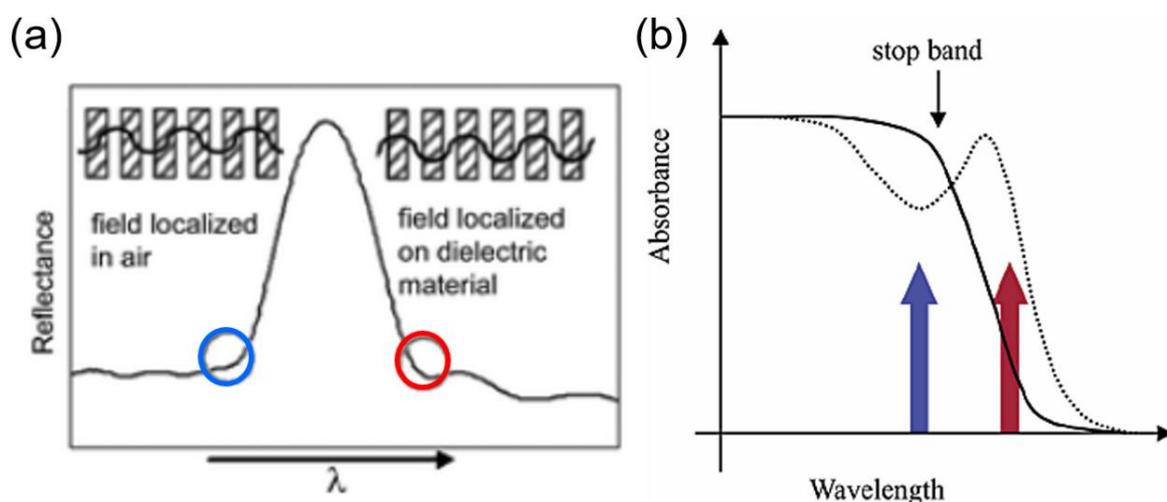

**Fig. 12** (a) A schematic diagram showing the blue and red-edge slow photon regions surrounding a photonic bandgap region[186]; slow photon frequency ranges are directly related to the photonic bandgap position. Adapted from ref[186]. Copyright 2008, Royal Society of Chemistry. (b) Schematic representation illustrating the effect of blue and red-edge slow photon regions on the absorbance of a semiconductor photocatalyst[180]. Standing wave localisation in the high index semiconductor material at the red-edge is shown to increase absorbance. Adapted from ref[180]. Copyright 2003, American Chemical Society.

Photocatalysts are materials used in light-generated photochemical reactions where the photocatalyst (typically a semiconductor material) will generate reactive oxygen species stemming from electron-hole formation under light illumination. The photo-catalysed production of reactive oxygen species from the semiconductor acts to facilitate rapid reactions with other species in the reaction vessel[187]. Major sectors where photocatalytic applications are explored are environmental waste/pollutant degradation[188 189 190 191 192], microbial disinfection[193 194 195] and solar-fuelled energy production[196 197 198 199 200]. A suitable electronic bandgap is a pre-requisite condition on the properties of semiconductors used as photocatalysts; redox potentials for the formation of reactive oxygen species and the evolution of hydrogen and oxygen from water should lie within the electronic bandgap of the semiconductor[201].

$TiO_2$ is one of the most prominently used semiconductors in photocatalysis owing to a combination of its suitable bandgap, relative abundance, low cost, stability, insolubility in water and nontoxicity[190 201]. The basic operational principle for any photocatalytic material is the absorption of light at energies above the bandgap of the semiconductor, forming electrons and holes. Photo-excited electrons promoted to the conduction band can react with oxygen species to form superoxide/hydroperoxide radicals and photo-generated holes in the valence band can oxidise water or hydroxyl groups to create hydroxyl radicals[202]. These reactive oxygen species act to degrade other species present in the vessel, such as environmental pollutants, through oxidation and decomposition. Figure 13 (a) displays a simple mechanism for photocatalysis oxidation and reduction reactions in a material such as $TiO_2$ with a bandgap below 390 nm[203].

Although $TiO_2$ is widely used as photocatalyst on merit of its material properties, certain aspects of its semiconductor performance limit its ability for photocatalysis. Visible light-activated (VLA) photocatalysis relates to photocatalytic material designs which harvest the solar spectrum for catalytic reactions. The large bandgap of rutile $TiO_2$ (3.2 eV) and anatase $TiO_2$ (3.0 eV) requires UV light excitation for electron-hole generation; less than 5% of the solar flux incident on the earth's surface is located within this spectral range[201]. Rapid charge recombination rates between photo-generated electron-hole pairs limits the quantum efficiency of photocatalytic processes in $TiO_2$[202]. Several strategies have been utilised by researchers in an attempt to move towards VLA photocatalysis by tuning the bandgap of $TiO_2$ towards the visible range, enhancing the absorption of the semiconductor in the visible range or suppressing the electron-hole recombination rate which limits the quantum efficiency.

Dye-sensitised photocatalysis relates to organic dyes acting as sensitiser molecules on the surface of the semiconductor, enabling absorption of visible light wavelengths by the dye molecules and the transfer of electrons from the dye molecules to the conduction band of the semiconductor. Colored, toxic organic dyes in wastewater, originating from industry sources like textile coloring[204], can be self-sensitising by creating pathways for their own photocatalytic degradation under visible light[205]. To enhance the performance of $TiO_2$ as a photocatalyst, other semiconductors have been interfaced with $TiO_2$ forming heterojunctions[206] which can act to limit charge recombination[207] and increase visible photocatalytic activity[208] for correct choice of material. Noble metal nanoparticles with a surface plasmon resonance such as Au[209] and Pt[210] incorporated on the surface of semiconductors have shown reports of increased photocatalytic activity attributed to electron injection from the metal to the semiconductor, effectively inhibiting charge recombination rates and extending the absorption of the system into the visible range[211]. Non-metals such as nitrogen[212], sulfur[213] and carbon[214], have been explored as dopants for semiconductors as $TiO_2$ with reports of enhanced visible light photocatalysis. Schematic diagrams representing the mechanisms behind photocatalyst semiconductor modifications can be seen in Figs. 13 (b), (c) and (d) for dye-sensitisation[215], heterojunction[206] and metal nanoparticle modifications[216], respectively.

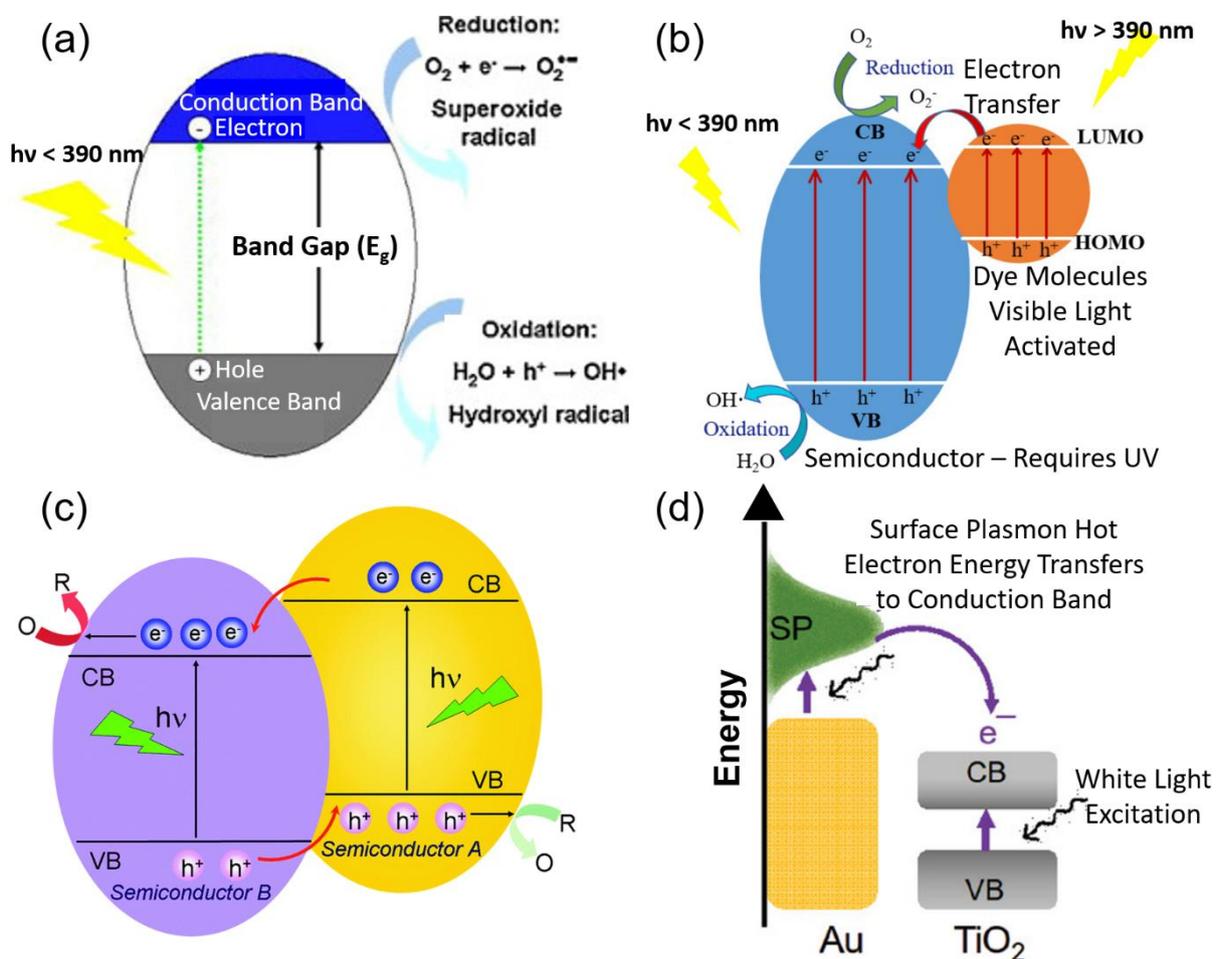

**Fig. 13** (a) Mechanism of photocatalysis oxidation and reduction reactions for a semiconductor photocatalyst with an electronic bandgap in the UV[203]. Adapted from ref[203]. Copyright 2012, Elsevier. (b) Dye-sensitised photocatalysis mechanism showing visible excitation of dye molecules for a semiconductor with a wider bandgap[215]. Adapted from ref[215]. Copyright 2020, MDPI. (c) Semiconductor heterojunction showing the differences in bandgap energies and the flow of charges between valence and conduction bands for improved charge separation[206]. Adapted from ref[206]. Copyright 2014, Royal Society of Chemistry. (d) Schematic diagram of gold metal nanoparticles on a $TiO_2$ semiconductor[216]. The localised surface plasmon resonance of the metal nanoparticles allow electrons generated in the visible region to enter the semiconductor conduction band. Adapted from ref[216]. Copyright 2020, Wiley-VCH.

IO PhC designs are frequently used in photocatalytic systems. The structural porosity and higher effective surface area of IOs makes them attractive candidates for photocatalysts with reports of improved photocatalytic performances arising due to the effects of the structure[211,217]. The optical effects associated with IOs, in particular the slow photon effect, has generated a lot of interest in photocatalysis research. As a controllable optical effect largely dependent on the structural size, the slow photon effect is often investigated in combination with other strategies, as described above, to enhance the photocatalytic performance of materials through a synergistic response[218,219,220,216]. The reduced group velocity of light accompanying the slow photon effect is a desirable quality in photocatalysis; a decrease in group velocity leads to an increase in the effective optical path length of light which enhances light-matter interactions in the structure[186]. Extending the active range of frequencies for photocatalytic semiconductor materials through engineered slow photon effects is also a possibility with IO designs. At first glance, it may seem counterintuitive to associate photocatalytic enhancement (improved or extended light-matter interactions) with a PhC material which acts to reflect light from a surface. A common strategy is to engineer the photonic bandgap/stopband frequency to coincide with regions of high absorption in the semiconductor, effectively suppressing the photonic reflection, while simultaneously locating the slow photon region in a spectral region of interest for enhancement[112,185].

For slow photon photocatalytic enhancement in IOs, some of the earliest investigations were focused on slow photon amplified dye-sensitised systems[180,221]. Dye-sensitised degradation of methylene blue on an anatase phase $TiO_2$ IO was investigated by varying the size (tuning the photonic stopband) of the IO periodicity and also the angle of the incident light, effectively shifting the slow photon region between different samples[221]. The largest enhancement factors for the photocatalytic degradation occurred when the slow photon red-edge overlapped with the incident monochromatic radiation, as seen depicted in Fig. 14 (a). A separate study found that matching the absorption peaks of organic dyes such as methylene blue, rhodamine B and methyl orange to the stopband regions of various $TiO_2$ IOs enhanced the photocatalytic activity of the system under solar light illumination[222]. The slow photon effect is credited with the enhancement of the dye-sensitisation, as seen from the increased kinetic constants in Fig. 14 (b).

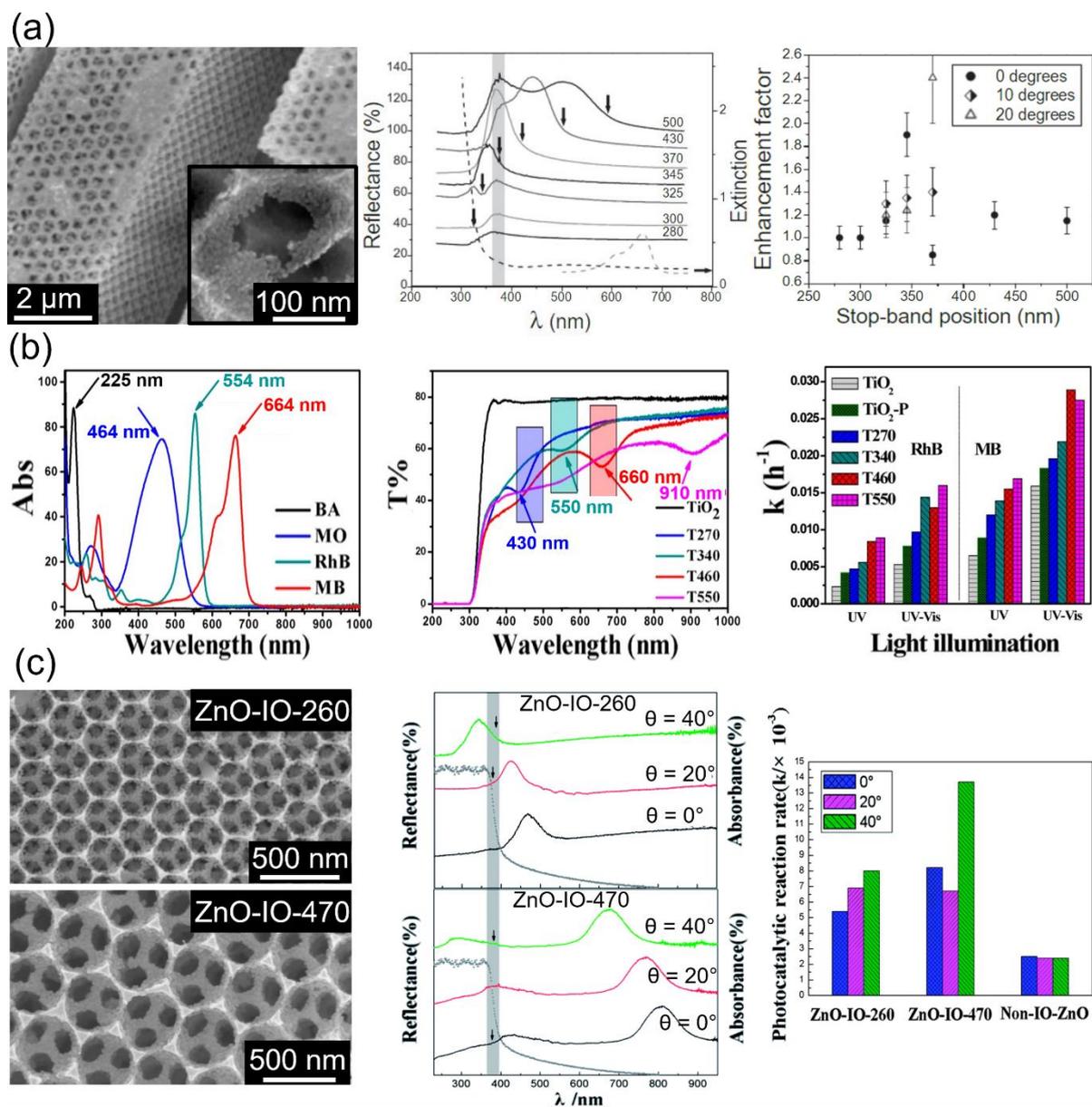

**Fig. 14** (a) Dye-sensitised photocatalysis with slow photon effects for the photocatalytic degradation of methylene blue on $TiO_2$ IOs[221]. The reflectance spectra of $TiO_2$ IOs, showing the stopbands, the red-edge of the slow photon regions (arrow) and the absorption region of $TiO_2$ are highlighted. The photocatalytic enhancement factor is highest under 370 nm illumination for an overlapping slow photon effect. Adapted from ref[221]. Copyright 2006, Wiley-VCH. (b) Dye-sensitised photocatalysis, matching the peak absorbance of organic dyes with the slow photon effect of $TiO_2$ IOs[222]. The stopbands of IOs are matched to the peak absorbance of methyl orange (MO), rhodamine B (RhB) and methylene blue (MB). For RhB and MB, overlapping the absorbance with slow photon effects are shown to give greater reaction rates in the UV-Vis. Adapted from ref[222]. Copyright 2013, American Chemical Society. (c) Dye-sensitised degradation of rhodamine B on ZnO IOs[223]. The reflection spectra show the variable angle used to align the slow photon effect with the electronic absorbance of ZnO (shaded region). The photocatalytic reaction rate is shown highest upon overlap with the red-edge of the slow photon region. Adapted from ref[223]. Copyright 2014, Royal Society of Chemistry.

Outside of $TiO_2$ other IO materials report similar slow photon enhancement effects. A ZnO IO was found to present the highest photocatalytic reaction rate for dye-sensitised rhodamine B photocatalysis when the red-edge of the photonic stopband overlapped with the ZnO electronic absorption band[223] at higher angles of incidence. For the ZnO IO, the slow photon effect at the red-edge of the photonic stopband outperformed the blue-edge, as seen depicted in Fig. 14 (c).

Hybrid structures of $SnO_2$ IOs coated with nanocrystalline $TiO_2$ layers reported the highest rates of photocatalytic degradation of rhodamine B when the photonic stopband of the $SnO_2$ IO closely matched the electronic bandgap of $TiO_2$[224]. A heterostructure design of a $TiO_2$ IO infiltrated with 10 nm $Cu_2O$ nanocrystals reported an increase in the photocatalytic degradation of rhodamine B and bisphenol A under UV-visible wavelengths when the slow photon region overlapped with the optical absorption of the $Cu_2O$ nanocrystals[225], as shown in Fig. 15 (a). Increased photocatalysis in this case was attributed to the synergy of the increased charge separation resulting from the heterostructure and the slow photon effect enhancing the light harvesting of the system. Similarly, a heterostructure of an $SnO_2$ IO with CdS quantum dots deposited uniformly over the surface of the IO reported increased rates for the removal of carbamazepine[226], again attributed to a combination of the heterostructure and the slow photon effect as seen in Fig. 15 (b).

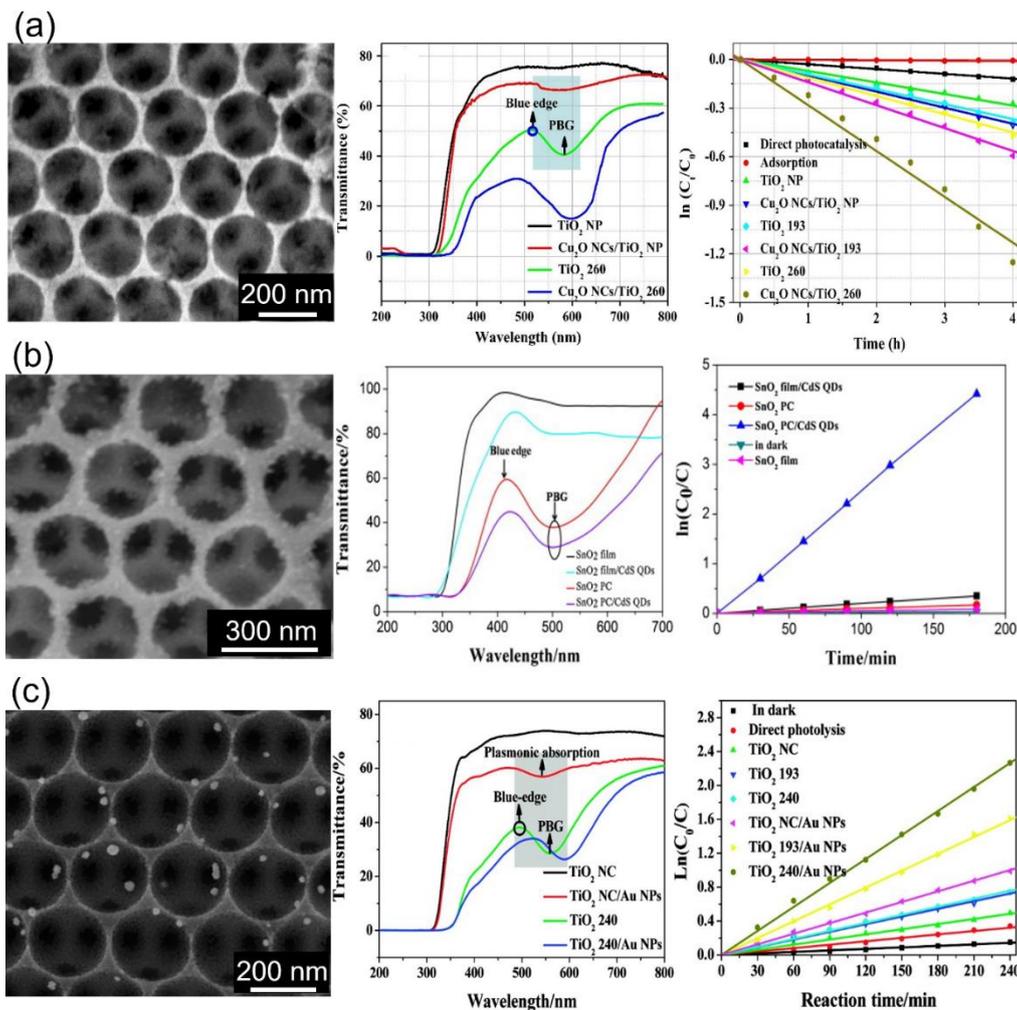

**Fig. 15** SEM images, transmission spectra and measured photocatalytic degradation kinetics for synergistic slow photon effects with other semiconductor modification methods. (a) For a $TiO_2$ IO with a $Cu_2O$ heterojunction, the highest degradation kinetics for bisphenol A occurs for overlap of the $Cu_2O$ visible absorption with the blue-edge of the IO slow photon effect[225]. Adapted from ref[225]. Copyright 2015, Elsevier. (b) In a $SnO_2$ IO with a CdS quantum dot heterojunction, the photocatalytic degradation kinetics for carbamazepine were highest for an overlap of the blue-edge of the slow photon effect with the absorption of the CdS quantum dots[226]. Adapted from ref[226]. Copyright 2014, Elsevier. (c) For plasmonic Au nanoparticles deposited over a $TiO_2$ photocatalyst, the plasmonic absorption of the Au particles was amplified by the blue-edge of the slow photon region yielding increased degradation kinetics for 2,4-dichlorophenol[211]. Adapted from ref[211]. Copyright 2012, American Chemical Society.

The photocatalytic synergy between noble metal modification of semiconductors and the slow photon effect has also been explored. 2.3 wt% of Pt nanoparticles deposited on the surface of a $TiO_2$ IO demonstrated an enhancement of the photo degradation efficiency of acid orange by a factor of four[218]. Neither the slow photon effect from the IO structure or the enhanced charge separation from the Schottky barriers at the $TiO_2$-Pt interface alone was sufficient to achieve this enhancement; the cooperative effects were needed to record the greatest enhancement factor. For improved performances in the visible spectrum Au nanoparticles have been deposited on $TiO_2$ IOs for photocatalytic degradation of 2,4-dichlorophenol using a combination of the TiO$_2$-Au heterojunction for charge separation, the slow photon effect of the IO and the increased light absorption in the visible from the localised surface plasmon resonance of the Au nanoparticles[211]. This combination effect can be seen from the data depicted in Fig. 15 (c). These strategies for enhancing the photocatalytic performance of semiconductors through a combination of the slow photon effect with other existing methods are currently of great interest for photocatalysis research.

Recent works with visible light and Au nanoparticles incorporated onto ZnO[219,227] or V$_2$O$_5$[216] IOs are further examples exploring this synergy of photocatalytic effects. The placement of Au particles in TiO$_2$ IOs and the role of slow photons in synergistic photocatalytic effects has also been studied[228]. Many emerging slow photon works focus on IO heterostructures, where two semiconductor materials are placed in contact to improve light harvesting effects in conjunction with slow photon effects. For photocatalytic CO$_2$ reduction, a rhenium-doped TiO$_{2-x}$/SnO$_2$ IO heterostructure[229] and CeO$_2$ nanolayers supported by a TiO$_2$ IO[230] have been investigated. Similarly, a TiO$_2$/graphitic-C$_3$N$_4$ IO heterostructure showed an enhanced performance for rhodamine B degradation[231] and Bi$_2$WO$_6$/WO$_3$ IO heterostructures with graphene quantum dots enhanced phenol degradation[232]. Recent prevailing opinions[233,234,235,236] emphasise that high quality, structurally scalable and mechanically stable inverse opal films with a strong optical signature are crucial for maximising the slow photon effect. It has even been suggested that higher order slow-light modes found only in high quality TiO$_2$ IO films may be more efficient at light trapping than traditional red and blue-edge of the photonic bandgap[234].

## 4 Photonic Crystals in Solar Cell Devices

Solar cell devices operate under the principle of photovoltaic energy conversion under which electromagnetic radiation is converted into electrical energy. The process of electrical energy conversion can be described as light absorption by an absorber material creating a transition from ground to excited state, production of an electron and hole from this transition, charge migration of the electron through an external circuit from cathode to anode and recombination of the electron with a hole to return the absorber to the ground state[237]. Thick, crystalline silicon-based solar cells have been the dominant photovoltaic devices in solar cell technology for quite some time. Silicon solar cells are popular due to the abundance of silicon available in the earth's crust, the non-toxicity of the material, the proven efficiency and stability of the technology and the semiconducting properties of silicon were widely studied.

Limitations on conventional silicon solar cells are known and were predicted[238][239]; the indirect bandgap of crystalline silicon necessitates thick layers of silicon for efficient cells and the high associated cost of manufacturing thick layers of crystalline silicon created an economic impediment. Research trends in solar cells have been focused on driving productions costs down, minimising energy usage and environmental $CO_2$ emissions[240] from crystalline silicon production while attempting to preserve or improve on the efficiencies of the silicon solar cell. Thin-film silicon solar cells[241], thin films of alternate materials like cadmium telluride or copper-indium diselenide[242], organic solar cells[243], perovskite solar cells[244], and dye-sensitised solar cells[245] are among some of the alternative classes of solar cell being researched for future implementation[246][247][248]. Here, we discuss the role of PhC materials, specifically their optical properties, which have been used to enhance the performance of certain classes of solar cell.

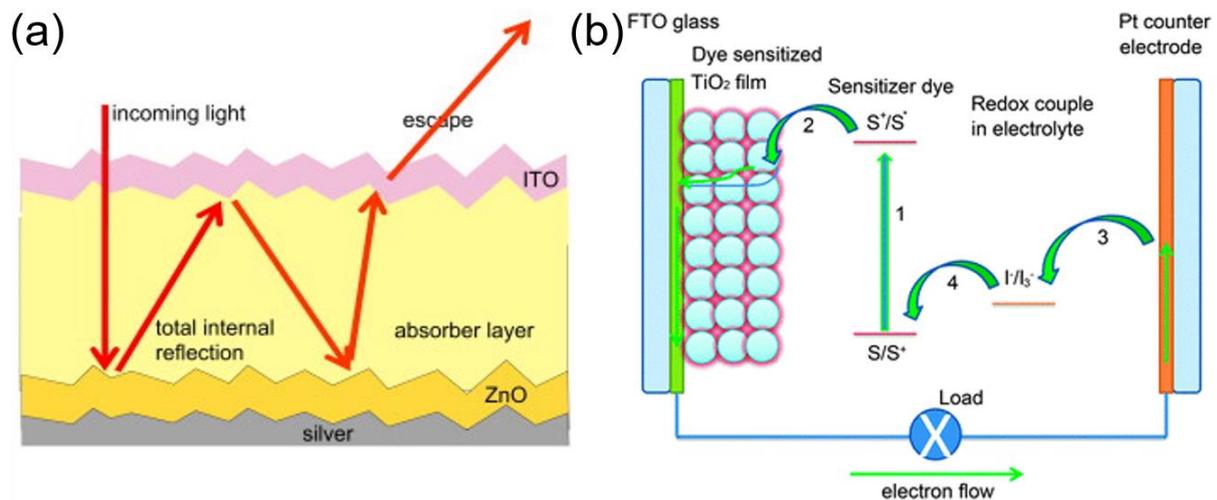

**Fig. 16** (a) Schematic diagram showing typical modifications made to layers in a thin film solar cell, including textured surfaces for light scattering and a metal back reflector, to increase the path length and absorption probability of light in the cell[249]. Adapted from ref[249]. Copyright 2014, Elsevier. (b) Schematic diagram depicting the electron transfer processes in a dye-sensitized solar cell for a $TiO_2$ film sensitized with dye molecules versus a Pt counter electrode[250]. Adapted from ref[250]. Copyright 2013, Royal Society of Chemistry.

Thin film solar cells have been researched as alternatives to the thicker crystalline wafer silicon solar cells. Hydrogenated amorphous silicon (a-Si:H) and hydrogenated microcrystalline silicon (μc-Si:H) are common examples of silicon thin films used in these solar cells; the thin films of these materials offer the potential advantages of cost reduction in material preparation and are much quicker to pay back the energy input required to fabricate these devices[251]. Direct bandgap semiconductors like cadmium telluride or copper-indium selenide species feature much larger absorption coefficients allowing for higher absorption of incident photons over a shorter distance, such as a thin film[252]. Amorphous semiconductors with an indirect bandgap (such as a-Si:H) tend to feature larger absorption coefficients[253] when compared to their crystalline counterparts due to the structural disorder in the amorphous material relaxing the quantum mechanical selection rules for absorption[251]. In spite of a larger absorption coefficient, amorphous silicon features a wider bandgap (~ 1.7 – 1.9 eV for a-Si:H) compared to crystalline forms of silicon (~ 1.1 eV), limiting the amount of the solar spectrum absorbed by the amorphous material[254]. Micromorph or tandem solar cells featuring heterojunctions of a-Si:H and μc-Si:H have been developed to promote increased absorption at longer wavelengths and limit light-induced degradation of the amorphous material[254,255].

Increasing the absorption of the semiconductor material is key to improving the performance of thin-film solar cells. A widely used technique in thin-film solar cells is light trapping, which uses material or layer modifications to increase the path length of light in the absorbing material thereby increasing the probability of absorption[256,257]. Light trapping techniques are essential for thin-film cells in particular as the absorbing layer is nominally short. The front surfaces of layers in the solar cell are often textured to promote scattering and total internal reflection effects which extend the path length of light in the material[257,258]. Back reflectors, textured metal surfaces separated from silicon layers by a thin dielectric layer[259], are also used to reflect incident light and create longer path lengths. Figure 16 (a) illustrates this effect where textured surfaces are utilised to confine light to the absorbing semiconducting layer. In thin film solar cells, PhCs find their application here by providing controllable and enhanced reflections from material layers to promote increased absorption.

A 1D PhC material of alternating dielectric layers of sputtered indium-tin oxide and $SiO_2$ layers has demonstrated potential for application in integrated photovoltaics, where semi-transparent solar cells for use in windows or skylights are designed using a back reflector of a conducting PhC in place of a metal layer for light trapping in thin-film cells[260]. The tunable reflections in the photonic structure can be modified by adjusting the layer thicknesses to selectively reflect specific wavelength ranges while transmitting the remainder of the spectrum for light or heating; in this case, wavelengths above the absorption edge of the a-Si:H thin-film were reflected to boost performance. The device structure and performance for these 1D photonic solar cell back reflectors can be seen in Fig. 17 (a)[260]. Another application involving alternating 1D PhC dielectric layers of silicon nitride and silicon oxynitride integrated into a crystalline silicon solar cell is based on aesthetically designing colored solar panels using the reflected wavelengths from the PhC for uses in smart design and green building[261]. Red, blue, green and white solar cell panels were fabricated by adjusting the thicknesses of the dielectric layers deposited on the glass to tune the reflectivity of the photonic bandgap, as seen in Fig. 17 (b)[261].

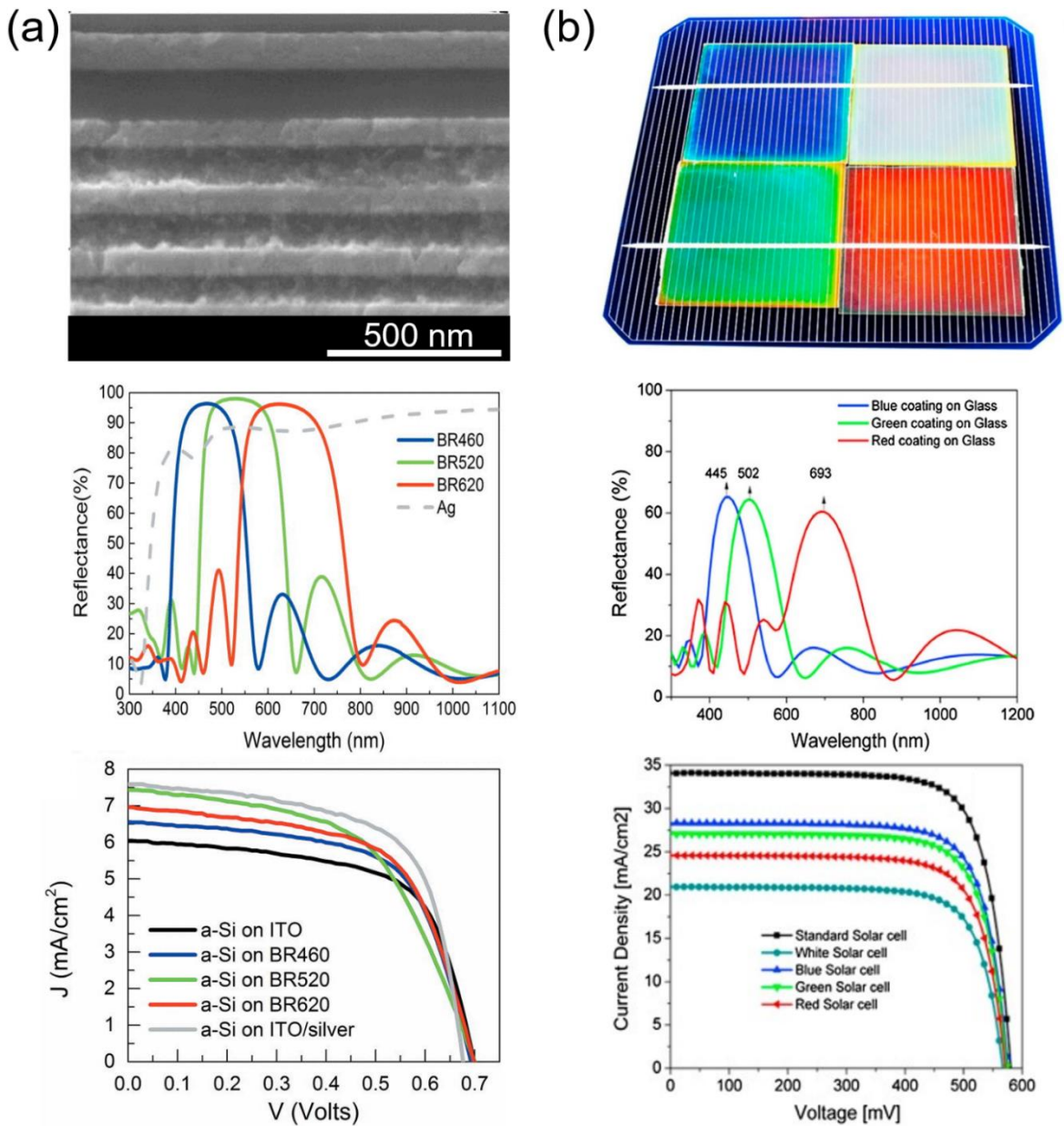

**Fig. 17** (a) Design of a 1D PhC back reflector in a semi-transparent solar cell with alternating layers of indium-tin oxide and SiO$_2$. Peak reflections are shown are shown for various PhC layer thicknesses alongside current densities for solar cells incorporating these designs[260]. Adapted from ref[260]. Copyright 2013, American Institute of Physics. (b) Aesthetic coloration for solar cell devices with visible transparency for non-reflective wavelengths based on a 1D PhC of alternating silicon nitride and silicon oxynitride layers. Peak reflections for different layer thicknesses are shown in addition to solar cell current densities for different designs[261]. Adapted from ref[261]. Copyright 2019, Elsevier.

3D PhCs in the form of silicon IOs have also been incorporated as back reflectors in crystalline silicon thin-film solar cells with a reported increase in the light trapping of the solar cell resulting from the IO back reflector[262]. Photonic bandgap reflections, pseudogap regions, diffuse back scattering and surface diffraction effects in the IO material were all factors believed to contribute to the performance enhancement of the IO back reflector versus a back reflector of aluminium; the IO reflection could also be tuned to the near-IR region where it is most needed for crystalline silicon solar cells. The IO back reflector solar cell was also combined with surface texturing, showing improved absorption using a combination of these effects as seen in Fig. 18 (a)[262].

PhC materials also find application in thin-film solar cells as intermediate reflector layers in tandem solar cells. Tandem solar cells often suffer from current mismatch between the absorbing layers in the solar cell and intermediate layers are often included in an attempt to equalise the current generated from each layer. In the case of a micromorph silicon tandem solar cell, the a-Si:H top layer emits a lower current compared to the µc-Si:H bottom layer. This limits the overall efficiency of the solar cell and intermediate layers are included in an attempt to equalise the current generated by spectrally splitting and reflecting the spectrum[263]. A ZnO:Al IO in a tandem solar cell between a-Si:H and µc-Si:H layers reported an enhancement factor of 3.6 in the external quantum efficiency of the limiting a-Si:H layer by acting as wavelength selective filter; high energy wavelengths were reflected back into the a-Si:H layer using the photonic bandgap and lower energy wavelengths were transmitted to the µc-Si:H layer for absorption[264]. The IO intermediate layer was reported to outperform other intermediate layer materials such as thin-film based intermediate layers, the results of which can be seen in Fig. 18 (b).

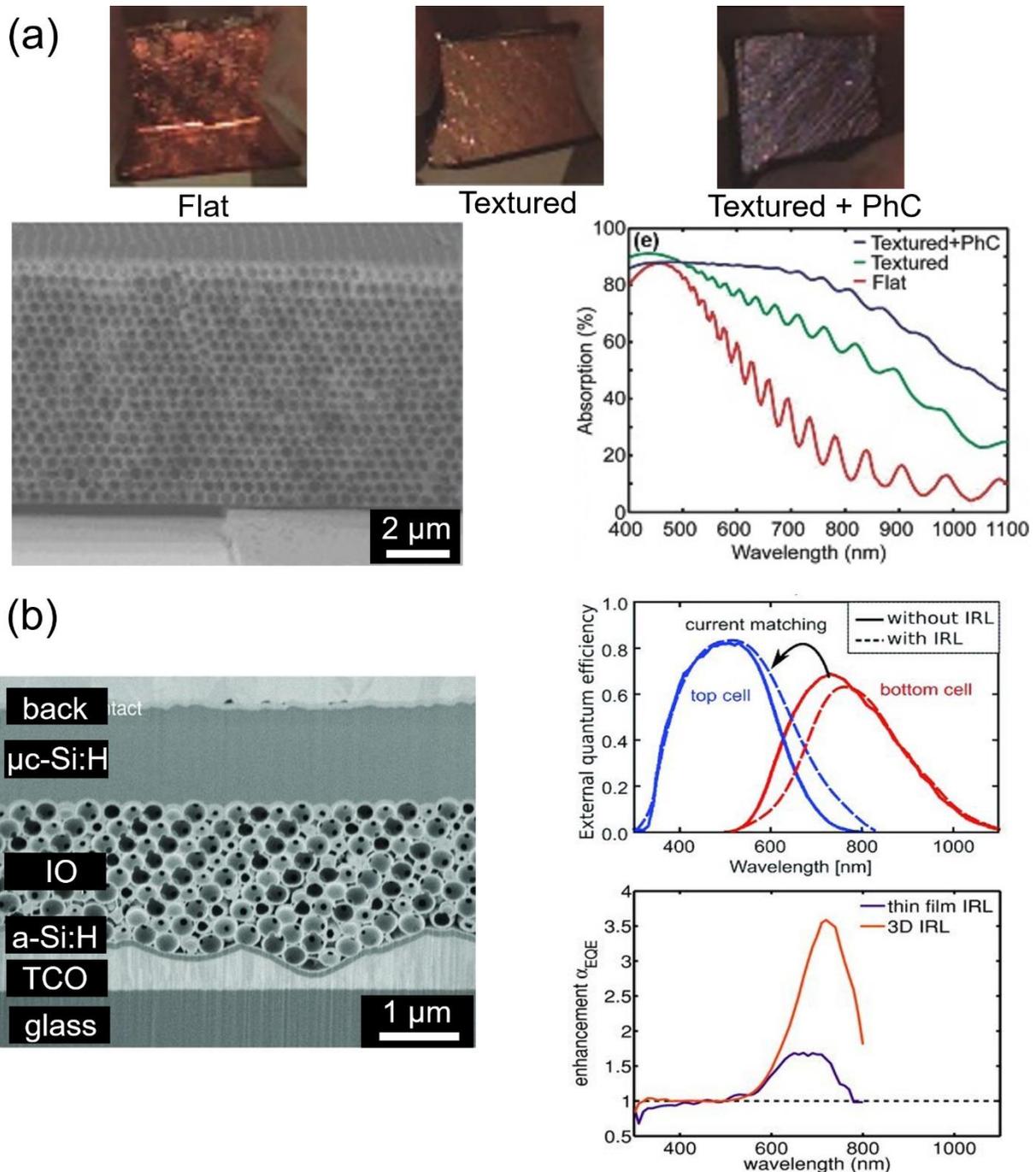

**Fig. 18** (a) A 3D silicon IO used to enhance back reflection and light path lengths in a thin-film crystalline Si solar cell. A combination of surface texturing and PhC reflections were found to result in the highest solar cell absorptions[262]. Adapted from ref[262]. Copyright 2013, Wiley-VCH. (b) A tandem silicon micromorph solar cell with a ZnO:Al IO acting as a wavelength selective intermediate layer between a-Si:H and μc-Si:H absorption layers. The current matching facilitated by the IO is shown via increased external quantum efficiency of the limiting top cell (a-Si:H) and an increased overall enhancement factor of the cell compared to other thin film layers[264]. Adapted from ref[264]. Copyright 2011, Wiley-VCH.

A separate class of solar cell devices which have seen an increased uptake of PhC materials are dye-sensitised solar cells (DSSCs). The operating principles of DSSCs are fundamentally similar to the dye-sensitised photocatalysis mechanism discussed in the Section 3. In DSSC systems, the production of electricity from light can be thought of in terms of a cycle of reactions as shown in Fig. 16 (b)[250]. Photoanodes, such as mesoporous $TiO_2$, are deposited on transparent conductive substrates and sensitized with dye molecules to harvest light. A circuit is connected from the photoanode to a counter electrode, such as platinum coated conductive glass, and an organic electrolyte containing a redox couple, commonly $I^-/I^{3-}$, separates the electrodes[245]. Photo-excited electrons generated from dye absorption are transferred to the conduction band of the semiconductor, collected by anode, transported through an external circuit to perform work and arrive at the counter electrode. Sensitised dye molecules, after transferring electrons, are reduced by the electrolyte and in-turn the electrolyte absorbs the electrons which arrive at the counter electrode[245,250].

Unlike the photocatalysis reactions which act to degrade or eliminate dye molecules or other species in the reaction vessel, DSSCs operate by regenerating the dye species using an electrolyte and closed circuit such that no material is lost or converted and the cycle can continue to generate electricity. Compared to conventional crystalline silicon solar cells, DSSCs are cheaper, easier to fabricate, more flexible, feature a higher sensitivity to low intensity environments and have the potential to possess a low environmental impact with respect to material choices[245,265]. Low efficiencies, long-term stability and material production costs of DSSCs hinder their economic viability[265]. The most successful DSSCs with highest efficiencies tend to feature Ru-bipyridyl compounds as sensitiser molecules which has a high associated cost, time consuming production steps and is a toxic material[250,266,267] and presently natural alternatives do not offer comparably high efficiencies[266,268]. For all manner of DSSCs research methods to improve efficiency of light absorption through cell design or material choice are an essential consideration. With regards to the photoanode material, methods to improve the light harvesting efficiency are based on increasing surface area, dye molecule pickup, reducing charge recombination, improving electron charge transport and increasing the light scattering ability to increase absorption[245,269,270].

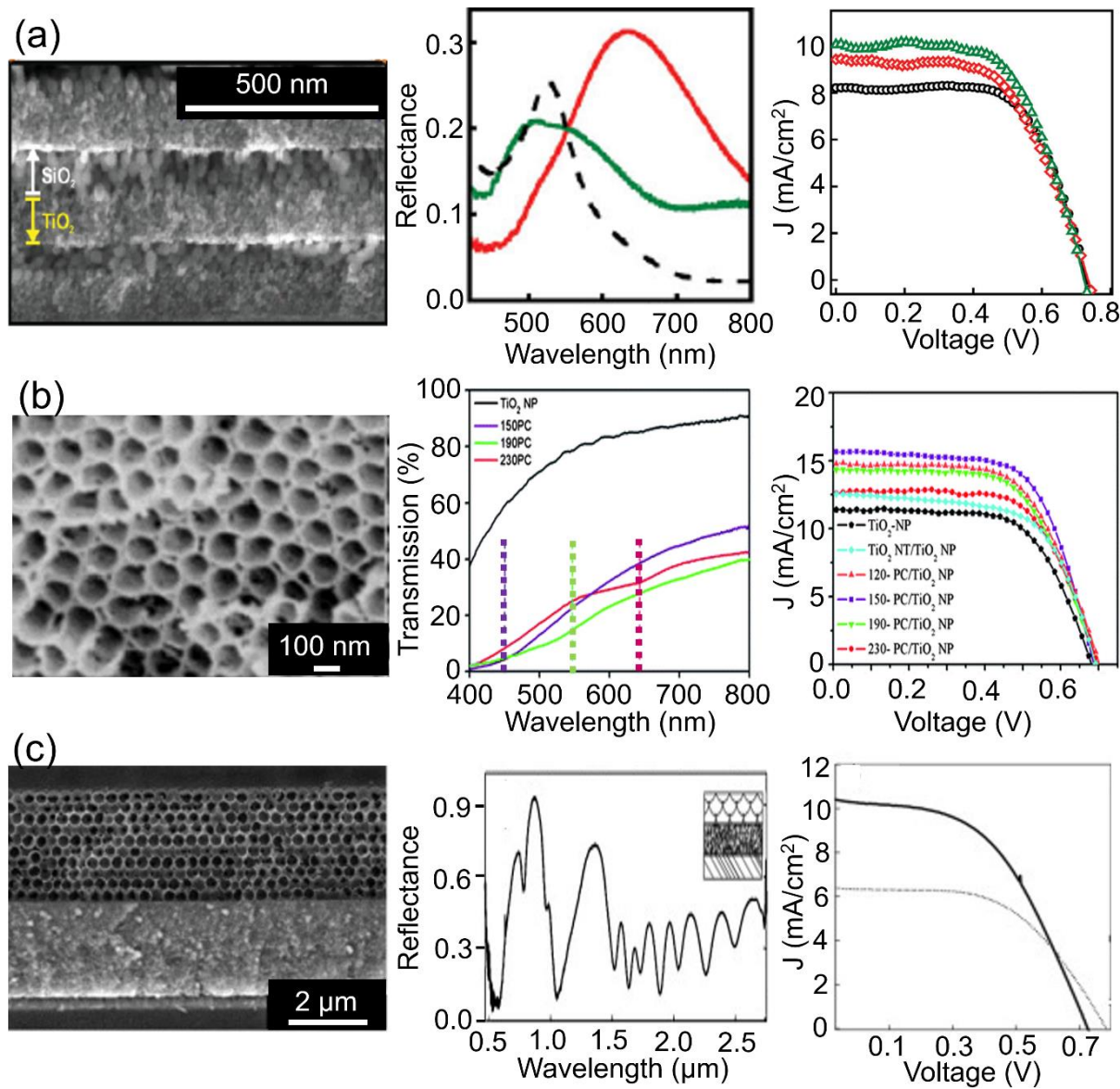

**Fig. 19** (a) Porous 1D PhC layers deposited on a nanocrystalline $TiO_2$ anode in a DSSC. Reflectance spectra with photonic stopband maxima shown for two different PhC materials. Current density for both structures shown versus a DSSC with no PhC reflector[271]. Adapted from ref[271]. Copyright 2009, Wiley-VCH. (b) 2D PhC $TiO_2$ nanotube reflecting layer deposited on a $TiO_2$ nanoparticle layer in a DSSC. Transmission spectra for different nanotube sizes highlighting photonic stopband locations is presented alongside current densities showing enhanced performances for nanotube PhC layers[272]. Adapted from ref[272]. Copyright 2012, Royal Society of Chemistry. (c) A 3D silicon IO used as a reflecting layer in a DSSC atop a nanocrystalline $TiO_2$ anode with a reflectance spectrum of the silicon IO included. Larger current densities are observed for DSSCs including an IO reflecting layer[273]. Adapted from ref[273]. Copyright 2011, Wiley-VCH.

PhC materials find application in DSSCs as a means of promoting scattering and reflection effects for light incident on the photoanode. Typical photoanodes in $TiO_2$ DSSCs consist of a layer of mesoporous $TiO_2$ nanoparticles with a large surface area loaded with dye molecules. $TiO_2$ IOs with photonic stopbands in the visible range acting as the electrode in DSSCs were found to underperform compared to nanoparticle anodes, attributed to the superior dye loading ability of nanoparticle layers[274]; however, the potential of the IOs to act as tunable dielectric mirrors to enhance dye absorption was highlighted. Different PhC configurations have been explored in conjunction with mesoporous $TiO_2$ electrodes to establish an improvement in the light harvesting ability of the photoanode using the optical properties of the PhC as a reflector or scattering region. Porous 1D PhCs, consisting of alternating layers of $SiO_2$ and $TiO_2$, deposited on the back surface of a nanocrystalline $TiO_2$ photoanode were shown to enhance power-conversion efficiency in DSSCs; the porous PhCs allowed the electrolyte to interact with dye molecules in the inner layer[271]. Both 1D PhC materials with different photonic stopbands were found to provide an enhancement of the current density; the largest enhancement was achieved when the peak reflection of the PhC matched closest to the absorption band of the ruthenium dye, as seen in Fig. 19 (a).

Similar effects were observed with 2D $TiO_2$ nanotube PhCs prepared on the back surface of a dye-sensitised $TiO_2$ nanoparticle layer[272]. All PhC nanotube structures were found to enhance the power-conversion efficiency when compared to non-PhC nanotubes or nanoparticle photoanodes fabricated at identical thicknesses; the enhancement effect was most pronounced for matching the PhC reflection with the maximum dye absorption[272]. Figure 19 (b) displays the structure and performance enhancement of the nanotube PhC. 3D silicon IOs when successfully deposited onto nanocrystalline $TiO_2$ and ZnO nanowire electrode layers also demonstrated enhanced light-trapping in DSSC systems[273]. Reflections from the pseudo-photonic bandgap of the electrolyte infiltrated silicon IO enhanced the current density recorded for the DSSC, as observed in Fig. 19 (c). A large impediment to 3D PhC incorporation into a DSSC is the fabrication of high quality films on the mesoporous $TiO_2$ surface, though successful examples do exist in the literature[273,275,276]. Multilayers of IOs with differing lattice constants and optical responses have also been suggested for further spectral light harvesting in DSSC systems[277] with a two-layer IO system reporting an enhancement in a $TiO_2$ DSSC[276].

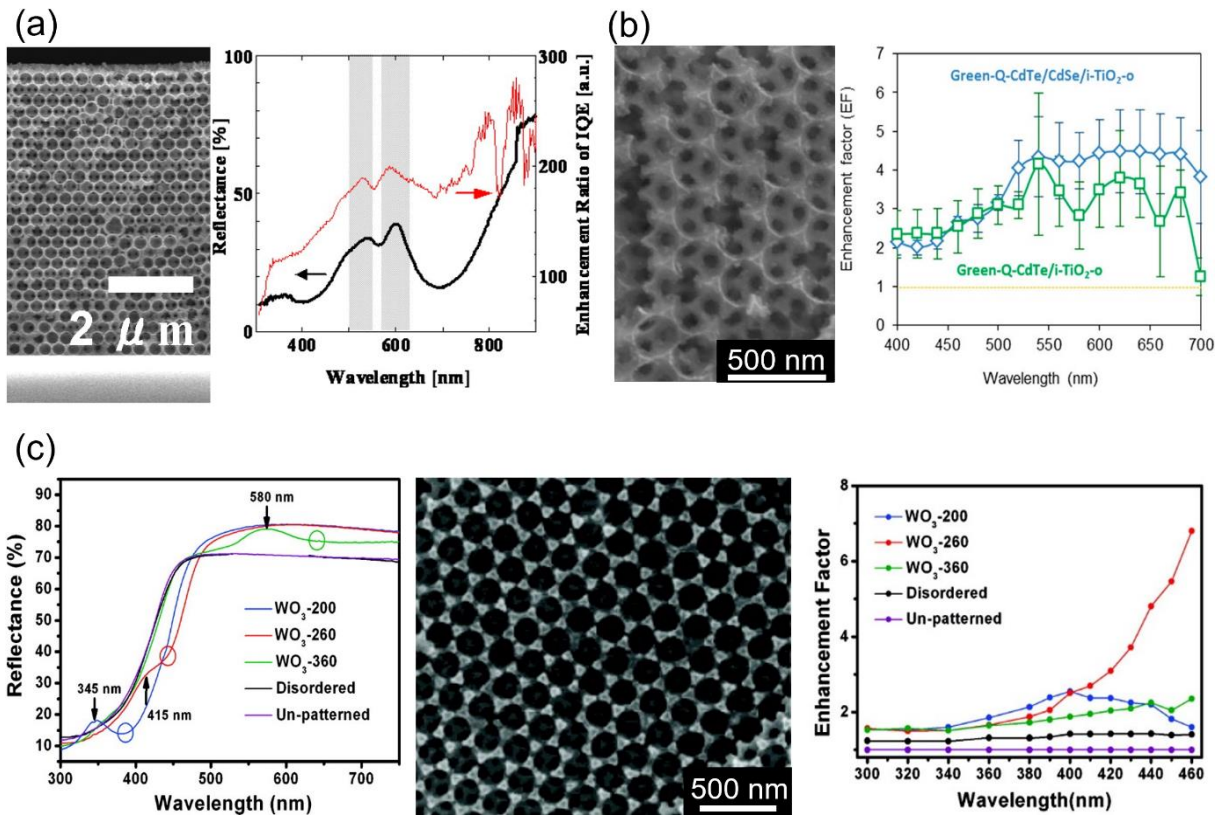

**Fig. 20** (a) An inverse crystalline silicon opal showing reflectance (black line) and solar cell enhancement ration (red line) with photonic bandgap reflections, slow photon edges and regions of enhancement highlighted as shaded regions[278]. Adapted from ref[278]. Copyright 2011, American Institute of Physics. (b) Enhancement factors of CdTe/CdSe (blue line) and CdTe (green line) quantum dot-sensitised $TiO_2$ IOs showing a 4-fold enhancement for slow photon effects overlapping with electronic absorption[279]. Adapted from ref[279]. Copyright 2020, American Chemical Society. (c) Reflectance spectra and solar cell enhancement factor for $WO_3$ IOs (sizes 200, 260 and 360) compared to disordered porous and unpatterned nonporous $WO_3$. Slow photon effects for $WO_3$-260 with the highest enhancement factor[280]. Adapted from ref[280]. Copyright 2011, American Chemical Society.

Finally, IO materials have also been explored in solar cell devices in order to exploit slow photon effects. The reduced group velocity of light at the photonic band edge has the potential to increase light-matter interactions, see Section 3 for more details. Inverse crystalline silicon opals treated with hydrogen plasma passivation demonstrated increases in the enhancement ratio of the internal quantum efficiency of the silicon solar cell at wavelengths close to the photonic bandgap and band edges[278]. This amplification in the photon-to-electron conversion efficiency around the photonic bandgap regions was attributed to slow light trapping effects, as seen in Fig. 20 (a). Large-area and high quality IO $WO_3$ photoanodes were found to exhibit large increases in the photon-to-electron conversion efficiency compared to disordered porous and unpatterned nonporous $WO_3$ electrodes[280]. Between the various IO sizes tested, the largest enhancement was observed when the red-edge of photonic bandgap overlapped with the $WO_3$ electronic absorption edge, as seen with the $WO_3$-260 sample in Fig. 20 (c)[280].

Quantum dot sensitised solar cells, where quantum dots are used as absorber sensitisers on semiconductor films, also feature reports of slow photon enhancement effects. TiO$_2$ IOs sensitized with quantum-confined CdSe films displayed a significant amplification of the photon-to-current conversion efficiency with an average enhancement factor of 6.7 ± 1.6 at 640 nm for an IO structure with a photonic bandgap at 700 nm compared to a nanocrystalline TiO$_2$ with similar CdSe sensitisation[281]. This enhancement was attributed to a blue-edge slow photon effect overlapping with CdSe absorption edge (600 – 650 nm). More recently, CdTe/CdSe quantum dot sensitized TiO$_2$ IOs were shown to feature a 4-fold enhancement factor of the photon-to-current conversion, compared to nanocrystalline TiO$_2$ films, when the blue or red-edge of the photonic stopband was tuned to match the quantum dot absorbance edge[279]. Smaller gains in the enhancement factor were noted for quantum dot absorbance edges centred in the photonic stopband compared to blue or red-edge effects, heavily suggesting a slow photon amplification effect. Figure 20 (b) displays the slow photon enhancement factor for both CdTe/CdSe and CdTe quantum dots on TiO$_2$ IOs.

These are just some of the strategies combining the optical performance of PhC layers with the light harvesting capabilities of semiconductors materials and sensitizer molecules present in solar cell devices. Optically, the PhC layers in solar cells are often used to promote light absorption through targeted reflectance of specific wavelengths and improving the path length of light through the cell. Some recent works[282,283,284] have attempted to determine the effects of the angular iridescence response of the photonic signature associated with PhCs on solar cell performance, simulating more realistic operating conditions with the changing position of the sun. Other works are focusing on the aesthetic capabilities of PhCs in solar cell devices with a wide range of tunable structurally colored designs options available for decorative purposes in solar cells which incorporate PhC layers[285,286,287,288]. Solar cells featuring controllable PhC structural color are often explored as candidates for smart window designs[261,289,290]. Current research trends appear point toward both a functional and aesthetic role for PhCs in solar cell devices.

# 5 Photonic Crystals in Waveguides and Optical Fiber Designs

Optical waveguides are structured materials designed to transport energies, typically in the visible or infrared regions, between a source and destination. Waveguides are designed to guide light based on how light interacts with different regions of the material. The theory and optics behind traditional optical waveguides are quite well understood[291]. In general, a traditional optical waveguide will consist of a high refractive index core material, surrounded by a lower refractive index cladding, all encompassed by a protective jacket. Waveguides with this design are commonly called optical fibers.

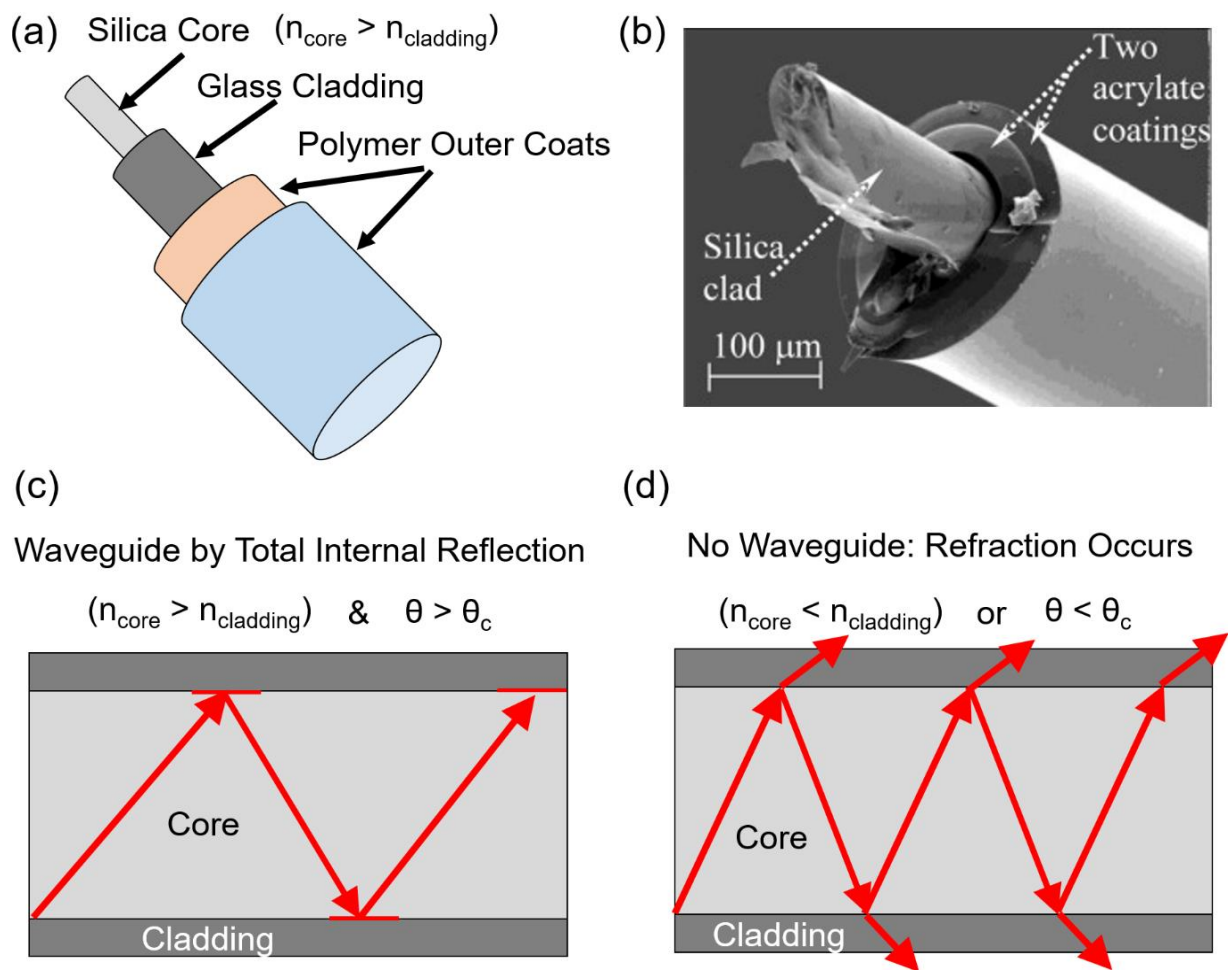

**Fig. 21** Labelled components of a typical silica optical fiber design highlighting the core, cladding and protective outer coating materials for (a) a schematic diagram and (b) an SEM image of a used multimode silica fiber with two protective acrylate coatings[292]. Reproduced from ref[292]. Copyright 2015, Elsevier. Material constraints and light guiding mechanism in an optical fiber operating (c) as a waveguide where total internal reflection occurs for chosen refractive indices and propagating angle and (d) without a waveguide where light refracts through the cladding due to a lower core refractive index or angles smaller the critical angle.

Light is guided from one end of the fiber to the other using the principle of total internal reflection (TIR) between different refractive index regions of the fiber. The constraint of a high index core material is introduced from this mechanism of light guiding; waveguides relying on TIR confine light to the core region of the fiber only if TIR can occur i.e. the core refractive index must exceed the cladding refractive index. The angle at which the light enters the core region must also exceed the critical angle for TIR effects to be observed. The typical structure and waveguide mechanism for light in an optical fiber are shown in Figs. 21 (a) – (d).

Optical fibers are a widely used technology across many different fields of study. Flexible fibers are ideal for imaging internal organs in medicine[293,294], optical fiber lasers have been developed with doped cores to control the refractive index and photon absorption[295,296] and optical fiber refractive index sensors and their applications were earlier discussed in Section 2. The most critical application of optical fiber waveguides is found in optical fiber communication technologies where nearly two billion kilometres of optical fiber are used for telecommunications and data transfer[297]. Internet connections and phone reception are dependent on high speed data transfer and the associated low cost of optical fibers. The conventional waveguide design by TIR is not without its challenges. Most optical fibers use silica glasses in their composition, linking the optical transparency and absorption to the material properties of silica. Silica is transparent in the visible and near-infrared regions yet prone to high attenuation in the mid-infrared and ultraviolet regions[298]. Silica based fiber optical cables of this design are also limited by small refractive index contrasts of the core and cladding material which need to be thermally compatible and the nonlinear optical response of the silica core limits the amount of light which can be transferred[299]. Conventional silica based optical fibers are designed with a careful balance between optical material losses and nonlinearity effects and years of refinement of fiber design has pushed the fabrication technology to close to the innovative limit for these structures[299,300].

PhC fibers are a separate class of optical fiber emerging in optical fiber research since the first demonstration of a silica core surrounded by a silica-air PhC fiber in 1996[301]. PhC fibers were proposed as alternatives to traditional optical fibers, specifically for applications outside of optical communications where standard fibers failed to meet demands such as higher power transport, optical sensing, engineered dispersion and nonlinearity[302]. PhC fibers make specific use of the photonic bandgap in ordered structures to guide light along the waveguide. Typically, PhC fibers are categorised as solid or hollow core fibers depending on the composition of the waveguide. The mechanisms for light guiding in these structures are drastically different, as are the associated applications with each fiber. Solid core fibers are reminiscent of the conventional optical fiber design and function, while hollow core fibers represent a departure from the traditional optical waveguide mechanics used.

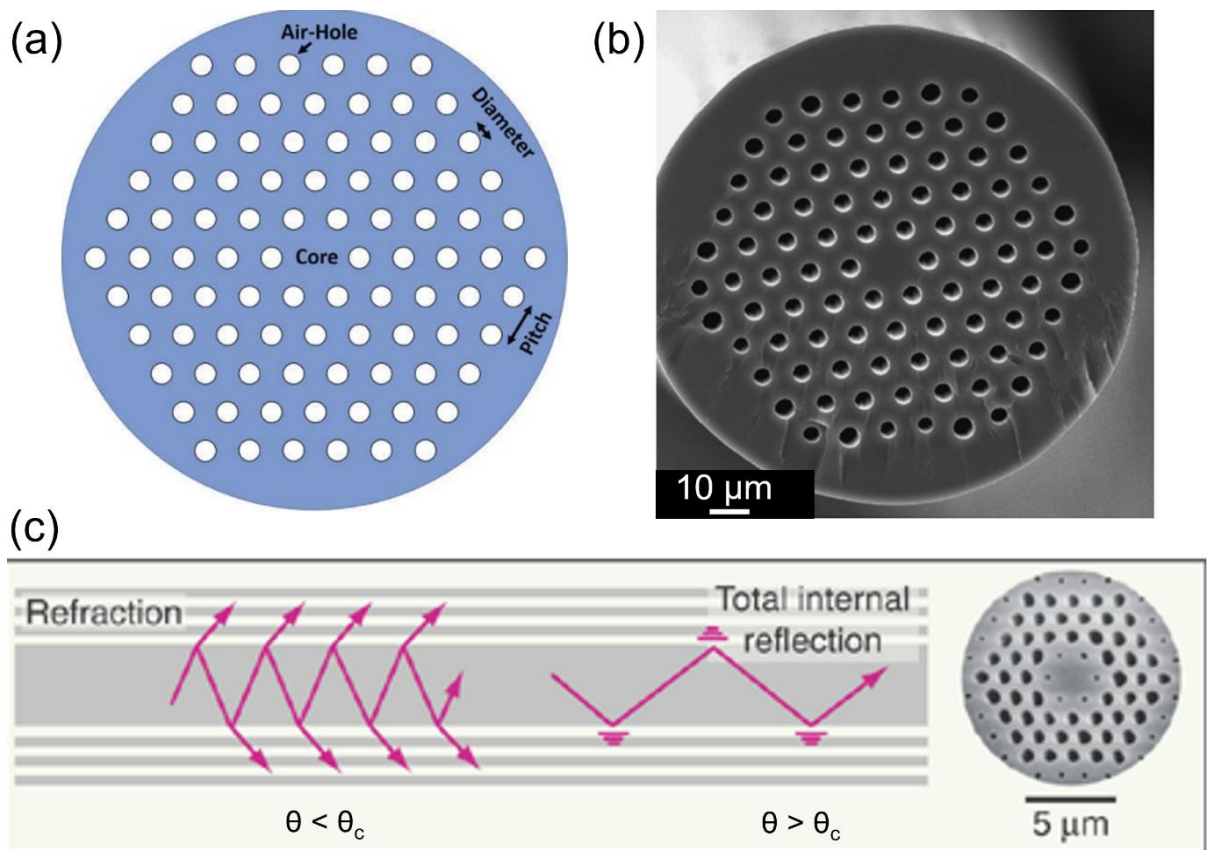

**Fig. 22** (a) A solid core silica PhC fiber cross sectional image, showing a hexagonal PhC arrangement of five rings of air holes surrounding a silica core in a labelled schematic diagram[303]. Reproduced from ref[303]. Copyright 2019, OPTICA. (b) A microscopy image of a solid core 125.3 μm fiber with an average hole diameter of 4.79 μm and an average pitch of 9.97 μm[304]. Adapted from ref[304]. Copyright 2006, OPTICA & IEEE Photonics Society. (c) The light guiding mechanism in solid core PhC fibers is shown governed by total internal reflection effects for angles larger than the critical angle[299]. Refraction effects prevent waveguide formation at angles below the critical angle. Adapted from ref[299]. Copyright 2002, American Association for the Advancement of Science.

First looking at solid core designs in PhC fibers, as depicted in Fig. 22, TIR governs the waveguide mechanism just as with standard optical fiber cables. The earliest PhC fiber designs were solid silica core structures surrounded by a PhC composed of a silica and air matrix with an incomplete photonic bandgap[301]. Light is confined to the core region of these fibers by TIR effects between the silica core of higher refractive index and the silica PhC with a much smaller effective refractive index due to the presence of air holes. The increase in refractive index contrast between the core and cladding affects the dispersion in these fibers, enabling the creation of a single-mode broadband optical continuum over the visible range[305].

Ultra-flattened, near zero dispersion has also been achieved in the infrared region of solid core structures with precise control over the hole shape, size and core diameter[306]. The birefringence of the fiber can also be engineered in the structure by altering the air capillary thicknesses surrounding the silica core, creating birefringent effects up to ten times stronger compared to conventional optical fiber designs[307]. Larger core diameters supporting only a single-mode are also possible, provided the shape and geometry of the PhC fiber are conserved[308]. Larger effective core diameter designs allow for higher power applications while reducing nonlinear effects[309][310]. Many of these effects can be engineered in fibers depending on the specific application. For example, supercontinuum light sources are feasible with PhC fibers due to the higher power output, broad spectral bandwidth and the control over nonlinear effects with applications in optical coherence tomography, spectroscopy and frequency metrology[311][312][313].

Hollow core PhC fibers operate on with a fundamentally different mechanism for guiding light. Hollow core fibers are generally formed as an air core surrounded by a 2D PhC structure. The first demonstration of a single mode hollow core fiber structure was of an air core surrounded by a silica-air PhC[314]. Waveguides in hollow core fibers are created using the complete photonic bandgap of the cladding around the central air core. Specific frequencies of light are confined to the core based on the coherent Bragg reflection from the PhC cladding. Importantly, TIR effects are not occurring in hollow core fibers as the refractive index of the air core is strictly lower than the surrounding material; light is guided by photonic bandgap reflections only. This distinction in waveguide mechanics has led to the term photonic bandgap fiber often being used to describe hollow core fibers. Light travelling along a hollow core fiber is unique compared to other fiber designs as it does not need to transport through a solid material, like silica. As a result of this, 99% of the optical power in these fibers travels through air instead of silica glass removing the fundamental scattering and absorption processes associated with glass waveguides[315]. The absence of the solid core loss mechanisms and reduced optical nonlinearity effects make hollow core fibers attractive candidates for low loss fiber designs[315][316]. Typical designs of hollow core PhC fibers can be seen depicted in Fig. 23.

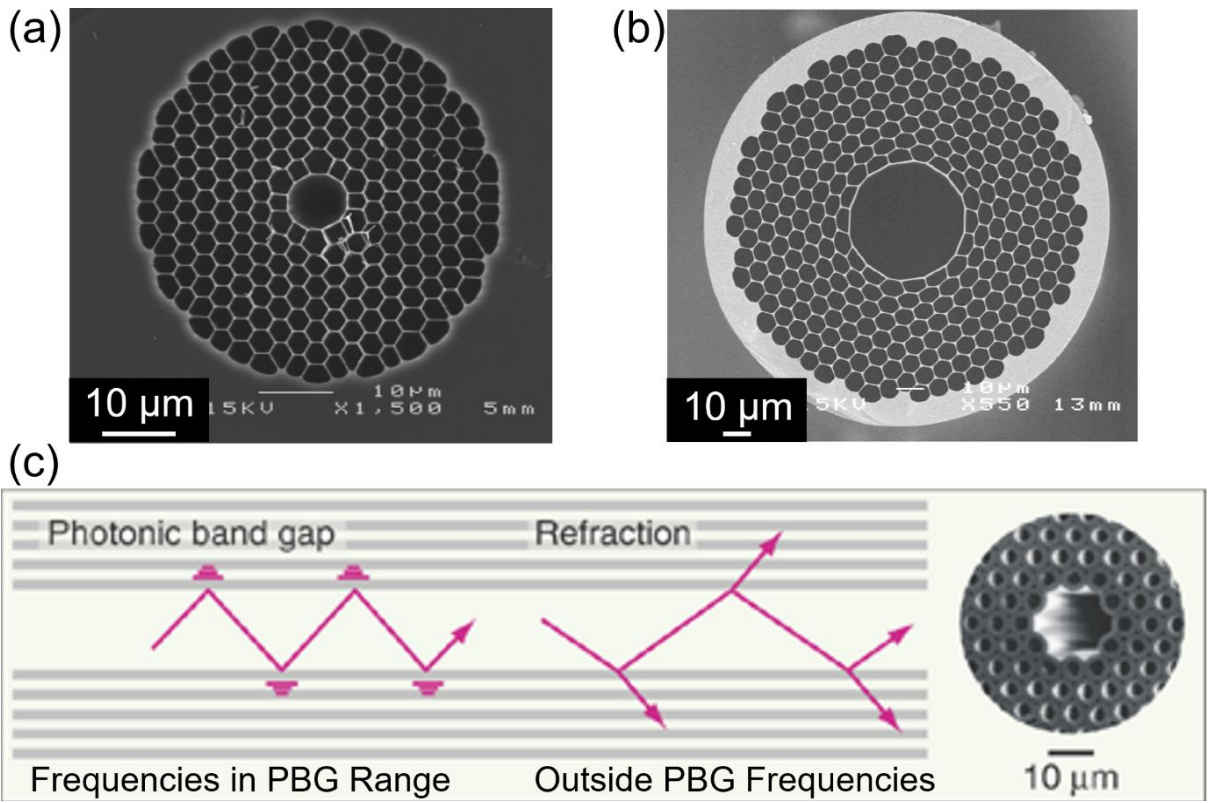

**Fig. 23** (a) Microscopy image of a hollow core silica PhC fiber designs engineered with an 8.2 μm diameter hollow core surrounded by a 7-layer silica PhC used as a waveguide for 1064 nm radiation[317]. Adapted from ref[317]. Copyright 2004, OPTICA. (b) A 40 μm diameter hollow core surrounded by a 7-layer silica PhC designed as a waveguide for Mid-IR wavelengths above 3 μm[318]. Adapted from ref[318]. Copyright 2005, OPTICA. (c) Schematic diagram showing the conditions for establishing a waveguide in hollow core PhC fibers[299]. Frequencies inside the span of the photonic bandgap are guided along the fiber; refraction occurs for frequencies outside the photonic bandgap. Adapted from ref[299]. Copyright 2002, American Association for the Advancement of Science.

Hollow core fiber designs present many opportunities and applications compared to traditional optical fibers. The reduced nonlinear response and absorption of an air core compared to a silica core is reported to reduce the Kerr response (nonlinear refractive index changes) of the fiber by a factor of up to 1000[319,320]. This reduction in nonlinear effects has led to the development of anti-resonant hollow core fibers used for transporting high power ultrashort laser pulses[298,317,320,321] without lengthening the pulse duration, reducing the peak intensity or incurring damage to the fiber, as experienced in conventional silica core waveguides. Ultrashort laser pulses have practical applications in areas such as micromachining[322] and biomedicine[323,324]. Hollow core PhC fibers are also ideal for transporting wavelengths of light which fall outside the silica transparency window. Silica is prone to damage from UV wavelengths; anti-resonant hollow core fibers can better transport UV light due to significantly reduced interactions between the guided light and solid material in the PhC cladding[298,325]. While mid-IR waveguides in solid core fibers can be achieved by using modified chalcogenide glasses in the fiber design, hollow core fibers have the advantage of being less fragile and transmitting deeper into the infrared[326].

One area which has exploited the photonic bandgap confinement in hollow core fibers is nonlinear optics[327]. The design of hollow core fibers allows a tailored nonlinear optical response depending on the application; air-filled fibers with low intrinsic nonlinearities are ideal for transporting high power pulses and liquid or gas-filled fibers can create high levels of nonlinearity for low power frequency conversion effects. Nonlinear materials, such as liquids and gasses, can also be introduced into the hollow core region for specific applications requiring increased light path lengths, high intensities, high optical damage threshold and a controlled group velocity dispersion by alteration of the pressure[328]. Argon gas-filled hollow core PhC fibers have demonstrated a capability of acting as bright, spatially coherent and wavelength tunable UV laser sources using the nonlinear response of a 27 μm gas-filled fiber when pumped with IR ultrashort pulses[329]. Variation of the pulse energy and Ar gas pressure allowed the wavelength of the UV light produced to be tuned between 200 and 300 nm with conversion efficiencies up to 8%.

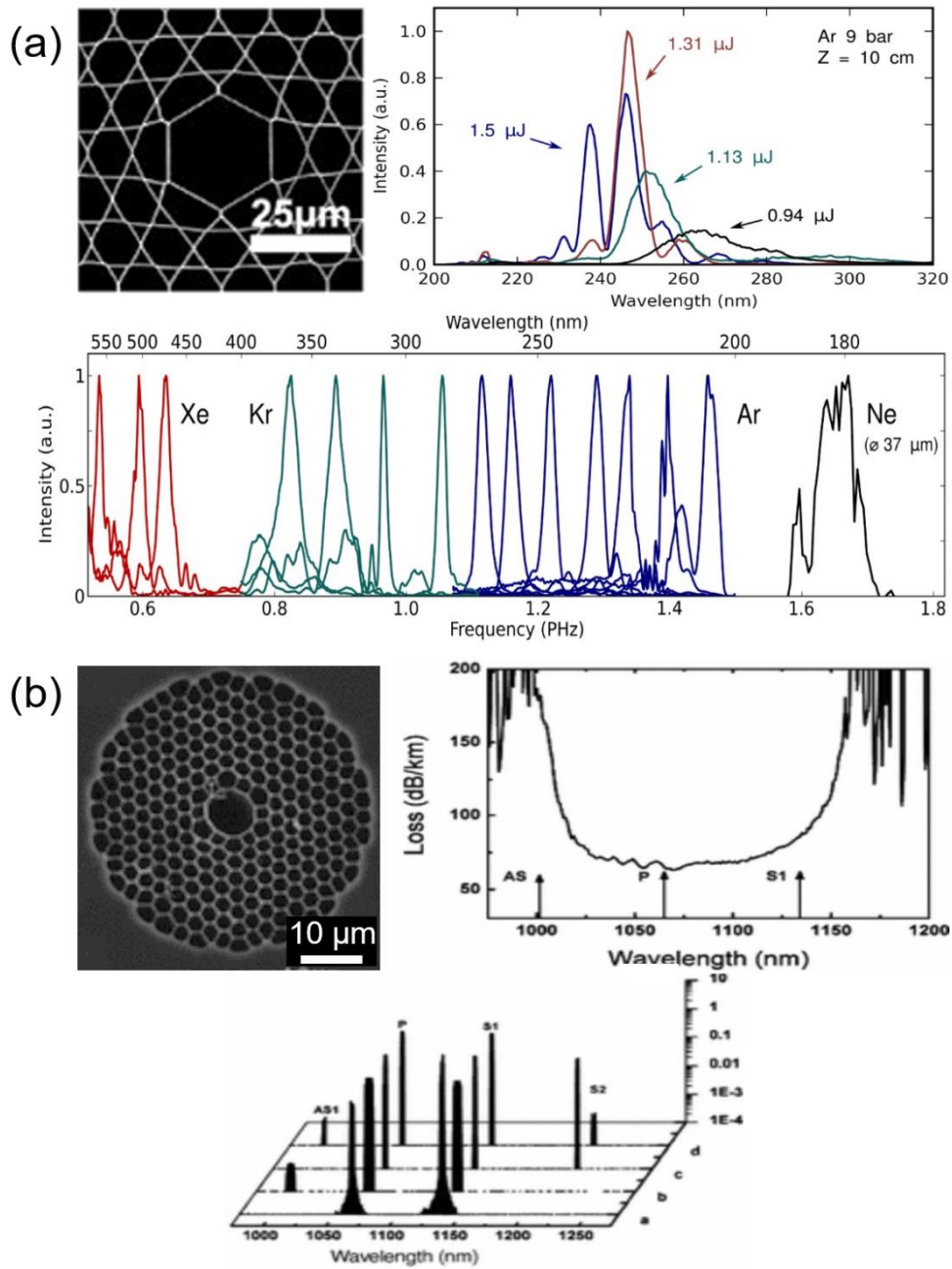

**Fig. 24** (a) A 27μm hollow core PhC fiber with a Kagomé design. The Ar-filled fiber pumped at 800 nm displays a range of different spectral shapes depending on the power of the input pulse. Different gas pressures and pump energies vary the spectral output for Xe and Kr gas in the same fiber. A deeper UV output is possible with Ne gas in a larger fiber[330]. Adapted from ref[330]. Copyright 2013, OPTICA. (b) A hollow core fiber design filled with $H_2$ gas for detection of rotational Raman bands with low loss across the fiber for the pump wavelength (P), Stokes (S1) and anti-Stokes (AS) signals. Output could be tuned by varying polarisation and input pump power[331]. Adapted from ref[331]. Copyright 2004, American Physical Society.

A follow up study[330] found that the gas could also be changed resulting in optical frequencies for Xe and Kr and a deeper UV signal for Ne, as seen in Fig. 24 (a). Low threshold stimulated Raman scattering is another example of an application of hollow core fibers. Using photonic bandgap guided light in hollow core fibers, frequencies were filtered in light interacting with $H_2$ gas, suppressing stronger vibrational modes and allowing the observation of weaker rotational bands at energy thresholds up to $10^6$ lower than previous observations[331][332]. Varying the pump polarisation further allowed the Raman output spectral components to be controlled, as seen in Fig. 24 (b). Lastly, there have even been reports of utilising the slow photon effect in conjunction with nonlinear optics in PhC waveguides[333]. Enhancements of self-phase modulation induced spectral broadening of output signals and two photon absorption[334] alongside delayed transmissions of picosecond optical pulses[335] are some examples of effects being attributed to slow photon behavior in 2D silicon waveguides.

Recent studies with PhC fibers, primarily hollow core designs, are exploring the vast range of analytes which can be detected using PhC fiber sensors. Sensors for chemical and biological species are consistently emerging from the literature. Gas sensors for $CO_2$[336], $N_2O$[337] and $CH_4/HF$[338], chemical sensors for kesosene concentration in fuel[339] and detecting illicit drugs[340], health monitoring biosensors for glucose[341] and cholesterol[342] and health diagnostic sensors for detecting various types of cancer cells[343][344][345] are among some of the emerging applications for PhC fiber sensors. Some PhC fiber sensor research aims to improve the selectivity and sensitivity of detections by combining the waveguide mechanism of the fiber with the surface plasmon effect[346][347][348], incorporating metal particles or thin metal layers into the fiber to enhance detection sensitivity. Other research works investigate the material used to fabricate the fiber to identify optimal operating conditions with thermoplastic polymer materials, such as Zeonex and Topas, showing promising applications for Terahertz frequency ranges[349][350]. Emerging graphene-based PhC fibers are being explored due to graphene's high surface to volume ratio, low associated cost with simple synthesis and its ability to protect against oxidation of metals (e.g. Ag) with its high electron density preventing molecules from passing through the ring-like structure for surface plasmon applications[351][352][353].

# 6 Photonic Crystal Structural Porosity and Interconnectivity Applications

Thus far, the wide array of optical applications for PhC materials have been discussed across several disciplines. When discussing optical implementation, the ability of the material to attenuate, tune or reflect incoming light is a major consideration for the incorporation of repeating structured dielectric layers; the role of the structure is to provide the required optical response. Beyond optical applications, PhC materials have found practical uses relating to their structure alone. 3D PhC structures, such as IOs in particular, are often prepared for the porous interconnected structure of crystalline material as opposed to any optical considerations. In many cases where the structure of ordered material is central to its implementation, IOs and other 3D PhCs are often classified as three-dimensionally ordered macroporous (3DOM) materials[82,354,355]. From a structural standpoint, some of the attractive qualities of 3DOM materials include an increased surface-to-volume ratio compared to bulk materials, a bi-continuous material with an interconnected solid material and an interconnected pore system, efficient mass transport trough the structure with low tortuosity and relatively large pores to allow for surface functionalisation of the material walls[83,356]. For porous structures, the simplicity and inexpensive size-controlled fabrication method for IOs alongside the wide range of compatible materials add to the appeal of using IO 3DOM materials versus other ordered porous materials[357]. In this section, applications involving the structure of 3DOM materials are discussed to shed some light on the potential benefits arising from utilising materials like IOs other than their ability to modulate light propagation.

IO materials have emerged as ideal structured candidates in certain areas of biomedicine[114,115]. In general, the scale of these types of IO materials are much larger than their photonic counterparts with pore sizes of around 100 μm. However, the colloidal crystal templating and inversion procedure is identical, producing structurally similar IO materials, albeit with larger repeating lattice sizes etc. The role of IO materials in biomedical research is to act as structural scaffolds or biomimetic supports with an ordered and interconnected surface; purely a structural function with no optical component. In biomedicine, the area of cell culturing and tissue engineering has explored the benefits of IO materials as structural supports. Tissue engineering involves transplanting a biofactor (cells, genes or proteins) into a porous degradable material scaffold; the scaffold acts to provide temporary mechanical support and deliver the biofactor throughout so that the regenerated tissue assumes function as the scaffold degrades[358]. Aside from initially supporting the cells and aiding in their proliferation, the architecture of the porous support ultimately decides the shape of the newly grown tissue in an ideally reproducible and economically viable large scale[359].

A major factor found to influence the reproducibility and the homogeneity of cells grown across the porous scaffold is the uniformity of the scaffold material, with uniform IOs outperforming structures with non-uniform pore sizes in terms of diffusion of macromolecules and the uniformity of the fibroblast cell distribution in the IO scaffold[360], as seen in Fig. 25 (a). Non-uniform pore structures can produce regions with insufficient concentration of nutrients for cells, inhibiting homogenous tissue formation. In this case, gelatin microspheres, prepared via sedimentation evaporation, were used as the sacrificial template and infiltrated with PGLA under vacuum suction to form the IO structure. A common IO material used in tissue engineering is poly(D,L-lactide-co-glycolide) (PLGA)[360,361,362], a polyester material approved for clinical trials with a controllable degradation rate depending on the lactic and glycolic acid ratios used in preparation[363]. A study of PLGA IOs found that the pore and window (interconnected circular openings between adjacent pores) size of uniform IOs had an effect on the physical properties (compressive moduli) and migration depth of cells in the IO scaffold[361]. Larger pores and windows featured a greater migration depth of cells with a lower compressive modulus. PLGA IO scaffolds with gradations in hydroxyapatite (HAp) nanoparticle minerals allowed for a spatial control of cells grown across the scaffold, potentially beneficial for emulating connective tissue-to-bone interfaces[362]. Outside of PLGA IOs, chitosan[364] and silk fibroin[365] IO scaffolds have also been investigated for tissue engineering. For the silk fibroin IO scaffolds, the ordered geometry of the IO structure was found to promote almost double the mineralisation of human mesenchymal stem cells compared to a silk fibroin scaffold prepared with a similar average pore diameter by salt-leached processes[365], as seen from calcium concentrations in Fig. 25 (b). In more recent works, hydrogel IO systems have been used as targeted drug delivery systems with drug release controlled by application of NIR radiation for wound healing[366] or spinal cord repair[367]. Anisotropic IO structures, deformed by stretching an axis in a particular direction, have been utilised to direct the orientation of cell growth for controlling nerve cell growth[368] and preventing skin scar formation[369].

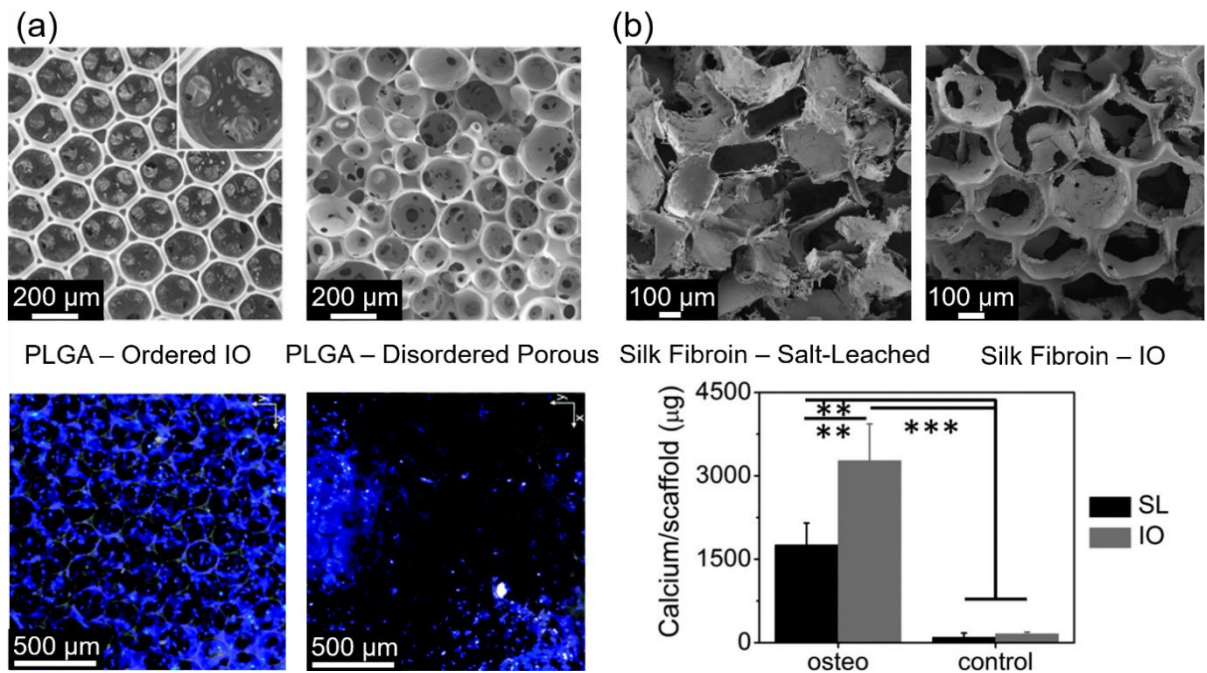

**Fig. 25** (a) Comparison between PLGA scaffolds using an ordered IO and a disordered porous scaffold with fluorescence micrographs showing fibroblast cell distributions throughout each structure after 7 days of cell culture[360]. Adapted from ref[360]. Copyright 2010, American Chemical Society. (b) Comparison between silk fibroin scaffolds prepared from a salt-leaching process and an IO. Greater mineralisation is detected for the IO with higher calcium content per scaffold[365]. Adapted from ref[365]. Copyright 2017, Wiley.

The macroporous structure of IO configurations has also attracted some attention for its potential to act as a separation or filtering membrane. Specifically, 3DOM IO structures with highly interconnected and regular pore sizes have been explored as microfiltration[370] or ultrafiltration[371] separation membranes. Porous membranes are often used in filtration systems where the membrane attempts to remove particles suspended in a liquid, with water purification being a prime example of the process. Porous membranes are often investigated in terms of the separation efficiency (the efficiency at which particles are removed) and the permeation flux (the stable rate at which a liquid feed can be processed). An enhancement of the separation efficiency often results in a decrease of the permeation flux and vice-versa. 3DOMs are appealing as separation membranes due to their high pore uniformity with the potential to act as highly selective separators for desired species[372]. The ability to tune pore size in fabrication allows the sizes of filtered particles to be controlled while an interconnected pore system in a membrane is advantageous for establishing a slower rate of flux decline from pore blockage[373]. Compared to other methods of fabricating 3DOMs, including non-solvent-induced phase separation of block co-polymers and microelectromechanical systems, IOs offer a simple solution of producing a uniform porous membrane without the need for highly specialised machinery for patterning or fine tuning the stochastic porous structures for block co-polymers.

IO membranes prepared from photo-initiated polymerisation of monomer solutions of hydroxybutyl methacrylate and 2-hydroxyethyl methacrylate at differing ratios were demonstrated to be capable of particle fractionation of silica particles in a feed solution[370]. For samples prepared from 835 nm silica templates, smaller interconnecting pore windows of about 200 nm in the IO membrane prevented larger particles (440 and 835 nm silica particles) from transmitting across the membrane while allowing smaller 60 nm silica particles to pass through. Many battery systems often contain a separator membrane in the electrolyte between the anode and the cathode to prevent electrical contact between the electrodes and maintain ionic flow. Uniform IO structures fabricated from UV curing of ethoxylated trimethylolpropane triacrylate (ETPTA) polymer in a $SiO_2$ nanoparticle matrix have shown promising applications as battery separator membranes relative to commercial trilayer polypropylene / polyethylene / polypropylene (PP/PE/PP) separators with randomly distributed pores[374]. In a $LiCoO_2$ cathode and graphite anode battery configuration, the ETPTA IO separator membranes showed similar performances to the commercial separators at low mass loadings and slow cycling rates and improved performances for high mass loadings and fast cycling rates, as seen in Fig. 26 (a). The improvement in performance was attributed to the increased ion transport through the well-ordered, nanoporous structure of the IO separator membrane.

Modifications to macroporous 3DOMs have allowed IO membranes to modify their particle selectivity. Particle-nested IO structures, consisting of polystyrene particles embedded in a poly(urethane acrylate) IO matrix, have been shown to possess excellent separation efficiency of nanoparticles with the potential to tune the size selectivity of the particle separation via alteration of the size ratio between the nested polystyrene particle and the IO pore window[375]. Nanoparticle separation is attributed to nanochannels formed between the nested polystyrene particle and the IO walls; Au nanoparticles exhibited high rejection rates when the estimated size of the nanochannels were smaller than the Au nanoparticle sizes, as seen in Fig. 26 (b). A hybrid structure combining the macroporous nature of a poly(urethane acrylate) IO matrix with mesoporous block-copolymer (polystyrene-block-polymethyl methacrylate) structures inside the interconnecting pore windows has demonstrated a size-selective nanoparticle separation without comprising the high permeability of the IO structure[376]. 18 nm nanosieves in the block-copolymer material of the hybrid membrane structure displayed excellent size selectivity in filtering Au nanoparticles with sizes above 18 nm showing complete separation, as seen Fig. 26 (c).

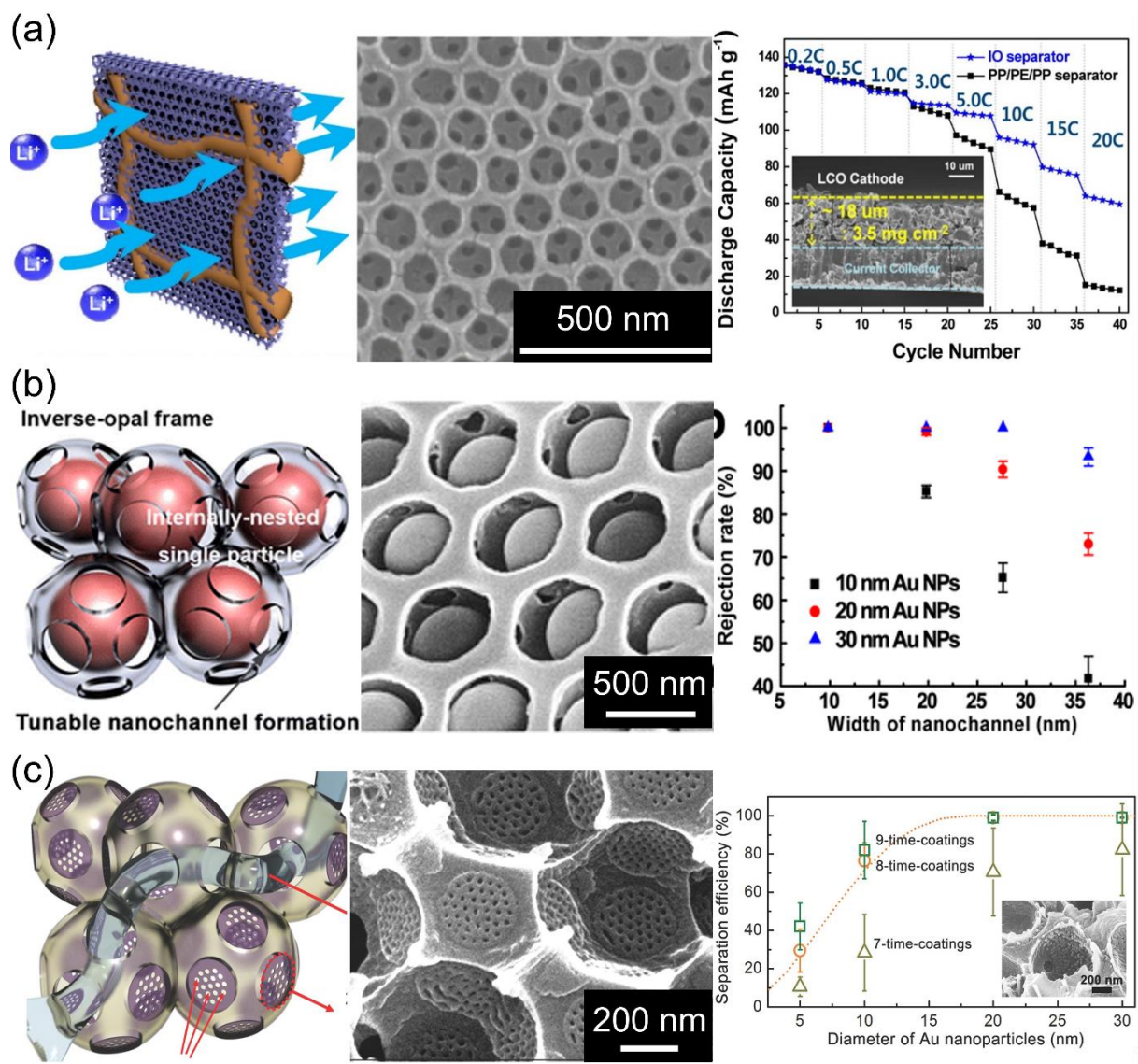

**Fig. 26** (a) SEM image and concept art of an ETPTA IO separator membrane used in a LiCoO$_2$ versus graphite battery[374]. Discharge capacities are compared for an IO separator versus a commercial randomly porous PP/PE/PP separator for various cycling rates. Adapted from ref[374]. Copyright 2014, American Chemical Society. (b) Schematic diagram and SEM image for a poly(urethane acrylate) IO with a nested polystyrene particle within the IO cavity[375]. The rejection rate of Au nanoparticles of different sizes versus estimated widths of nanochannels is also shown. Adapted from ref[375]. Copyright 2014, American Chemical Society. (c) Schematic diagram and SEM image for a hybrid macroporous poly(urethane acrylate) IO with mesoporous block-copolymer nanosieves. The separation efficiency of the structure with 18 nm nanosieve dimensions is shown for Au particles of various diameters[376]. Adapted from ref[376]. Copyright 2014, Wiley-VCH.

The interconnectivity and increased surface area exposed in 3DOMs has attracted attention for catalytic applications where catalytic reactions are increased due to the macroporous nature of the structured material. Ceria-zirconia IOs ($Ce_{0.5}Zr_{0.5}O_2$) were investigated in relation to the role of the macroporous interconnected geometry of the 3DOM in relation to the catalytic oxidation of propane and compared to $Ce_{0.5}Zr_{0.5}O_2$ catalysts produced from crushed IOs or featuring no template[377]. The macroporous IO structures featured the best performances, attributed to the increased tortuosity of the gas transport path in the other structures investigated, as seen in Fig. 27 (a).

The ordered IO structure can also function as an excellent catalyst support for loading of metal particles onto a semiconductor surface. The high surface area and porosity of an $SnO_2$ IO was utilised for functionalisation of the IO material with monodispersed immobilised Pd nanoparticles which were well dispersed throughout the structure and easily removed post reaction[378]. Compared to commercially available Pd/C structures, the Pd/$SnO_2$ IOs demonstrated formic acid oxidation with an onset voltage ~180 mV below the commercial standard, as depicted in Fig. 27 (b). This enhancement effect was attributed to the uniform distribution of particles and the porosity of the IO structure facilitating increased access to catalytically active sites. Some recent works choose to adopt IO structured materials primarily for their ability to easily increase the specific surface available through their natural porosity, thus increasing the number of active catalytic sites for the photocatalyst material[379,380]. Other works choose IO materials for their ability to easily accommodate metal particles onto a semiconductor surface due to the open network of material alongside the intrinsic structural porosity, such as noble metal decoration onto $TiO_2$[381,382] and $ZnO$[227,383] photocatalysts. Double inverse opal structures have also been proposed as a method to confine particulate photocatalysts to a structure without decreasing the active surface area[384]. From examples in the literature, photocatalysts are frequently structured as 3DOMs, both for the strucutural advantages of the macroporous structure and optical phenomenon such as the slow photon effect, see Section 3 for more details.

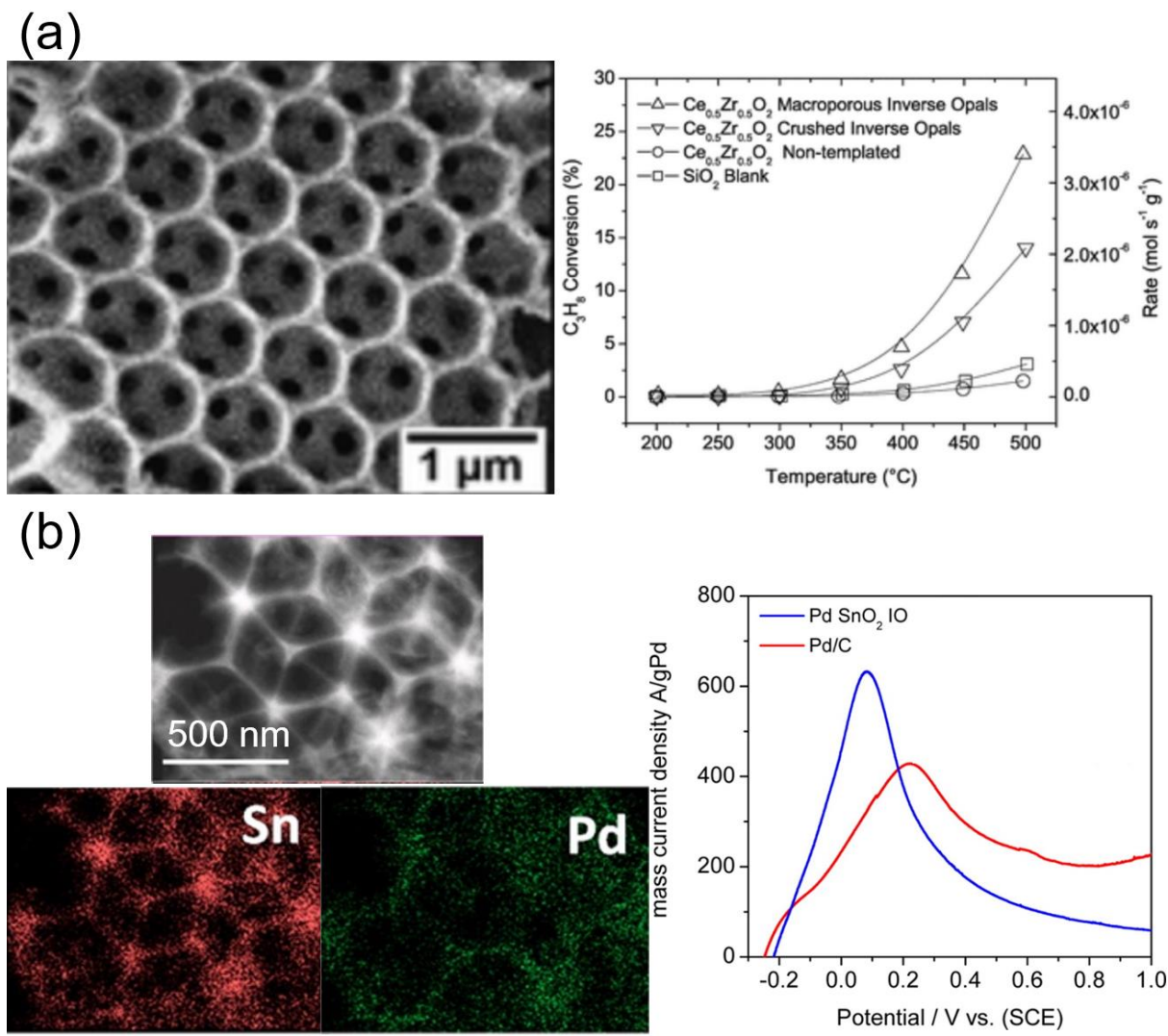

**Fig. 27** (a) Ce$_{0.5}$Zr$_{0.5}$O$_2$ IOs with a macroporous structure showing an enhanced catalytic oxidation of propane compared to non-templated and crushed IO structures[377]. Adapted from ref[377]. Copyright 2008, Royal Society of Chemistry (b) SnO$_2$ IOs with Pd nanoparticles dispersed throughout the structure[378]. The peak potential for catalytic formic acid oxidation is compared for the Pd SnO$_2$ IO and commercially used Pd/C structures. Adapted from ref[378]. Copyright 2013, American Chemical Society.

IO 3DOMs have seen widespread use in battery electrode research in recent years with the macroporosity of the interconnected electrode being investigated for lightweight, high cycling rate and high energy density applications. A vast array of different material candidates have been structured with an IO geometry for evaluation of the electrochemical performance as an anode or cathode. On the anode side silicon[385], carbon[386], germanium[387], TiO$_2$[111], GeO$_2$[388] and Co$_3$O$_4$[389] are among some of the materials for which the incorporation of the IO structure has been investigated. IO structures of LiFePO$_4$[390], V$_2$O$_5$[391] and sulfur[392] (encapsulated by a polypyrrole IO framework) have been explored as cathodes. From the earliest application of a 3DOM to an electrode[393], the electrochemical benefits of an IO structure were proposed as a reduction in electrode polarisation compared to bulk materials where mass transport is inhibited by the high tortuosity of ion diffusion paths. Lithium-ion battery electrodes with a 3D structure are thought to reduce the overall cell resistance by improving the electrolyte transport, creating a fast charging capability and reducing the risk of lithium plating in the cell[394]. For IOs, the macroporous structure facilitates electrolyte access to the electrode material at interstitial points. The IO material walls are also nominally thin relative to the IO pore size, often just several nanometres thick, allowing for short ion diffusion paths and creating high rate capabilities.

In terms of the specific energy associated with an electrode, the specific capacities (capacity per mass of material) for a range of different electrode designs in lithium-ion batteries have been investigated relative to IO designs. For flexible $TiO_2$ electrodes deposited on carbon cloth, two anatase $TiO_2$ IO electrodes (3T-CC – annealed in air and 3T-C-CC – annealed in Ar) were investigated versus an anatase $TiO_2$ nanoparticle slurry on carbon cloth (NP-CC) with carbon black and polyvinylidene difluoride additives[395]. Figure 28 (a) displays the performance of the electrodes compared at a high cycling rate of 10 C with the superior performance of the IO materials displaying larger specific capacities compared to the nanoparticle coating; a larger surface area in contact with the electrolyte and fast lithiation kinetics were proposed for the superior IO 3DOM performance. A separate study of rutile $TiO_2$ IOs[111] found that the IO electrode composition was maintained alongside a stable specific capacity across 5000 cycles indicating long-term structural stability and cycle life, even when tested at high cycling rates.

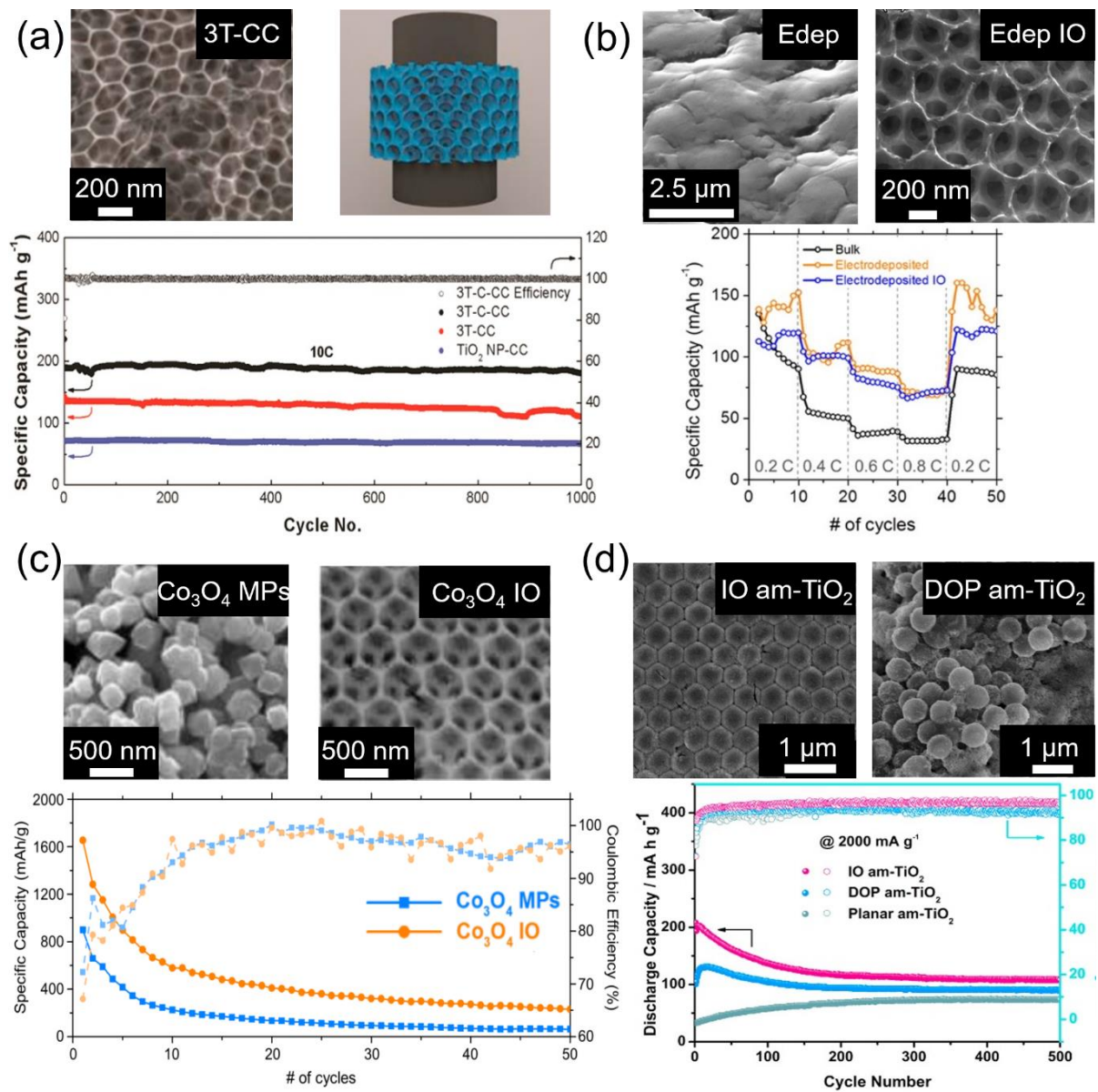

**Fig. 28** (a) Flexible anatase TiO$_2$ IO electrodes prepared on carbon cloth (red and black data) showing superior specific capacity compared to anatase TiO$_2$ nanoparticle (blue data) slurry[395]. Adapted from ref[395]. Copyright 2016, Elsevier. (b) SEM images of electrodeposited V$_2$O$_5$ films for non-structured and IO-structured electrodes with higher specific capacities at rates below 1 C for electrodeposited films compared to bulk V$_2$O$_5$ films[391]. Adapted from ref[391]. Copyright 2015, American Chemical Society. (c) SEM images of Co$_3$O$_4$ micro particles and Co$_3$O$_4$ IO anodes. The specific capacity for the electrodes is higher with the IO structure[389]. Adapted from ref[389]. Copyright 2017, IOP Publishing. (d) SEM images of hollow amorphous TiO$_2$ spheres in IO and disordered porous architectures. The IO architecture shows the highest specific capacity compared to TiO$_2$ disordered porous and planar films in a sodium-ion battery system[396]. Adapted from ref[396]. Copyright 2017, Elsevier.

At low cycling rates (< 1 C) electrodeposited V$_2$O$_5$ cathodes, structured with and without an IO architecture, were found to boost improved specific capacities relative to bulk V$_2$O$_5$ films, as seen in Fig. 28 (b)[391]. The V$_2$O$_5$ IO cathodes were also found to maintain a dominant intercalation mode response at higher scan rates. The performance of a conversion mode $Co_3O_4$ anode was compared for a 400 nm IO structure and an electrode formed from 400 nm $Co_3O_4$ micro particles, both deposited on stainless steel[389]. The ordered IO structure offered significantly higher specific capacity compared to the disordered collection of micro particles, as seen in Fig. 28 (c). Outside of lithium-ion batteries, sodium-ion batteries have also explored the IO structure for their electrodes[396][397][398]. amorphous $TiO_2$ IO architectures, prepared as hollow spheres using atomic layer deposition, were explored for use in sodium-ion batteries and compared to amorphous disordered $TiO_2$ and planar $TiO_2$ films[396]. The ordered IO architecture featured the largest specific capacity and highest rate performance, as depicted in Fig. 28 (d), which was attributed to good electrolyte wettability enhancing the effective surface ion availability, an important consideration in sodium-ion batteries due to the larger ion size relative to lithium-ion batteries.

The interconnected electrode material may also achieve better electrical conductivity compared to other material designs aimed at improving surface area interactions. In comparison to interconnected 3DOMs, energy densities in composite nanoparticle coatings are often limited by the requirement of additional binder and conducting agents to ensure electronic conductivity between particles despite the higher relative surface areas nanoparticle electrodes could provide[399]. As mentioned previously, the porous IO network allows for greater electrolyte access and easier ion insertion/removal. A common observation with IO electrode materials is an impressive C-rate for the battery material; high specific currents are often shown to be fully reversible while maintaining an impressive specific capacity for the system, an important consideration for high energy applications. The macroporosity of the IO structure has also been proposed to help reduce structural stresses in materials with large volume changes during lithium insertion and removal[385]. Some of these features of IO electrodes are examined in Fig. 29.

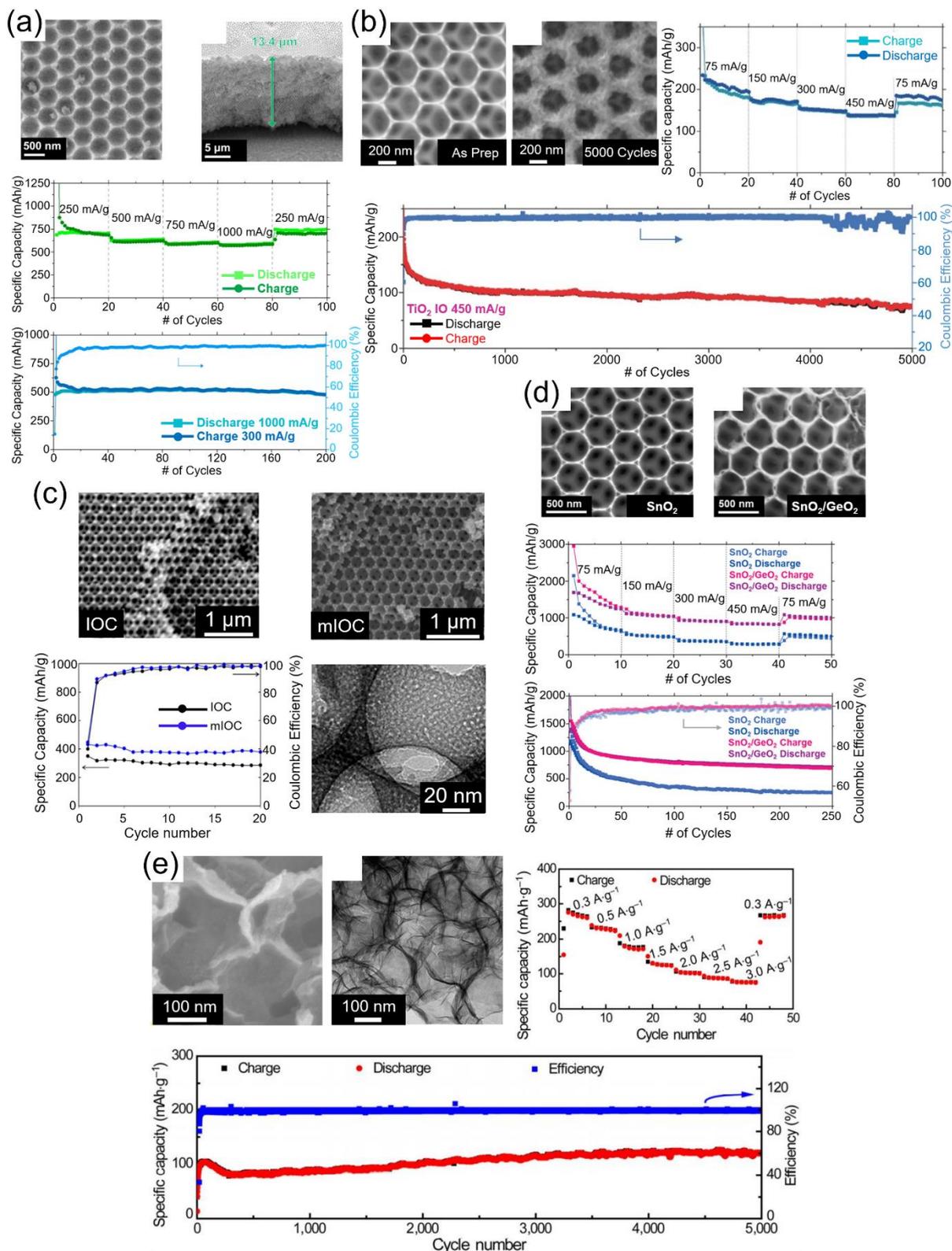

**Fig. 29** (a) GeO$_2$ IO anode exhibiting large specific capacities, high rate capabilities (highly reversible capacities up to 1000 mA/g) and stable asymmetric galvanostatic cycling[388]. Adapted from ref[388]. Copyright 2018, Elsevier. (b) TiO$_2$ IO anode with high rate capability and long term performance stability of up to 5000 cycles[111]. Adapted from ref[111]. Copyright 2017, Wiley. (c) Carbon IO anode materials with a bare carbon IO (IOC) compared to a mesoporous carbon IO (mIOC)[400]. Mesoporous carbon combined with the IO shows superior performance. Adapted from ref[400]. Copyright 2013, American Chemical Society. (d) Nanocomposite SnO$_2$/GeO$_2$ IO anodes with strong rate capability and higher specific capacities when compared to a bare SnO$_2$ IO anode[401]. Adapted from ref[401]. Copyright 2020, Wiley-VCH. (e) MnO2 IO cathode used in an aqueous zinc-ion battery. The stable electrode performance (up to 5000 cycles) and high rate capability is attributed to the stacked IO template[402]. Adapted from ref[402]. Copyright 2019, Springer Science + Business Media.

Figure 29 (a) shows a germanium(IV) dioxide ($GeO_2$) inverse opal explored as an anode candidate in a lithium-ion battery[388]. The highly interconnected IO structure of the electrode does not require additional binders or conducting additives and delivers a high specific capacity of 531 mA h g$^{-1}$ even after 1000 cycles at a specific current of 300 mA g$^{-1}$. Excellent rate capabilities are also demonstrated with fully reversible cycling present for higher specific currents up to 1000 mA g$^{-1}$. The potential for asymmetric discharge (1000 mA g$^{-1}$) and charge (300 mA g$^{-1}$) rates are also explored. A long term study for rutile phase $TiO_2$ IOs is present in Fig. 29 (b), where a $TiO_2$ IO is investigated as an anode material in lithium-ion batteries for up to 5000 cycles[111]. The SEM images illustrate the structural stability is maintained in the IO even after long-term cycling. Different specific currents ranging from 75 to 450 mA g$^{-1}$, approximately yielding C-rates of 0.45 C & 2.68 C respectively, were found to be highly reversible for the material. The IO electrode demonstrated an impressive cycle life, yielding a specific capacity of 95 mA h g$^{-1}$ after 5000 cycles.

Bare carbon inverse opals are compared to mesoporous carbon inverse opals in Fig. 29 (c) in relation to their performance as anode materials in lithium-ion batteries[400]. The extra surface area provided by the mesoporous carbon was thought to establish shorter solid state diffusion lengths. This effect, in combination with the inherent high interconnectivity of the IO template, gave larger specific capacities (32% larger after 20 cycles with a specific current of 100 mA g$^{-1}$) for the mesoporous carbon IO material compared to the bare carbon IO. Composite IO materials consisting of a combination of different active materials have also been explored as electrodes. A $SnO_2$/$GeO_2$ composite anode for use in lithium-ion batteries has been explored as a method of mitigating capacity fade from material volume variations commonly experienced in $SnO_2$ electrodes[401]. Comparisons in electrochemical performance between bare $SnO_2$ IOs and composite $SnO_2$/$GeO_2$ IOs can be observed in Fig. 29 (d). Most notably, a much higher specific capacity and an improved capacity retention can be observed for the composite IO material. Similar IO works with composite materials of have explored similar strategies to improve the electrochemical performance and stability; Some examples include SnO2/Bi2O3 composite IOs prepared on nickel foam[403] and $Li_2FeSiO_4$/C composite nanofibers incorporated into an IO design[404]. The enhanced rate capabilities and impressive cycle life of IO materials have also been explored for non-lithium-ion systems, such as a aqueous zinc-ion battery system with a $MnO_2$ cathode[402] shown in Fig. 29 (e). Even after completing 5000 cycles the aqueous system with the layered $MnO_2$ IO structure shows an impressive specific capacity of 121 mA h g$^{-1}$ at a high current density of 2000 mA g$^{-1}$, effects which were attributed to the structure of the cathode.

Conversely, in spite of the relatively large specific energies associated with IO electrodes, the inherently high porosity of the IO structure (26% material, 74% pore space) creates a lower volumetric energy density, when compared to other electrode structures[399]. Some efforts to enhance the volumetric energy density of IO electrodes have focused on additional material coatings deposited onto the IO scaffold[399 405].

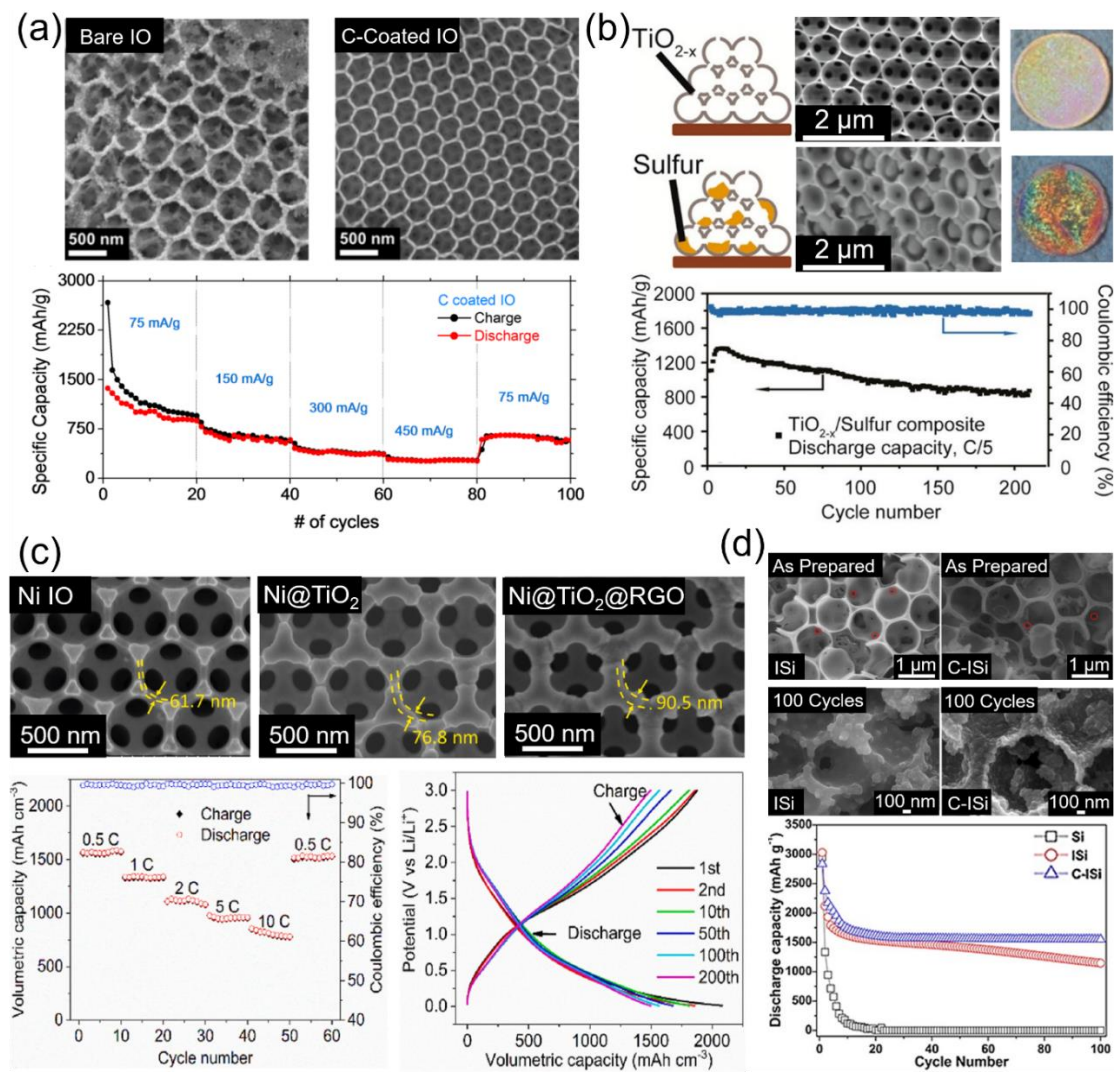

**Fig. 30** (a) A conversion mode Ni-Mn-Co-O IO anode coated with carbon displaying high-rate performance and high specific capacities[406]. Adapted from ref[406]. Copyright 2017, Nature. (b) A hydrogen reduced TiO$_2$ IO structure infiltrated with sulfur to create a lithium-sulfur battery cathode[407]. Sulfur infiltration can be visually seen via structural color changes to the electrode and high reversible capacities are obtained after 200 cycles. Adapted from ref[407]. Copyright 2014, American Chemical Society. (c) A Ni IO anode coated with TiO$_2$ and further coated with reduced graphene oxide (RGO) showing high volumetric capacities and an excellent rate capability and an overall increased active material loading of the coated IO[408]. Adapted from ref[408]. Copyright 2021, Elsevier. (d) Silicon inverse opals (ISi) compared with carbon-coated silicon inverse opals (C-ISi) as anode materials[409]. SEM images and specific capacity are shown after 100 cycles. Adapted from ref[409]. Copyright 2015, Elsevier.

The inherent high interconnectivity and electrical conductivity, combined with the open porous network, of the materials used in IO electrodes makes them ideal candidates for hosting materials within their structure. Figure 30 explores some examples of how the IO structure has been utilised as a structural template support for other materials and the synergistic performance observed.

A conversion mode IO anode (Ni-Mn-Co-O) consisting of mixed oxides of nickel, manganese and cobalt used in a lithium-ion battery, both in a half-cell configuration and full cell versus a $V_2O_5$ cathode, is shown in Fig. 30 (a)[406]. The mixed oxide IO material was further altered via a carbon coating applied to the IO structure with the additional carbon thought to help maintain the structure during cycling. A lithium-sulfur system can be seen in Fig. 30 (b), where the sulfur material for the cathode is infiltrated into a hydrogen-reduced $TiO_2$ IO network acting as a host material to accommodate volume changes and strain during cycling[407]. The IO network acts to confine the sulfur material to the cathode and the surrounding hydrogen-reduced $TiO_2$ material prevents diffusion to the anode by binding with the polysulfide material generated, delivering a high specific capacity of 890 mA h g$^{-1}$ after 200 cycles at a rate of 0.2 C. Similar strategies are used in other lithium-sulfur works with IOs of carbon[410] and polypyrrole decorated with ZnO nanocrystals[411] investigated for confinement.

A Ni IO structural template was modified with two successive material coatings in an attempt to improve the volumetric energy capacity of porous electrode systems, where specific mass performances are often high yet the volume fraction of active material is often low due to high porosity. Figure 30 (c) displays Ni IO coated with $TiO_2$ material through atomic layer deposition and then further coated with reduced graphene oxide (RGO) through spray coating for an anode in a lithium-ion battery[408]. High volumetric capacities are obtained for the electrode system which were thought to arise from the synergistic effects of all active materials. The inner nickel material was stable while highly conductivity and the similar Li-ion intercalation mechanism in $TiO_2$ and RGO helped to maintain structural integrity. Silicon IOs are often investigated as anode materials for lithium-ion batteries due to the large specific capacities associated with silicon electrodes. Volumetric expansion of silicon when alloying with lithium-ions often necessitates the need for a porous, structural network to account for volume changes. Figure. 30 (d) shows a comparison between a silicon IO (ISi) and a carbon-coated silicon IO (C-ISi)[409]. The carbon material deposited onto the IO displayed provided stronger specific capacity retention and helped to maintain the structure of the IO network after 100 cycles.

# 7 Conclusions and Outlook on the Role of Photonic Crystal Materials

The significance of PhCs across a multitude of different areas has been discussed with particular emphasis on the optical application of the structure. The repeating lattice of alternating dielectric materials found in a PhC can be designed for 1D, 2D and 3D structures with plenty of applications incorporating the optical response of these materials. The properties of the photonic bandgap/stopband associated with a PhC are an essential consideration for any optical application of these structures. As such, the desired wavelength position of the photonic bandgap/stopband for a particular application will dictate the size, geometry and material of the PhC utilised. For creating waveguides using the reflection from a photonic bandgap, such as in hollow core photonic crystal fibers, the highest possible reflectivity from the photonic bandgap is desirable, often utilising silica fibers with a full photonic bandgap to accomplish this. The degree of reflectivity is also a concern for PhC reflectors used in solar cell designs, yet not the only consideration. The specificity of the reflection from the PhC can be designed for selectively reflecting certain wavelength ranges in tandem thin film silicon solar cells. Likewise, the porosity of the PhC back reflectors utilised in DSSC systems is a crucial consideration for enabling the electrolyte to interact with dye-sensitised molecules on the electrode surface.

Certain applications act to suppress the peak reflection altogether, treating PhC reflection as an undesirable effect, such as in photocatalysis. Photocatalysts structured as 3DOM, reap the structural benefits of increased surface area and porosity for catalytic reaction sites and the optical benefits of the slow photon effect at the photonic band edges for increased light path lengths and photocatalytic reaction rates. However, peak reflections from PhCs in photocatalysts are often suppressed using the regions of strong electronic absorption of the semiconductor PhC to ensure that light which may be used for initiating photocatalysis is not wasted by reflections from the photocatalyst surface. Conversely, for applications involving PhC sensors, well-defined, easily-observable and tunable photonic bandgaps/stopbands are of paramount concern for optically detecting changes in the sensor system. The sensitivity of the peak reflection/transmission dip to minor changes in the material environment or modifications to the structure is the core function of the PhC sensor. The key point illustrated from the many successful examples of implementation from the literature is that the strength of the PhC structure is inherently tied to the flexibility of its signature optical response. Depending on the application, the photonic bandgap can be maximised for reflection, supressed and tuned for slow photon effects or monitored for material changes simply by control over the dimensions of the repeating structure.

Beyond optical applications, the ordered structure of photonic crystals has attracted attention in certain fields, particularly IO 3DOMs. High surface-to-volume ratios, bi-continuous networks of scaffold material and pores, efficient mass transport and relatively easy functionalisation of the material walls have established IO PhCs as prominent 3DOMs for a range of structural applications[83][356]. The interconnecting scaffold and porous structure of the IO has found applications in tissue engineering, selectively porous membranes, catalysis and battery electrode architectures. Certain areas, such as photocatalysis, have already begun to capitalise on the structural and optical benefits offered by IO PhCs. Slow photon effects can be introduced into the photocatalyst material via tuning of the optical properties of the IO, while exploiting the structural porosity for increased surface area or incorporation of dye molecules and metal particles. There are numerous examples of IO materials being utilised as photocatalysts to capitalise on synergistic effects between the slow photon effect of the IO structure and existing semiconductor modifications e.g. dye-sensitisation and plasmonic nanoparticle resonances. The slow photon optical activity of the IO structure is used to improve the performance of the semiconductor material.

Moving forward, the potential of PhCs will most likely be realised from identifying areas where the inclusion of an ordered architecture with a signature optical response can enhance or add functionality to existing technologies. Recently, in-situ and operando analysis techniques are a prime example of an emerging application involving IO structures. Structuring existing successful materials as an IO has provided an opportunity to obtain real-time data using a variety of different techniques. When Cu-In alloys used in the electrochemical reduction of $CO_2$ adopted an IO structure, the Cu-In IO displayed efficient $CO_2$ reduction while also allowing a sensitive detection of the $CO_2$ reduction intermediates using surface-enhanced Raman scattering[412]. The inclusion of the IO structure dramatically enhanced the Raman signal of the Cu-In alloy when compared to a structure without an IO design, enabling in-situ Raman analysis due to the enhanced signal. This effect is thought to be facilitated by an electric field enhancement where the high curvature of the IO structure concentrates the electric field creating "hot edges". A nickel IO used as a structural scaffold for silicon material has demonstrated a capability of operando identification of strain in an anode used for a lithium-ion battery[413]. Silicon is known to experience a volume expansion upon lithiation; on an IO architecture the silicon is free to expand into the IO pores but constrained from expanding into the nickel scaffold, creating a compression on the Ni material. Operando x-ray diffraction monitoring of the strain on the Ni scaffold during lithiation was accomplished using diffraction peak position and peak broadness with the possible identification of amorphous silicon delamination from the scaffold. A similar study[414] has since adapted this in-situ strain measurement technique for $Ni_3Sn_2$ coated Ni IO anodes.

PhC optical sensors have long demonstrated their capability of acting as in-situ monitors on merit of the sensitivity of the photonic bandgap/stopband. Several of these sensing applications have been highlighted throughout this review. PhC sensors have been developed for identification of solvents[117,129], vapors[145], oils[140], ions[155], biomolecules[154,158] and nerve agents[157] to name just a few. The photonic bandgap of the PhC material defines the sensing response of the structure with systems calibrated to interpret changes to the photonic bandgap position to changes in concentration or structure. Extending this optical sensing ability to incorporate into other functional materials has the potential to establish operando monitoring for different techniques. Recently, a reduced oxide graphene film deposited on the surface of an IO acetylcellulose structure displayed a dual functionality for in-situ monitoring[415]. The electrical resistance of the graphene sheet displayed a sensitivity to the tensile/compressive strain applied, allowing human motion to be tracked in-situ via changes in resistance. The IO component of the composite sensor allowed NaCl concentrations to be tracked as function of peak reflection position, with the potential to act as a glucose sensor.

Certainly, the work summarized here acts to further extend the sensing applications of PhC materials in the energy storage field using the optics from solid-state physics, and knowledge of advances in photoelectrochemistry and related fields. Lithium-ion battery electrodes are now being designed in the form of an IO PhC with reasonably good electrochemical behavior in battery cells, with the added function of an optical sensor arising from the electrode architecture. The presence of an observable photonic stopband allows an operando monitoring of the electrode during battery operation. For lithium-ion batteries, operando characterisation techniques are desirable for the advancing the development of electrode materials due to their potential to provide in-depth information on reaction processes, degradation mechanisms, side reactions, structural evolution, SEI formation and redox mechanisms[416,417]. X-ray diffraction, microscopy techniques, neutron analysis, scanning probe microscopy, Raman and FTIR spectroscopy are among some of the techniques currently utilised for operando monitoring of lithium-ion batteries. Current research shows ongoing efforts to improve the quality, scale and design of photonic crystal films through mixed oxide precursors[128], precursor co-assembly with crystalline nanoparticles[418,236], increased film wettability during assembly[419], particle-nested double inverse opal structures[384,420,421] and photonic crystal heterostrucutures with varying lattice size[126,422,423] or material composition[231,424,425]. These various strategies show a concerted effort to achieve large scale crack-free IO photonic crystal films with a strong optical signature. This focus on higher quality films, often with simplified fabrication procedures, should only act to expand the prevalence of IO photonic crystal films in other research areas, particularly in areas which may benefit from an added optical sensor function in addition to the ordered porous network of material present.

**Appendix I**

Here, the Bragg-Snell model for light diffraction will be derived for the maximum wavelength of destructive interference for light transmission through photonic crystal structures which is indicative of the photonic bandgap position. Technically speaking the condition for maximum constructive interference in the reflected wave is shown, as this corresponds to maximum destructive interference in the transmitted wave. Figure A.1 displays a geometric diagram showing the diffraction angles involved in constructive interference for reflected light waves incident from a medium of refractive index $n_1$ at some angle $\theta_1$ into a photonic crystal medium of composite refractive index $n_{eff}$ (individual particles of refractive index $n_2$) with an interplanar spacing d between the photonic crystal planes.

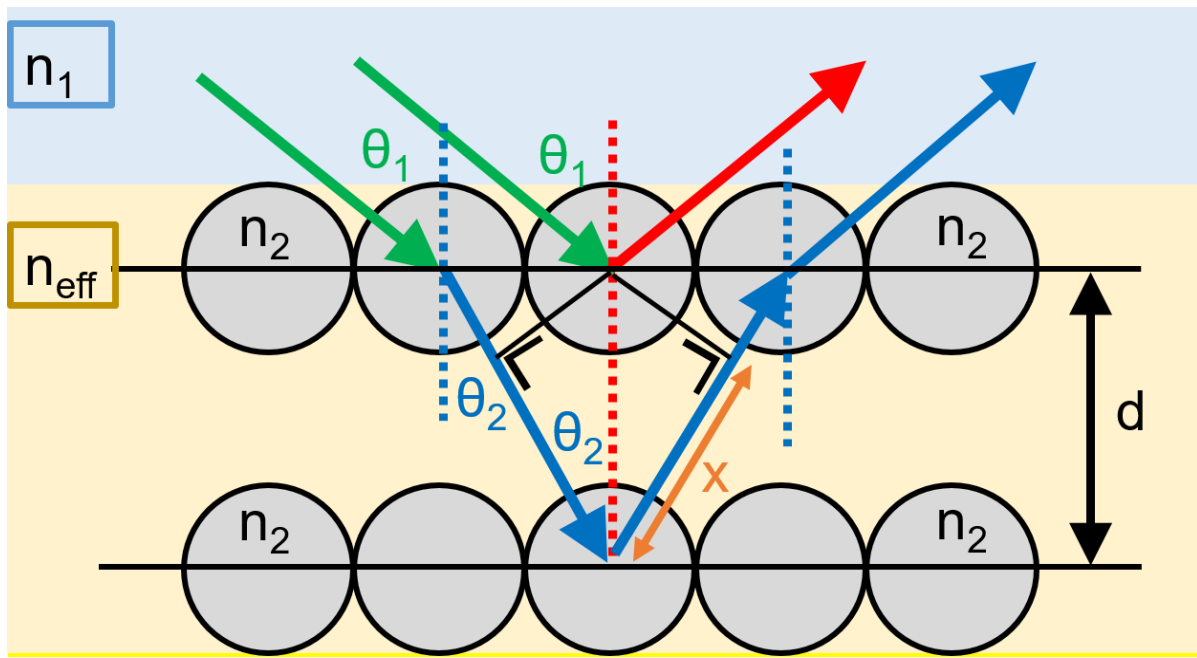

**Figure A.1** Schematic diagram showing diffraction from crystal planes in photonic crystal structures as per the Bragg-Snell relation.

With reference to Fig. A.1, two reflected waves (red and blue arrows) from the photonic crystal surface will be in phase and interfere constructively if the path difference of the refracted wave through the medium is an integer number of wavelengths. The path difference created by refraction is labelled $x$ and the light wave travels this distance twice. For some integer $m$ at a wavelength $\lambda'$, the condition for constructive interference in the reflected wave, as per thin film interference, is then:

$$m\lambda' = 2x \qquad (A.1)$$

The path difference $x$ can be found by taking the cosine of the refracted angle as:

$$x = d \cos \theta_2 \qquad (A.2)$$

It is very common to see the refractive index of the photonic crystal medium represented by $n_{eff}$ or $n_{avg}$, indicating the dependence of the refractive index in the photonic crystal region as some combination of refractive indices of the constituent materials (photonic crystal material + interstitial void space filling). There are different approaches used to estimate the effective refractive index for a photonic crystal medium in the literature. Two of the more common methods used to calculate this parameter for a two material photonic crystal medium (material 1 with refractive index $n_1$ and volume fraction $\varphi_1$ and material 2 with refractive index $n_2$ and volume fraction $\varphi_2$) are given by the models:

$$n_{eff} = n_1 \varphi_1 + n_2 \varphi_2 \qquad (A.3)$$

$$n_{eff} = \sqrt{n_1^2 \varphi_1 + n_2^2 \varphi_2} \qquad (A.4)$$

Unfortunately, there appears to be an unclear consensus on which method is most suitable for estimating $n_{eff}$ for a given system. Here for simplicity, a general $n_{eff}$ term will be used to derive the final form of the Bragg-Snell equation.

The wavelength $\lambda'$ corresponds to the wavelength of light in the photonic crystal medium as it passes through a refractive index $n_{eff}$. This can be related to the vacuum wavelength of light $\lambda$ exterior to the photonic crystal medium with:

$$\lambda' = \frac{\lambda}{n_{eff}} \qquad (A.5)$$

Substituting equations (A.2) and (A.6) back into equation (A.1) yields:

$$\frac{m \lambda}{n_{eff}} = 2 d \cos \theta_2 \qquad (A.6)$$

This can also be written as:

$$m \lambda = 2 d\, n_{eff} \cos \theta_2 \qquad (A.7)$$

This can be further modified using Snell's law to create a dependence on $\theta_1$, which is more easily measured/controlled than $\theta_2$, with Snell's law giving:

$$\sin \theta_2 = \frac{n_1}{n_{\text{eff}}} \sin \theta_1 \qquad (A.8)$$

Squaring both sides of the equation and subtracting them from 1 allows equation (A.8) to be rewritten as:

$$1 - \sin^2 \theta_2 = 1 - \frac{n_1^2}{n_{\text{eff}}^2} \sin^2 \theta_1 \qquad (A.9)$$

Recognising the left-hand side of equation (A.9) as being equivalent to $\cos^2 \theta_2$, we now have:

$$\cos \theta_2 = \sqrt{1 - \frac{n_1^2}{n_{eff}^2} \sin^2 \theta_1} \qquad (A.10)$$

Equation (A.10) is often further simplified as:

$$\cos \theta_2 = \frac{1}{n_{\text{eff}}} \sqrt{n_{\text{eff}}^2 - n_1^2 \sin^2 \theta_1} \qquad (A.11)$$

Substitution of this form of $\cos \theta_2$ into equation (A.7) creates a constructive interference condition for reflected light as:

$$m\,\lambda = 2\,d\,\sqrt{n_{\text{eff}}^2 - n_1^2 \sin^2 \theta_1} \qquad (A.12)$$

Equation (A.12) is a general form for the maximum reflected constructive interference wavelength in photonic crystal structures which depends on interplanar spacing $d$, composite refractive index in the photonic crystal medium $n_{\text{eff}}$, refractive index of the surrounding medium $n_1$ and the angle of incidence for light entering the structure. For opal and inverse opal structures the interplanar spacing of the (111) reflection, the most prominent reflection at normal incidence, using the unit cell parameter can be found as:

$$d_{hkl} = \frac{a}{\sqrt{h^2 + k^2 + l^2}} \qquad (A.13)$$

$$d_{111} = \frac{\sqrt{2}\,D}{\sqrt{3}} \qquad (A.14)$$

The parameter $D$ represents the centre-to-centre distance between adjacent spheres/IO pores for photonic crystal structures. For reflections from the (111) plane in FCC structured photonic crystals the condition for constructive interference can be written as:

$$m\lambda = \sqrt{\frac{8}{3}} D \sqrt{n_{\text{eff}}^2 - n_1^2 \sin^2\theta_1} \tag{A.15}$$

This form of the constructive interference condition is often presented when working with photonic crystal structures. If the photonic crystal is placed in an air medium ($n_1 = 1$) and a first order resonance ($m = 1$) is being measured, this equation can also be seen represented as:

$$\lambda = 1.633 D \sqrt{n_{eff}^2 - \sin^2\theta_1} \tag{A.16}$$

**Acknowledgments**

We acknowledge support from the Irish Research Council Government of Ireland Postgraduate Scholarship under award no. GOIPG/2016/946. We also acknowledge funding from the Irish Research Council Advanced Laureate Award under grant no. IRCLA/2019/118.